\documentclass{scrartcl}

\usepackage[utf8]{inputenc}
\usepackage[T1]{fontenc}
\usepackage{authblk}

\usepackage{graphicx}
\usepackage{amsmath,amssymb,amsthm}
\usepackage{esint} 
\usepackage{enumitem,color}
\usepackage{comment}
\usepackage{hyperref}
\usepackage{makecell}
\numberwithin{equation}{section}

\usepackage{xstring}

\newtheorem{thm}{Theorem}

\theoremstyle{definition}

\theoremstyle{remark}
\newtheorem{rem}[thm]{Remark}

\newcommand{\fig}[2]{\includegraphics[width=#1\textwidth]{#2}}

\newcommand{\crit}{\hbox{\tiny cr}}
\newcommand{\gs}{\gamma_{\hbox{\tiny{S}}}}
\newcommand{\gsp}{\gamma'_{\hbox{\tiny{S}}}}

\newcommand{\Proba}{\mathrm{Pr}}
\newcommand{\SF}{\mathcal S}
\newcommand{\tc}{t_{\crit}}
\newcommand{\gc}{g_{\crit}}
\newcommand{\Mc}{M_{\crit}}
\newcommand{\uc}{u_{\crit}}
\newcommand{\block}{\hbox{\tiny block}}
\newcommand{\rhob}{\rho_{\block}}
\newcommand{\rhobb}{\bar{\rho}_{\block}}

\DeclareMathOperator{\Tr}{Tr}

\title{Liouville Quantum Duality \\ and Random Planar Maps II}
\author[$$]{Bertrand Duplantier\thanks{bertrand.duplantier@ipht.fr}}
\author[$$]{Emmanuel Guitter\thanks{emmanuel.guitter@ipht.fr}}
\affil[$$]{{\normalsize Université Paris-Saclay, CEA, CNRS, Institut de Physique
    Théorique, \par 91191 Gif-sur-Yvette,  \textsc{France} }}

\date{\today}

\begin{document}
\maketitle 
\begin{abstract} 
This is Part II of our project on block-weighted planar maps and Liouville quantum duality. 
Focusing on the scaling properties at the dual critical point, 
we derive the conditional distribution of the root block size given the total size, as well as, conversely, the distribution 
of the total size for a fixed root block size. We show that these laws are in perfect agreement with the results of 
Liouville quantum gravity (LQG), obtained by modifying the standard \mbox{Liouville} random measure with additional \emph{atomic} 
contributions representing localized quantum areas.
 The ratio of dual and direct partition functions with punctures is shown to be universal, its explicit LQG expression
exactly matching its combinatorial analogue.
We also investigate the block \emph{distance} profile for doubly rooted maps, which is here rigorously related 
to the distance profile of maps consisting of a single block. 
Finally, we analyze the \emph{multifractal} properties of the usual and dual Liouville measures, predicting
the associated spectra, from both quantum and Euclidean perpectives. 
We illustrate our results through specific realizations of block-weighted planar maps, \emph{i.e.}, quadrangulations 
decomposed into simple blocks, tree-like structures formed by attaching quartic maps, and bicubic maps decomposed 
into 3-connected blocks. For each model, we give the single non-universal constant which uniquely determines the strength of the corresponding atomic Liouville measure.
  \end{abstract}

\section{Introduction}
\label{sec:intro}
In a recent work \cite{FS24}, Fleurat and Salvy identified a remarkable phase transition in so-called \emph{block-weighted planar maps}, a general class of models where maps are decomposed into elementary components (blocks) connected through bottlenecks and forming a tree-like structure. Various block decomposition schemes for several families of maps have been studied \cite{ZS23,ZSPhD,AFZ24,DG25}, including maps decorated by statistical models, 
and it was found that a universal pattern emerges when assigning a weight $u$ per block.

For maps of large size $n$, three regimes appear. In the subcritical regime ($u<u_{\crit}$), the map contains a single macroscopic block of size $\mathcal{O}(n)$, completed by finite outgrowths made of smaller blocks. As $u$ increases, the size of
the macroscopic block decreases until one reaches the critical regime (at $u=u_{\crit}$), where the largest blocks 
have size $\mathcal{O}(n^{\alpha'})$ with $\alpha'<1$. In the supercritical regime ($u>u_{\crit}$), all blocks become small, of size $\mathcal{O}(\log n)$, and the map behaves essentially as a Continuum Random Tree (CRT) built out of these small blocks.

It was then recognized in \cite{DG25} that a number of universal relations could be established between properties in the subcritical and critical regimes, which coincide precisely with the \emph{duality} predictions from Liouville quantum gravity (LQG).
Recall that LQG arises in the study of random two-dimensional geometries \cite{MR623209}, with deep connections with conformal
field theory, random matrix models \cite{Ginsparg-Moore,MR1320471} and Schramm-Loewner evolution (SLE) \cite{MR3551203,DMS14}, in particular through the Knizhnik-Polyakov-Zamolodchikov  (KPZ) relation \cite{MR947880}, relating critical exponents in the plane and on a two-dimensional random surface \cite{MR981529,MR1005268,2008arXiv0808.1560D,PSS:8474530}. 

Standard LQG comes with a parameter $\gamma \leq 2$, associated with 
the central charge $c \leq 1$ of the underlying critical theory, and describes the continuous limit of random surfaces 
with a non-positive string susceptibility exponent, $\gs=1-4/\gamma^2$, as those constructed from standard matrix models. 
An extension of this construction to \emph{modified matrix models} \cite{1990MPLA....5.1041D,1992PhLB..286..239J,1992PhLB..296..323K,1992MPLA....7.3081K,1993NuPhB.394..383A,1994NuPhB.426..203D,1994MPLA....9.1221A} revealed the existence of
a different regime for random surfaces, now with positive string susceptibility exponent, associated with the 
proliferation of \emph{``baby universes,''}  \emph{i.e.}, small components connected through bottlenecks. 
A formal extension of LQG with parameter $\gamma'>2$ (the ``dual branch'') was proposed to describe this new regime  \cite{1995PhRvD..51.1836K,1995NuPhB.434..264K,1995NuPhB.440..189B,1996NuPhS..45..135K} and,
even though it initially lacked a rigorous measure-theoretic foundation, this led to conjectured \emph{duality relations between critical exponents} of $\gamma-$LQG and ${\gamma'}-$LQG for $\gamma'=4/\gamma$, describing two regimes of
decorated random maps associated with the same central charge  $c \leq 1$. This duality was rediscovered independently in the context of SLE$_\kappa$ exponents and Liouville quantum gravity, with the correspondence of SLE and LQG  parameters, $\kappa=\gamma^2 \leq 4$ and $\kappa'=\gamma'^2=16/\kappa\geq 4$ \cite{2000PhRvL..84.1363D,MR2112128}.  
 
A mathematical construction of the dual regime ${\gamma'}-$LQG with $\gamma'=4/\gamma$ was proposed by
modifying the standard $\gamma-$LQG measure with additional \emph{atomic} contributions, representing \emph{localized} quantum area \cite{2009arXiv0901.0277D,2008ExactMethodsBD,pre06228485}. In this framework, duality relations between critical exponents arose naturally, as well as an extension of the KPZ relation to this dual regime \cite{2009arXiv0901.0277D,2008ExactMethodsBD}, made rigorous in \cite{pre06228485}.

In this context, the goal of Ref.~\cite{DG25}, of which the present paper is a continuation, was to show that block-weighted 
maps in their sub-critical and critical regimes provide 
precisely a realization of Liouville duality. More precisely, this article aimed to establish, by standard methods of
analytic combinatorics, universal relations between the 
critical exponents in the subcritical and critical regimes, and to show that these relations coincide with the duality predictions from Liouville quantum gravity.

In the present paper, we continue this study by now focusing on the joint distribution of the size $k$ of the root block and the
total size $n$ in block-weighted rooted maps at their critical point $u=\uc$, from now on called the \emph{dual critical point}. We derive 
in particular the universal law for $k$ at large fixed $n$, and conversely that of $n$ at large fixed $k$.
These distributions depend on the universality class of the maps at hand, characterized for instance by the exponent $\alpha$ describing the 
singular behavior of the generating function for maps made of a single block. As it turns out, the laws for the root block and total
size are in perfect agreement with those obtained via the above mentioned ``atomic'' LQG measure, 
thus providing a full justification of the  mathematical a priori construction of dual Liouville quantum gravity of Refs. \cite{2009arXiv0901.0277D,2008ExactMethodsBD,pre06228485}.

\medskip  
The paper is organized as follows:
Section~\ref{sec:generalities} recalls some basic facts about block-weighted planar maps. 
Section~\ref{sec:substitution} explains how the
combinatorial description of planar maps in terms of a root block and its outgrowths translates into a fundamental \emph{substitution relation}, Eq.~\eqref{eq:mapsubst}, between appropriate generating functions. The singularity analysis of these generating functions then allows us to identify different large size asymptotic regimes, in particular the dual critical point, which is the subject of our study in all the subsequent sections. 
We prove in Section~\ref{sec:singularity} a 
crucial relation, Eq.~\eqref{eq:ttogcrit},  which relates the distance from singularity measured in the variable $t$ conjugate to
the size $k$ of the root block to that measured in the variable $g$ conjugate to
the total size $n$ of the map. 

Section~\ref{sec:distkfixedn} computes the distribution for the root block size $k$ at fixed total size
$n$, first for simply rooted maps in Section~\ref{sec:simproot}, then for doubly rooted maps in the appropriate scaling
regime of large $k$ and $n$ in Section~\ref{sec:distkfixednscal}. The case of $p$-rooted maps is considered
in Section~\ref{sec:ratios}, where we obtain a remarkable relation, Eq.\eqref{eq:ratio}, for the ratio of the partition function of $p$-rooted maps at fixed $k$ to and that at fixed $n$. Section~\ref{sec:quadrangulations} illustrates all the previous results
by explicit calculations in the specific case of planar quadrangulations weighted according to their number of simple blocks.
 
Section~\ref{sec:distnfixedk} deals with the distribution for the total size $n$ at fixed root block size $k$ in the 
appropriate scaling regime at the dual critical point. It culminates with the Laplace transform \eqref{eq:limitlawbis} of the properly scaled ratio $n/k^\alpha$,  when the root block size $k$ tends to infinity, and with the fondamental exponent $\alpha:=\gamma'/\gamma=4/\gamma^2$. It is given by a universal stretched exponential, depending only on the Laplace conjugate variable, and a unique coefficient $D$ associated with the particular discrete model at hand. This is again illustrated by considering block-weighted quadrangulations.

We then address in Section~\ref{sec:distance} the question
of the \emph{block distance profile} of block-weighted maps, that is the distribution for the distance between two marked edges
in the same block. We obtain in Section~\ref{sec:distdual} a general \emph{convolution relation}, Eq.\eqref{eq:rho0rhob}, 
between the block distance profile
at the dual critical point and that for maps formed of a single block. We then verify in Section~\ref{sec:distdualquad}
this relation for the particular case of block-weighted quadrangulations, for which fully explicit formulas 
of the distance profiles are known. 

Section~\ref{sec:DASetal} presents a detailed analysis of 
the problem of \emph{tree structures made of attached quartic maps}, the model introduced by  
Das, Dhar, Sengupta and Wadia as early as 1990 in \cite{1990MPLA....5.1041D} as the first realization of a discrete
surface model displaying ``baby universes'' and in effect Liouville quantum duality. We revisit this model in Section~\ref{sec:DDSW} in a purely combinatorial way
and show how to incorporate it in our general formalism of block-weighted maps. We also reformulate it
in Section~\ref{sec:stuffed} as a model of \emph{stuffed maps}, for which we can compute the block distance
profile explicitly.

Section~\ref{sec:LQD} describes the main continuum properties of Liouville quantum duality. It starts with a brief  recall in 
Section~\ref{sec:LQM} of the construction of the Liouville quantum measure for $\gamma\leq 2$, followed by that of the notion of so-called \emph{quantum balls} \cite{2008arXiv0808.1560D} in  
 Section~\ref{QBS}. The next Section~\ref{sec:DQM} deals with the definition of the dual quantum measure for $\gamma'>2$ in terms of the original one for $\gamma <2$, via 
 the modification of the latter by a $(1/\alpha)-$stable subordinator of arbitrary strength $D$. Its main Laplace transform properties and its moments are studied in detail. 
 A fundamental continuum Laplace identity \eqref{Laplace4} is thus obtained, which exactly matches result \eqref{eq:limitlawbis} rigorously derived from the combinatorial approach. 

In Section~\ref{LQGsphere} the \emph{dual} Liouville partition functions in the sphere geometry and with multiple punctures are studied in detail with the help of the results of Ref. \cite{MR3465434}. We obtain in particular the universal ratio \eqref{ratiodual2} of the dual and direct, $p$-punctured, continuum partition functions, which remarkably exactly reproduces the combinatorial analogous relation \eqref{eq:ratio} mentioned above. 

The final Section~\ref{sec:multifractality} lays out the \emph{multifractal} properties \cite{1986PhRvA..33.1141H,PhysRevA.34.1601,FP} of the random Liouville measures. We start in   
 Section~\ref{Sec:1stspectrum} with those derived from the \emph{expected} moments of the $\gamma-$LQG measure for $\gamma<2$, reproducing here stronger rigorous \emph{almost sure} (\emph{a.s.}) results recently obtained by Bertacco \cite{10.1214/22-EJP893,Bertacco_2025}.  The \emph{dual expected} moments of the $\gamma'-$LQG measure for $\gamma'>2$ 
 are derived in Section~\ref{sec:secspectrum} from the rigorous results obtained in Section \ref{sec:DQM}, allowing us to predict all the \emph{a.s.}\ multifractal spectra of the dual Liouville measure.  

The last two sections deal with the \emph{reverse} perspective, obtained when looking at the multifractal properties of the random Euclidean balls associated with a given Liouville measure content. Section~\ref{Sec:gammaqballs} is concerned with the multifractality of $\gamma-$quantum balls for $\gamma<2$, whereas Section~\ref{sec:GammaprimeQballs} deals with that of $\gamma'-$quantum balls for $\gamma'>2$. Novel almost sure multifractal spectra are predicted in this way, which ultimately result from the rigorous validity of the KPZ relation, as applied to a \emph{continuum} of critical exponents. 

Appendix~\ref{app:Salpha} gives a number of properties of a special function appearing in our various
size distributions, a particular instance of the Wright function.

Appendix~\ref{app:bicubic} presents a study of block-weighted bicubic maps decomposed into $3$-connected blocks, 
yet another example for which explicit calculations can be done.

\section{Block-weighted maps at their dual critical point}
\label{sec:generalities}
\subsection{Substitution relation and dual critical point}
\label{sec:substitution}
Let us start with a few generalities about block-weighted planar maps. Recall that a \emph{planar map} is a finite
connected planar (multi-)graph embedded in the two-dimensional sphere without edge crossings, and 
considered up to continuous deformations. Various particular families of planar maps may be considered, 
subject to various additional requirements (\emph{e.g.}, on the degrees of the faces or vertices) and possibly decorated 
by extra degrees on freedom (loops, spins, hard particles, colors, ...). 
A map is characterized by its \emph{size} $n$, typically defined as its number of edges, or sometimes 
of faces or of vertices, with the implicit
requirement that the number of planar maps of fixed size $n$ is finite. 
A map is \emph{rooted} if it has a marked oriented edge, called the root edge.

\begin{figure}[h]
  \centering
  \fig{.6}{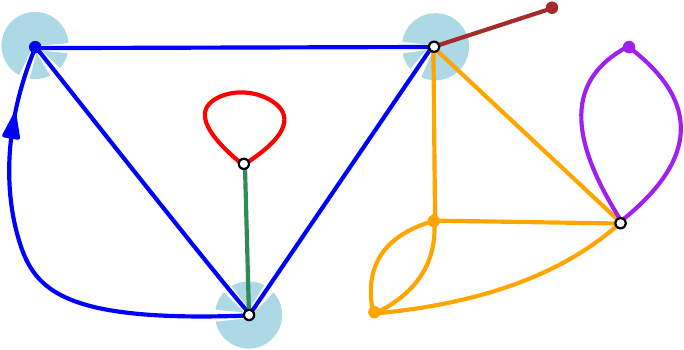}
   \caption{Decomposition of a rooted planar map into $6$ non-separable blocks. Separating vertices are indicated by small 
   circles and the different blocks are indicated by different colors.
   Here the root block size is $4$ (blue block made of $4$ edges) and there are $8$ positions around the root block (one in each of its corners, 
   in light blue) 
   where to attach 
   possible outgrowths. Here only $2$ of these positions are occupied by a non trivial outgrowth: that formed by the
    green and red blocks, and that 
   formed by the brown, orange and
    purple blocks. }
    \label{fig:blockgenmap}
\end{figure}

Many families of planar maps have a natural decomposition into \emph{blocks} which correspond to connected
sub-components of the map satisfying some restrictive property (say, of being 2-connected, or 3-connected, or simple, ...),
and whose sizes add up to the size of the map. 
The whole map is then recovered as a particular arrangement of these blocks, attached to each other by pinch 
points or small bottlenecks and organized into a global tree-like structure. 
As a guiding example, the reader may have in mind the case of \emph{general planar maps} whose size is 
measured by their \emph{number of edges}  and which are naturally decomposed into \emph{non-separable} (or 2-connected) blocks: 
recall that a vertex is a 
\emph{separating vertex} if its removal disconnects the map. A map with no separating vertex is called non-separable.
An arbitrary general planar map can be canonically decomposed into blocks formed by all its maximal 
non-separable sub-maps; see Figure~\ref{fig:blockgenmap} for an illustration.

Assuming that the map is rooted, its block decomposition selects a \emph{root block} (sometimes called the 
core of the map) which is the block containing the root edge. 
The other blocks are then arranged into a number of \emph{outgrowths} attached to the root block, each
outgrowth corresponding to a connected tree-like arrangement of blocks. 
We will call the \emph{root block size} of a map the size of its root block. In this paper, we will consider the 
archetypical situation of maps where (i) the number of positions where
outgrowths can be attached to the root block is equal to $2k$ if 
the root block size is $k$,  and where (ii) each outgrowths may itself be described as
a rooted map of the same type as the original decomposed map. This is the case for our guiding example of 
general planar maps decomposed into non-separable blocks: if the root block has $k$ edges, an outgrowth
can be attached at each of the $2k$ corners\footnote{Recall that a corner is the angular sector between two
consecutive edges around a vertex.} of the root block, and each outgrowth is itself a general map which can be
rooted canonically at its attachment point. Ref.~\cite{DG25} gives a number of explicit sample realizations
of other block decompositions satisfying (i) and (ii) above, for various map families, lying in various universality classes.

\medskip
The statistics of \emph{block-weighted planar maps} consists in considering (possibly decorated) planar maps together with 
a natural decomposition in blocks and in assigning \emph{a weight $u$ per block}.
Following \cite{DG25}, we denote by $m_n^{(u)}$ the generating function of rooted decorated 
maps of size $n$, 
enumerated with a weight $u$ per block, and with a conventional first term $m_0^{(u)}=1$ which we interpret
as accounting for an ``empty'' trivial map: these quantities are 
gathered in the generating function $$M_u(g)=\sum_{n\geq 0}m_n^{(u)} g^n\, .$$  Consider now the generating function 
$$B(t)=1+\sum_{k\geq 1}b_k t^k\, ,$$ where $b_k$ enumerates rooted decorated maps 
\emph{formed of a single block} of size $k$. The above description of maps in terms of their root block and outgrowths
translates into the following 
fundamental \emph{substitution relation} between the generating functions $M_u$ and $B$:
\begin{equation}
M_u(g)-1= u\left[B\left(g\, M_u(g)^2\right)-1\right]\ .
\label{eq:mapsubst}
\end{equation}
This relation simply states that a rooted (non empty) map, enumerated by $M_u(g)-1$, can be bijectively described by the data 
of its root block, enumerated by $u\, b_kg^k (k\geq 1)$ if it has size $k$, and of $2k$ possibly empty 
attached outgrowths, each enumerated by $M_u(g)$, hence the right-hand-side, equal to 
$\sum_{k\geq 1} u\, b_k g^k (M_u(g))^{2k}$.  
\medskip
We shall now assume that the generating function $B(t)$ for maps made of a single block is singular at 
$t=\tc$, with the expansion
\begin{equation}
B(t)=B(\tc)-(\tc-t)B'(\tc)+K_B(\tc-t)^\alpha+ o((\tc-t)^\alpha)
\label{eq:Bsing}
\end{equation}
where $1<\alpha<2$. This assumption simply amounts to demanding that the numbers $b_k$ 
have a large $k$ asymptotics of the form
\begin{equation}
b_k \sim c \frac{\tc^{-k}}{k^{1+\alpha}}\ ,\quad c= \frac{\alpha (\alpha-1) K_B \tc^\alpha}{\Gamma(2-\alpha)}\ ,
\label{eq:asympk}
\end{equation}
which is indeed a quite common behavior for decorated maps. By analyzing the substitution relation \eqref{eq:mapsubst}, 
it was shown in \cite{DG25} that the above singular behavior \eqref{eq:Bsing} induces a singular behavior for $M_u(g)$
at the value $g=\gc(u)$ such that $\gc(u)M_u(\gc(u))^2=\tc$, namely
\begin{equation}
\gc(u)=\frac{\tc}{(1+u(B(\tc)-1))^2}\ ,
\label{eq:guB}
\end{equation}
as long as $u$ is not larger than the critical value $\uc$, fixed by
\begin{equation}
\uc=\frac{1}{1-B(\tc)+2 \tc\, B'(\tc)}\ .
\label{eq:equcrit}
\end{equation}
This singular behavior implies in turn the large $n$ asymptotics for $m_n^{(u)}$:
\begin{align}
\label{eq:subcritical}&m_n^{(u)}\propto \frac{\gc(u)^{-n}}{n^{1+\alpha}}&\hbox{for\ \ } u<\uc\ ,\\
\label{eq:critical}&m_n^{(u)}\propto \frac{\gc(u)^{-n}}{n^{1+1/\alpha}} &\hbox{for\ \ } u=\uc\ .
\end{align}
As explained in \cite{DG25}, we read on the asymptotics \eqref{eq:subcritical} for $u<\uc$ the value of the 
\emph{string susceptibility exponent} $\gs$ of the theory via the identification $2-\gs=1+\alpha$, namely
\begin{equation}
\gs=1-\alpha\ ,
\label{eq:valgs}
\end{equation}
corresponding to the expected behavior for $\gamma-$LQG with $\gamma=2/\sqrt{\alpha}<2$
(so that $\gs=1-4/\gamma^2$). When $u=\uc$ instead, we read on the asymptotics \eqref{eq:critical}
another value for the string susceptibility exponent 
\begin{equation}
\gs'=1-\frac{1}{\alpha}\ ,
\label{eq:valgsp}
\end{equation}
corresponding now to the expected behavior for $\gamma'-$LQG with
$\gamma'=2\sqrt{\alpha}>2$ (so that $\gsp=1-4/\gamma'^2$). The passage from $\gamma-$LQG to $\gamma'-$LQG
with $\gamma\gamma'=4$ is 
precisely the so-called \emph{Liouville duality} and we will therefore refer to the $u=\uc$ situation as the \emph{dual 
critical point}.
The purpose of this paper is precisely to describe, in the map language, some properties of the dual critical point
and to compare them to similar results in the Liouville quantum gravity language.

The first property, already discussed in \cite{DG25} and immediately  read off \eqref{eq:valgs} and \eqref{eq:valgsp}, is 
the duality relation
\begin{equation}
(1-\gs)(1-\gs')=1\ 
\label{eq:dualgs}.
\end{equation}
We will discuss in Section \ref{LQGsphere} below how to obtain this relation in the LQG formalism.

\subsection{Singularity analysis at the dual critical point}
\label{sec:singularity}
From now on, we will concentrate our study to the dual critical point by setting $u=\uc$. 
We shall use the shorthand notations $\Mc(g):=M_{\uc}(g)$ and $\gc:=\gc(\uc)$,
so that, in particular, $\gc M^2_{\crit}(\gc)=\tc$. From \eqref{eq:equcrit} and the relation \eqref{eq:guB} taken 
at $u=\uc$, we also deduce (by eliminating $B(\tc)$) the alternative expression
\begin{equation}
\uc=\frac{1}{2 (\tc \gc)^{1/2} \, B'(\tc)}\ .
\label{eq:equcritbis}
\end{equation}
In order to estimate the large size asymptotics of various quantities, it proves useful to study the behavior 
when $g\to \gc$ of the quantity 
\begin{equation}
t=t(g):= g\, M^2_{\crit}(g)\ . 
\end{equation}
For $g\to \gc^-$, we have $t\to \tc^-$ and we may write
\begin{equation}
\begin{split}
g&=\frac{t}{M^2_{\crit}(g)}=\frac{t}{(1+\uc(B(t)-1))^2} \\
&=\frac{\tc}{(1+\uc(B(\tc)-1))^2}-\frac{\tc-t}{(1+\uc(B(\tc)-1))^3}
\overbrace{\left(1+\uc(B(\tc)-1-2 \tc B'(\tc)\right)}^{\overset{\eqref{eq:equcrit}}{\displaystyle{= }}\displaystyle{0}}\\
&\quad \quad \quad \quad \quad \quad \quad \quad \quad \quad \quad \quad \quad \quad \quad \quad \quad \quad-\frac{2 
K_B \tc \uc (\tc-t)^\alpha}{(1+\uc(B(\tc)-1))^3}+ o((\tc-t)^\alpha)\\
&=\gc-\frac{2 K_B \tc \uc}{(\tc/\gc)^{3/2}}(\tc-t)^\alpha+ o((\tc-t)^\alpha)
\end{split}
\end{equation}
where we used the identity $1+\uc(B(\tc)-1)=\Mc(g_c)=(\tc/\gc)^{1/2}$.
By inversion, this yield, using \eqref{eq:equcritbis},
\begin{equation}
\tc-t= C (\gc-g)^{\frac{1}{\alpha}} +o\left((\gc-g)^{\frac{1}{\alpha}} \right)\ , \quad
C=\left(\frac{\tc B'(\tc)}{\gc K_B}\right)^{\frac{1}{\alpha}}\ .
\label{eq:ttogcrit}
\end{equation}

\section{Distribution for the root block size at fixed total size}
\label{sec:distkfixedn}
\subsection{Simply rooted maps: discrete regime}
\label{sec:simproot}
Here we wish to discuss the law for the random root block size $X_n$ in rooted block-weighted maps \emph{of fixed size} $n$
at the dual critical point. These maps are enumerated by $m_n^{(\uc)}=[g^n]\Mc(g)$ and each rooted map 
$\mathcal{M}$ of size $n$ is drawn with probability $\uc^{\mathfrak{b}(\mathcal{M})}/ m_n^{(\uc)}$, where
$\mathfrak{b}(\mathcal{M})$ is its number of blocks.
For $n\geq 1$ and $k\geq 1$, we have the probability
\begin{equation}
\Proba(X_n=k)=\frac{\uc b_k [g^n]\left(g\, \Mc^2(g)\right)^k}{[g^n]\Mc(g)}
= \frac{b_k [g^n]\left(g\, \Mc^2(g)\right)^k}{[g^n]B\left(g\, \Mc^2(g)\right)}\ ,
\label{eq:probadef}
\end{equation}
since maps having block-root size $k$ are precisely enumerated by  $\uc b_k \left(g\, \Mc^2(g)\right)^k$.
Here we also used the relation $\Mc(g)=\uc B\left(g\, \Mc^2(g)\right)+1-\uc$ to go from the first to the second expression. 
From the expansion \eqref{eq:Bsing}, $B(t)=B(\tc)-(\tc-t)B'(\tc)+o(\tc-t)$,
we deduce, via \eqref{eq:ttogcrit}, that 
\begin{equation}
B\left(g\, \Mc^2(g)\right)=B(\tc)-B'(\tc) C(\gc-g)^{\frac{1}{\alpha}}+o\left((\gc-g)^{\frac{1}{\alpha}}\right)\ .
\end{equation}
This immediately leads to the large $n$ asymptotics 
\begin{equation}
[g^n]B\left(g\, \Mc^2(g)\right)\sim  \frac{\gc^{-n}}{n^{1+\frac{1}{\alpha}}} 
\times \frac{B'(\tc)\, C \gc^{\frac{1}{\alpha}}}{\alpha\, \Gamma\left(1-\frac{1}{\alpha}\right)}\ ,
\label{eq:saympM}
\end{equation}
where we recover the prediction \eqref{eq:critical} as we should.
Similarly, from the expansion $t^k= \tc^k-k\, \tc^{k-1}(\tc-t)+o(\tc-t)$, we get
\begin{equation}
\left(g\, \Mc^2(g)\right)^k=  \tc^k- k\, \tc^{k-1} C(\gc-g)^{\frac{1}{\alpha}}+o\left((\gc-g)^{\frac{1}{\alpha}}\right)
\end{equation}
which leads to the large $n$ asymptotics
\begin{equation}
[g^n]\left(g\, \Mc^2(g)\right)^k \sim  \frac{\gc^{-n}}{n^{1+\frac{1}{\alpha}}} 
\times \frac{ k\, \tc^{k-1}\, C \gc^{\frac{1}{\alpha}}}{\alpha\, \Gamma\left(1-\frac{1}{\alpha}\right)}\ .
\label{eq:saympgMk}
\end{equation}
From the estimates \eqref{eq:saympM} and \eqref{eq:saympgMk}, we obtain the discrete law for the root block
size $X_n$ at large $n$: 
\begin{equation}
\Proba(X_n=k)\underset{n \to \infty}{\rightarrow} p_k\ , \quad p_k = \frac{b_k\, k\, \tc^{k-1}}{B'(\tc)}\ .
\label{eq:probdiscret}
\end{equation}
In particular, we see that $\sum_{k\geq 1}p_k=1$, which means that the size of the root block is finite with probability $1$
when $n\to \infty$. This is expected from the following heuristic argument: as opposed to the case $u<\uc$ where the 
largest block has a size
proportional to $n$ (see for instance \cite{BFSS,FS24}), the largest blocks at $u=\uc$ have a size of order $n^{1/\alpha}$. The probability to have 
chosen a map having its root in such a given large block is of order $\mathcal O\left(n^{1/\alpha-1}\right)$ 
(since there are $\mathcal O(n)$ choices for the root edge in a map of size $n$) while the probability to have chosen a map
with root in an arbitrary finite block 
is of order $n\times (1/n) = \mathcal O(1)$ since the number of finite blocks
is of order $n$ and each contains the desired root edge with a probability of order $1/n$. Since $\alpha>1$, the former case vanishes at large $n$, hence the root block is finite with probability $1$ at the dual critical point.

\subsection{Doubly rooted maps: scaling regime}
\label{sec:distkfixednscal}
We may insist in capturing the contribution to $\Proba(X_n=k)$ of configurations of rooted maps with a large 
root block size, by letting $n\to \infty$ and $k\to \infty$ while keeping the ratio
\begin{equation}
x=k\, n^{-1/\alpha}
\label{eq:defx}
\end{equation}
fixed. We may write
\begin{equation}
[g^n]\left(g\, \Mc^2(g)\right)^k=\frac{1}{2\hbox{i}\pi}\oint \frac{dg}{g^{n+1}}\left(g\, \Mc^2(g)\right)^k\ ,
\label{eq:contourint}
\end{equation}
where the integration contour is a little circle around $g=0$.
\begin{figure}[h!]
  \centering
  \fig{.6}{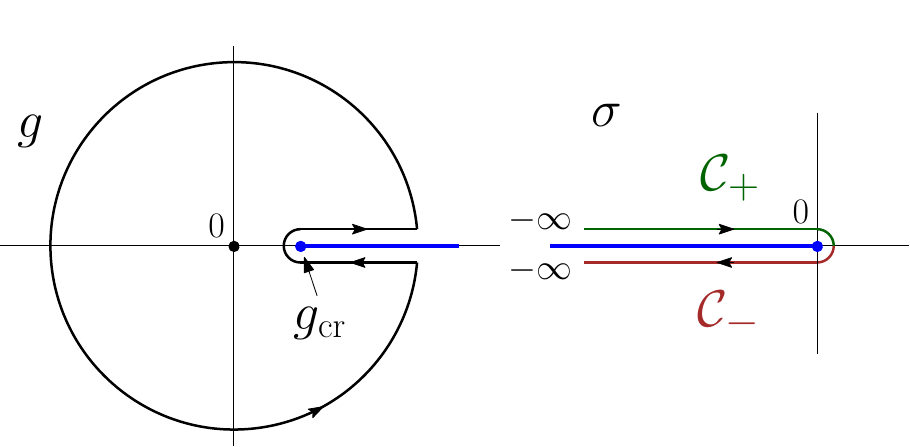}
   \caption{Left: in the integral \eqref{eq:contourint}, we deform the contour into a circle of radius strictly larger than $\gc$ (which contributes $0$ at large $n$) with a notch that comes back near and to the left of 
 $g=\gc$. Right: after the change of variable \eqref{eq:gtonu}, this contour in the $\sigma$ complex plane
 encircles the real half line $\Re(\sigma)\leq 0$.}
  \label{fig:contournu}
\end{figure}
Following \cite{FSbook}, we then deform this integration contour for $g$ into that displayed in Figure~\ref{fig:contournu}-left, and set
\begin{equation}
g=\gc \left(1-\frac{\sigma}{n}\right)\ ,
\label{eq:gtonu}
\end{equation}
with $\sigma$ running over the contour $\mathcal{C}_+\cup\mathcal{C}_-$ of Figure~\ref{fig:contournu}-right 
at large $n$. With this change of variable, we have
\begin{equation}
\begin{split}
\tc^{-k}\left(g\, \Mc^2(g)\right)^k&=\left(1-\frac{\tc-t}{\tc}\right)^k= \left(1-\frac{C}{\tc} (\gc-g)^{\frac{1}{\alpha}}+o\left((\gc-g)^{\frac{1}{\alpha}}\right)\right)^k\\
&\sim \left(1- \frac{C\gc^{\frac{1}{\alpha}}}{\tc}\times \frac{\sigma^{\frac{1}{\alpha}}}{n^{\frac{1}{\alpha}}}\right)^{x\, n^{\frac{1}{\alpha}}}
\underset{n\to \infty}\sim e^{-D\, \sigma^{\frac{1}{\alpha}}\, x}\\
\end{split}
\end{equation}
with $D= C\gc^{\frac{1}{\alpha}}/\tc$, namely
\begin{equation}
D=\frac{1}{\tc}\left(\frac{\tc B'(\tc)}{K_B}\right)^{\frac{1}{\alpha}}\ .
\label{eq:formD}
\end{equation}
We therefore get the equivalent
\begin{equation}
\tc^{-k} [g^n]\left(g\, \Mc^2(g)\right)^k  \underset{n\to \infty}\sim \frac{\gc^{-n}}{n} \frac{(-1)}{2\hbox{i}\pi} \int_{\mathcal{C}_+\cup\,\mathcal{C}_-}  \!\! 
d\sigma\, e^{\sigma -D\, \sigma^{\frac{1}{\alpha}}\, x}\ .
\label{eq:conti}
\end{equation}
 Setting
$\sigma=\mu e^{\mathrm{i}\pi}$ along $\mathcal{C}_+$ and $\sigma=\mu e^{-\mathrm{i}\pi}$
along $\mathcal{C}_-$, with $\mu$ real positive, we obtain finally
\begin{equation}
\begin{split}
\tc^{-k}[g^n]\left(g\, \Mc^2(g)\right)^k & \sim \frac{\gc^{-n}}{n} \frac{1}{\pi} \int_0^\infty
d\mu\, e^{-\mu} \frac{1}{2\mathrm{i}}\left(e^{ -D\, \mu^{\frac{1}{\alpha}}\, x\, e^{-\mathrm{i}\frac{\pi}{\alpha}}}-
e^{ -D\, \mu^{\frac{1}{\alpha}}\, x\, e^{\mathrm{i}\frac{\pi}{\alpha}}} \right)\\
&\sim  \frac{\gc^{-n}}{n}  \times \SF_{\frac{1}{\alpha}}(D\, x) \\
\end{split}
\label{eq:P1}
\end{equation}
with
\begin{equation}
\begin{split}
\SF_{\frac{1}{\alpha}}(x)&:=\frac{1}{\pi}\int_0^\infty d\mu e^{-\mu-\mu^{\frac{1}{\alpha}} x \cos\left( \frac{\pi}{\alpha} \right)}
\sin\left(\mu^{\frac{1}{\alpha}} x  \sin\left( \frac{\pi}{\alpha}\right) \right) \\
&= \frac{1}{\pi}\sum_{m\geq 1}(-1)^{m-1} x^m \frac{\Gamma\left(1+\frac{m}{\alpha}\right)}{m!} \sin\left( \frac{\pi m}{\alpha} \right)\ . \\
\end{split}
\label{eq:formS}
\end{equation}
We recover here the result of \cite[Eq.~(102) p.\ 709 with $\lambda \to 1/\alpha$]{FSbook}, where 
the scaling function $\SF_{\frac{1}{\alpha}}$ is related to the Wright function $W_{\lambda,\mu}$ by
$$\SF_{\frac{1}{\alpha}}(x)=W_{-\frac{1}{\alpha},0}(-x),\quad W_{\lambda,\mu}(z):=\sum_{m\geq 0}
\frac{z^m}{m!\, \Gamma(\lambda m+\mu)}\, .$$ 
Known properties of the function $\SF_{\frac{1}{\alpha}}$ are discussed in Appendix~\ref{app:Salpha}.
Note that this function satisfies $\SF_{\frac{1}{\alpha}}(x)=\mathcal{O}(x)$ at small $x$. Its 
large $x$ behavior is given in \eqref{eq:largx}.

Combining \eqref{eq:P1} with \eqref{eq:saympM} and \eqref{eq:asympk} in \eqref{eq:probadef}, with $k \sim x n^{\frac{1}{\alpha}}$, we 
get
\begin{equation}
\begin{split}
\Proba(X_n=\lfloor x\, n^{\frac{1}{\alpha}}\rfloor)&\sim\frac{1}{n} \times  \frac{\alpha (\alpha-1) 
K_B \tc^\alpha}{\Gamma(2-\alpha)}
\times  \frac{\alpha\, \Gamma\left(1-\frac{1}{\alpha}\right)}{B'(\tc)\, C \gc^{\frac{1}{\alpha}}}\times  \frac{1}{x^{1+\alpha}}
\SF_{\frac{1}{\alpha}}(D\, x)\\
&\sim \frac{1}{n}\tau(x)\ , \quad \tau(x)=\frac{\alpha^3 \Gamma\left(2-\frac{1}{\alpha}\right)}{\Gamma(2-\alpha)}\frac{1}{(D\, x)^{1+\alpha}}\SF_{\frac{1}{\alpha}}(D\, x)\ .\\
 \end{split}
\label{eq:probscaling}
\end{equation}
If we now wish to estimate the probability that $X_n>\epsilon\, n^{\frac{1}{\alpha}}$, we find
\begin{equation}
\begin{split}
\Proba(X_n >\epsilon\, n^{\frac{1}{\alpha}})&=\sum_{k>\epsilon\, n^{\frac{1}{\alpha}}}
\Proba(X_n=\lfloor x\, n^{\frac{1}{\alpha}}\rfloor)\\
&\underset{n\to \infty}\sim \frac{1}{n^{1-\frac{1}{\alpha}}}
\int_\epsilon^{\infty} \tau(x) dx\ .\\
\end{split}
\end{equation}
which is of order $n^{1/\alpha-1}$, hence is wiped out at large $n$. This is consistent
with our heuristic remark in Section~\ref{sec:simproot} that the root block size $X_n$ remains finite with probability $1$ 
at large $n$. Note also that, since $\SF_{\frac{1}{\alpha}}(D\, x)=\mathcal O(x)$ for small $x$, the above integral diverges
as $\epsilon^{1-\alpha}$ for $\epsilon \to 0$, in agreement with the fact that 
$\Proba(X_n >\epsilon\, n^{1/\alpha})$
should remain of $\mathcal O(1)$ at large $n$ if we let $\epsilon$ tend to $0$ as $n^{-1/\alpha}$.

\bigskip
\begin{figure}[h!]
  \centering
  \fig{1.}{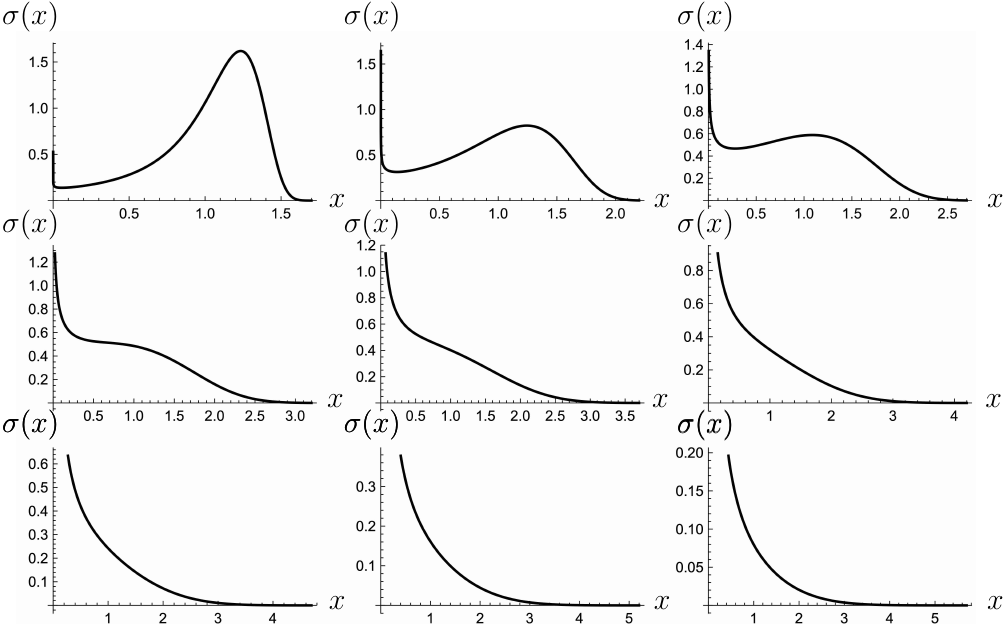}
   \caption{Plot of $\sigma(x)$ versus $x$ for $D=1$ and $\alpha=1+\frac{i}{10}$, $i=1,2,\ldots,9$ (from upper left to bottom right, line by line). The central plot corresponds to the $\alpha=3/2$ case.}
  \label{fig:Sigmaquad}
\end{figure}
A simple way to enhance the contribution of maps with a large root block size so as to obtain a proper probability
density is to consider the statistics of 
\emph{doubly-rooted} maps, that is rooted maps with an extra marked edge (different from the root edge, say) 
\emph{lying in the same block as the root edge}. Again we fix the total size $n$ and each doubly-rooted map 
$\mathcal{M}$ is drawn with a probability proportional to $\uc^{\mathfrak{b}(\mathcal{M})}$, where
$\mathfrak{b}(\mathcal{M})$ is its number of blocks.
The law for the root block size $X^{(2)}_n$ in doubly-rooted block-weighted maps of fixed size $n$
at the dual critical point is now obtained from the following expression, valid for
$n\geq 1$ and $k\geq 1$:
\begin{equation}
\Proba(X^{(2)}_n=k)
= \frac{(2k-1) b_k [g^n]\left(g\, \Mc^2(g)\right)^k}{[g^n]C^{(2)}\left(g\, \Mc^2(g)\right)}
\label{eq:probadef2}
\end{equation}
where \cite[Eq.~(6.10)]{DG25}
\begin{equation}
C^{(2)}(t)=1+2t B'(t)-B(t)
\label{eq:Coft}
\end{equation}
is the generating function for \emph{doubly-rooted maps\footnote{Here we assume 
that, in such a the rooted map of size $k$ (made of a single block), there are 
exactly $2k-1$ possible choices for the second marked edge. This is indeed the case for all the 
examples discussed in \cite{DG25} and here.} formed of a single block} (with by convention no weight $\uc$ for this single block). 
Using now the expansion
$C^{(2)}(t)=C^{(2)}(\tc)-2 \tc K_B \alpha (\tc-t)^{\alpha-1}+o((\tc-t)^{\alpha-1})$, we obtain that
\begin{equation}
C^{(2)}\left(g\, \Mc^2(g)\right)=C^{(2)}(\tc)-2 \tc K_B \alpha C^{\alpha-1}(\gc-g)^{1-\frac{1}{\alpha}}+o\left((\gc-g)^{1-\frac{1}{\alpha}}\right)
\end{equation}
and therefore
\begin{equation}
[g^n]C^{(2)}\left(g\, \Mc^2(g)\right)\sim \frac{\gc^{-n}}{n^{2-\frac{1}{\alpha}}} 
\times \frac{2 \tc K_B C^{\alpha-1} (\alpha-1)\gc^{1-\frac{1}{\alpha}}}
{\Gamma\left(\frac{1}{\alpha}\right)}\ .
\label{eq:asympC}
\end{equation}
Combining \eqref{eq:P1} with \eqref{eq:asympC} and \eqref{eq:asympk} with $k \sim x n^{\frac{1}{\alpha}}$, we 
eventually get for \eqref{eq:probadef2}, at large $n$, 
\begin{equation}
\Proba(X^{(2)}_n=\lfloor x\, n^{\frac{1}{\alpha}}\rfloor )\sim
\frac{1}{n^{\frac{1}{\alpha}}} 
\times  \frac{\alpha (\alpha-1) 
K_B \tc^\alpha}{\Gamma(2-\alpha)}
\times  \frac{\Gamma\left(\frac{1}{\alpha}\right)}{2 \tc K_B C^{\alpha-1} (\alpha-1)\gc^{1-\frac{1}{\alpha}}}\times  \frac{2}{x^{\alpha}}
\SF_{\frac{1}{\alpha}}(D\, x),
\end{equation}
hence
\begin{equation}
 n^{\frac{1}{\alpha}} \Proba(X^{(2)}_n=\lfloor x\, n^{\frac{1}{\alpha}}\rfloor )\underset{n \to \infty}{\rightarrow} \sigma(x)\ , \quad \sigma(x)=\frac{\alpha \Gamma\left(\frac{1}{\alpha}\right)}{\Gamma(2-\alpha)}\frac{D}{(D\, x)^{\alpha}}\SF_{\frac{1}{\alpha}}(D\, x)
\label{eq:probscaling2}
\end{equation}
with $D$ as in \eqref{eq:formD} and $\SF_{\frac{1}{\alpha}}$ as in \eqref{eq:formS}. We therefore find that, for large $n$,
the probability that $X^{(2)}_n$ lies in the range $[n^{\frac{1}{\alpha}}x, n^{\frac{1}{\alpha}}(x+dx)]$ is given by
$\sigma(x)dx$. It is easily checked that $\int_0^\infty \sigma(x)dx=1$ (see \eqref{eq:normal}) so that $\sigma$ 
is a probability density.
This in turn tells us that, for doubly-rooted maps of large size $n$, the two marked edges lie in a large block
of size $\propto n^{1/\alpha}$ with probability $1$. This is consistent with the following heuristic argument:
for a given map of size $n$, the probability that we chose the two marked edges in a given large block is proportional to
$(n^{1/\alpha}/n)^2=n^{2/\alpha-2}$ while the probability that we chose the two marked edges 
in any small block is proportional to $n\times(1/n)^2=n^{-1}$. Since $-1<2/\alpha-2<0$ for $\alpha<2$, the 
former situation is selected at large $n$.
Figure~\ref{fig:Sigmaquad} presents plots of the probability densities $\sigma$ for $D=1$ and $\alpha=1.1,1.2,\ldots,1.9$.

\subsection{Fixed-size partition functions}
\label{sec:ratios}
We introduced in \eqref{eq:Coft} the generating function $C^{(2)}(t)$ for doubly-rooted maps formed of a single block.
This is the first term of a  general family of generating functions $C^{(p)}(t)$, $p\geq 2$, 
for \emph{$p$-rooted maps formed of a single block}, \emph{i.e.}, maps with $p$ distinct marked edges (the first one being oriented).
We have
\begin{equation}
C^{(p)}(t)= \sum_{k>{\lfloor\frac{p-1}{2}\rfloor}} (2k-1)(2k-2)\cdots(2k-p+1) b_k t^k
\end{equation}
and, from \eqref{eq:Bsing}, the singular part of $C^{(p)}(t)$ for $t\to \tc$ reads
\begin{equation}
C^{(p)}(t)\Big\vert_{\hbox{\tiny{sing.}}}=(-2)^{p-1}\tc^{p-1}K_B\, \alpha(\alpha-1)\cdots(\alpha-p+2)(\tc-t)^{\alpha-p+1}\ .
\end{equation}
Let us introduce the \emph{fixed size partition function} $Z^{(p)}_k$ of $p$-rooted maps formed of a single block of size $k$, defined
as the quantity  $Z^{(p)}_k:=[t^k]C^{(p)}(t)$. We deduce from the above singularity the asymptotic behavior 
\begin{equation}
Z^{(p)}_k\underset{t\to \infty}{\sim} \frac{\tc^{-k}}{k^{\alpha-p+2}}(-2)^{p-1}\tc^{p-1}K_B\, \alpha(\alpha-1)\cdots(\alpha-p+2)\times
\frac{1}{\Gamma(p-1-\alpha)}\tc^{\alpha-p+1}\ .
\label{eq:Zasymp}
\end{equation}
Returning now to maps with an arbitrary number of blocks, \emph{at the dual critical point $u=\uc$}, we may consider
similarly the generating function $\widetilde{Z}^{(p)}_n$ of maps with $p$ distinct marked edges (the first one being oriented) \emph{in the same (root) block} 
 and with a fixed total size $n$. We have the simple substitution relation
\begin{equation}
\widetilde{Z}^{(p)}_n=[g^n]C^{(p)}\left(g\, \Mc^2(g)\right)
\end{equation} 
where $C^{(p)}\left(g\, \Mc^2(g)\right)$ is now singular at $g=\gc$ with, from \eqref{eq:ttogcrit},
\begin{equation}
C^{(p)}\left(g\, \Mc^2(g)\right)\Big\vert_{\hbox{\tiny{sing.}}}=(-2)^{p-1}\tc^{p-1}K_B\, \alpha(\alpha-1)\cdots(\alpha-p+2)C^{\alpha-p+1}(\gc-g)^{\frac{\alpha-p+1}{\alpha}}\ .
\end{equation}
We now deduce from this singularity the asymptotic behavior
\begin{equation}
\begin{split}
\widetilde{Z}^{(p)}_n\underset{n\to \infty}{\sim} \frac{\gc^{-n}}{n^{1+\frac{\alpha-p+1}{\alpha}}}(-2)^{p-1}\tc^{p-1}K_B\, 
&\alpha(\alpha-1)\cdots(\alpha-p+2)\\
&\qquad \qquad \times C^{\alpha-p+1}
\frac{1}{\Gamma\left(\frac{p-1-\alpha}{\alpha}\right)}\gc^{\frac{\alpha-p+1}{\alpha}}\ .\\
\end{split}
\label{eq:Zprimeasymp}
\end{equation}
Comparing \eqref{eq:Zasymp} with \eqref{eq:Zprimeasymp}, we obtain the asymptotic ratios for $p\geq 2$,
\begin{equation}
\frac{\gc^n\, \widetilde{Z}^{(p)}_n}{\tc^k\, Z^{(p)}_k}\underset{n,k\to \infty}{\sim} \mathcal{R}^{(p)}(n,k):=\frac{k}{n}\left(\frac{n^{1/\alpha}}{D\, k}\right)^{p-1-\alpha} \frac{\Gamma(p-1-\alpha)}{\Gamma\left(\frac{p-1-\alpha}{\alpha}\right)}
\label{eq:ratio}
\end{equation}
with $D= C\gc^{\frac{1}{\alpha}}/\tc$ as in \eqref{eq:formD}.

\subsection{The example of block-weighted quadrangulations} 
\label{sec:quadrangulations}
\begin{figure}[h]
  \centering
  \fig{.8}{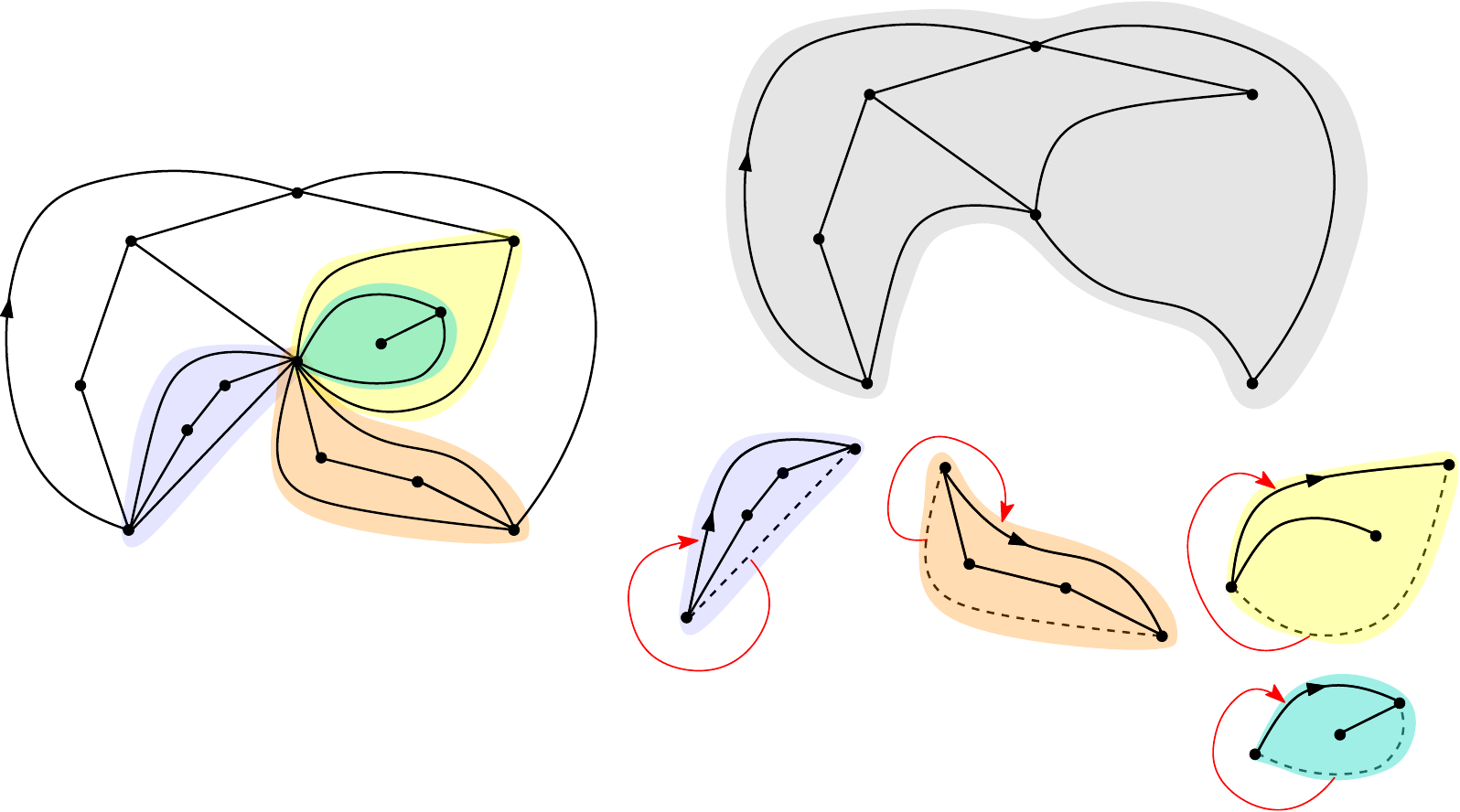}
   \caption{Decomposition of a rooted planar quadrangulation (left) into $5$ blocks which are rooted planar simple 
   quadrangulations. The blocks are obtained by cutting out recursively the maximal cycles of length $2$ 
   and transforming the extracted rooted maps with a boundary of length $2$ into rooted quadrangulations
   by identifying their two sides (red arrows); see \cite{DG25} for details.}
    \label{fig:blockquad}
\end{figure}

\begin{figure}[h!]
  \centering
  \fig{1.}{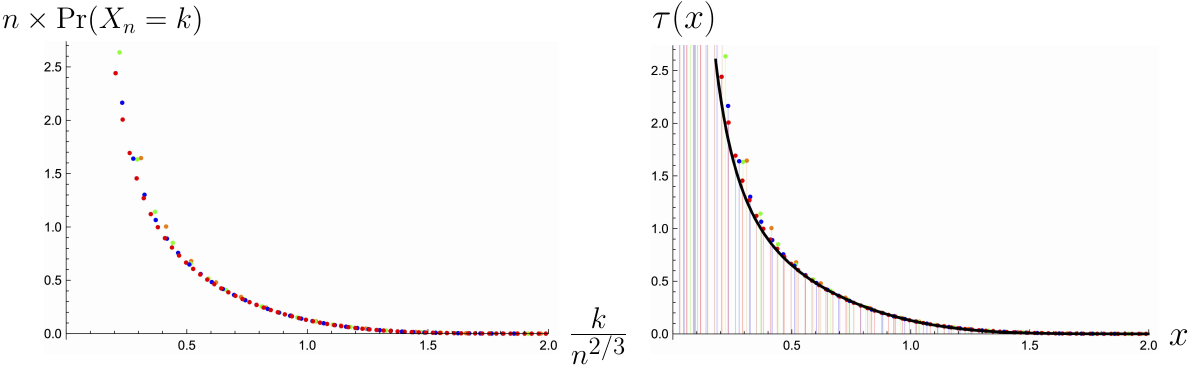}
   \caption{Left: the properly rescaled probability $\Proba(X_n=k)$ for $n=30, 50, 100, 200$ (orange, green, blue, red)
   in the case of block-weighted quadrangulations. 
   The points all fall on a well  defined curve. Right: the comparaison with the scaling function $\tau(x)$
of \eqref{eq:probscaling} with $\alpha=3/2$ and $D=3/2^{2/3}$. Vertical bars under the discrete points have been added for  
better visualization. 
}
  \label{fig:Proba1Xquad}
\end{figure}

\begin{figure}[h!]
  \centering
  \fig{1.}{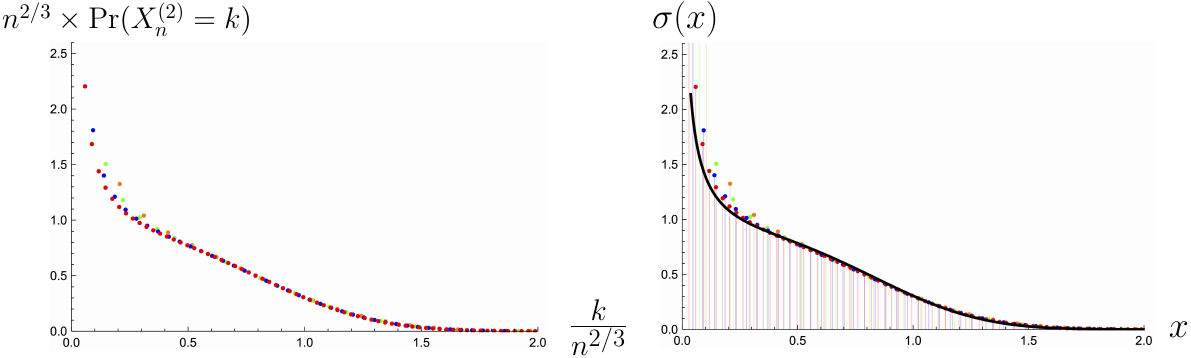}
   \caption{Left: the properly rescaled probability $\Proba(X^{(2)}_n=k)$ for $n=30, 50, 100, 200$ (orange, green, blue, 
   red) in the case of block-weighted quadrangulations. Right: the comparaison with the probability density $
   \sigma(x)$ of \eqref{eq:probscaling2} with $\alpha=3/2$ and $D=3/2^{2/3}$. }
  \label{fig:Proba2Xquad}
\end{figure}

Let us illustrate our results with the simple example introduced in \cite{FS24}, also studied in \cite{DG25},
of planar quadrangulations (maps whose all faces have degree $4$) weighted according to
their number of \emph{simple blocks}. Recall that these blocks are obtained by cutting the quadrangulation along its 
multiple edges in a canonical way: each of the cut pieces has outer degree $2$ and is transformed, by gluing its two external sides, into a rooted 
planar quadrangulation which is  
\emph{simple}, \emph{i.e.}, has no multiple edges (see \cite{FS24}, and Figure~\ref{fig:blockquad}, reproduced from 
\cite{DG25} for illustration). The generating function $M_u(g)$ for block-weighted quadrangulations
with a weight $g$ per face and $u$ per simple block satisfies the substitution relation \eqref{eq:mapsubst} where
\begin{equation}
B(t)=1+\sum_{k\geq 1} \underbrace{2\frac{(3k-3)!}{(2k-1)!k!}}_{\displaystyle{b_k}}t^k= 1+
\frac{9 t-2 \sqrt{3t} \sin \left(\frac{1}{3} \arcsin\left(\frac{3 \sqrt{3t}}{2}\right)\right)}{1+2 \cos
   \left(\frac{2}{3} \arcsin \left(\frac{3 \sqrt{3t}}{2}\right)\right)}
   \label{eq:Boft}
\end{equation}
is the generating function for planar rooted simple quadrangulations with a weight $t$ per face. In particular, the function \
$B(t)$ has a singular behavior of the form \eqref{eq:Bsing} with $\alpha=3/2$ and
\label{sec:bwquads}
\begin{equation}
\tc=\frac{4}{27}\ , \ \  B(\tc)=\frac{4}{3} \ , \ \ B'(\tc)=K_B=3 \ .
\end{equation}
We deduce in particular from these values that
\begin{equation}
 \uc=\frac{9}{5} \ ,\quad
 \gc=\frac{25}{432} \ , 
 \quad  C=\frac{2^4}{5^{4/3}}\ , 
\quad  D=\frac{3}{2^{2/3}} \ .
\label{eq:valguDquad}
\end{equation}
Figure~\ref{fig:Proba1Xquad} displays the properly rescaled probability $\Proba(X_n=k)$ for various
values of $n$ ranging from $n=30$ to $n=200$ and compares the result with the predicted scaling function $\tau(x)$
in \eqref{eq:probscaling} (with $\alpha=3/2$ and $D=3/2^{2/3}$). 
Similarly, Figure~\ref{fig:Proba2Xquad} displays the properly rescaled probability $\Proba(X^{(2)}_n=k)$ for the same values of $n$ and compares the result with the probability density $\sigma(x=k\, n^{-1/\alpha})$
in \eqref{eq:probscaling2}.

Note that, for $\alpha=3/2$, we have (see for instance \cite[Eq.~(16)]{math8060884})
\begin{equation}
\SF_{\frac{1}{\alpha}}(x)=2\frac{x}{3^{2/3}}e^{-\frac{2}{27}x^3} \left(\frac{x}{3^{2/3}}
   \text{Ai}\left(\frac{x^2}{3^{4/3}}\right)-
   \text{Ai}'\left(\frac{x^2}{3^{4/3}}\right)\right)=\frac{x}{3^{2/3}}\mathcal{A}\left(\frac{x}{3^{2/3}}\right)\ ,
\end{equation}
where $\mathcal{A}(x)$ is the probability density of the so-called standard Airy probability distribution introduced in \cite{BFSS}.
\section{Distribution for the total size at fixed root block size}
\label{sec:distnfixedk}
\begin{figure}[h!]
  \centering
  \fig{1.}{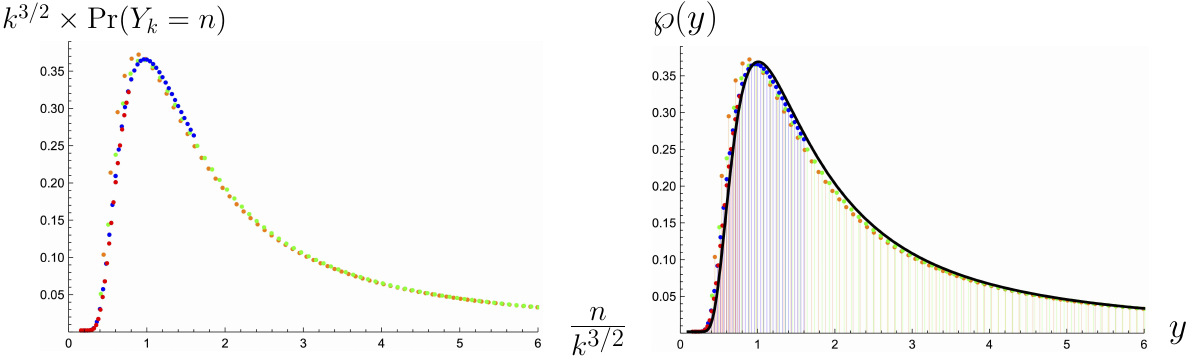}
   \caption{Left: the properly rescaled probability $\Proba(Y_k=n)$ for $k=5, 10, 25, 40$ (orange, green, blue, red) and 
   $n$ limited to $200$ in the case of block-weighted quadrangulations. These points form a well defined scaling curve. 
   Right: the comparison with the probability density $\wp(y)$ of \eqref{eq:probscaling3} for $\alpha=3/2$ and 
   $D=3/2^{2/3}$.}
  \label{fig:ProbaYquad}
\end{figure}

Let us now consider (simply-)rooted maps \emph{with a fixed root block size} $k$ and compute the law
for the random total size $Y_k$ of the map at the dual critical point. As, for fixed $k$, the value $n$ of the total size can 
now take arbitrarily large values, and since the number of maps of size $n$ typically grows like $\gc^{-n}$, 
we now draw a map $\mathcal{M}$ in the set of all rooted maps of fixed root block size $k$ with a probability proportional to
$\uc^{\mathfrak{b}(\mathcal{M})} \gc^{\mathfrak{n}(\mathcal{M})}$ where $\mathfrak{b}(\mathcal{M})$ is its number
of blocks and $\mathfrak{n}(\mathcal{M})$ its total size. In this ensemble, we then get
\begin{equation}
\Proba(Y_k=n)=\frac{(\gc)^n\ [g^n]\left(g\, \Mc^2(g)\right)^k}{\left(\gc\, \Mc^2(\gc)\right)^k}=(\gc)^n \tc^{-k} [g^n]\left(g\, \Mc^2(g)\right)^k\ .
\label{eq:probadef3}
\end{equation}
For large $k$, we are now interested in the scaling regime where $n$ scales as $k^\alpha$. We thus
set 
\begin{equation}
n=y\, k^\alpha
\label{eq:defy}
\end{equation}
with $y$ fixed.
Starting again from the contour integral representation \eqref{eq:contourint} with $g$ integrated
over the contour of Figure~\ref{fig:contournu}-left, we now set 
\begin{equation}
g=\gc \left(1-\frac{\sigma}{k^\alpha}\right)\ ,
\label{eq:gtonu2}
\end{equation}
with $\sigma$ running over the contour $\mathcal{C}_+\cup\mathcal{C}_-$ of Figure~\ref{fig:contournu}-right 
at large $k$. With this change of variable, we have
\begin{equation}
\begin{split}
\tc^{-k}\left(g\, \Mc^2(g)\right)^k&=\left(1-\frac{\tc-t}{\tc}\right)^k= \left(1-\frac{C}{\tc} (\gc-g)^{\frac{1}{\alpha}}+o\left((\gc-g)^{\frac{1}{\alpha}}\right)\right)^k\\
&\sim \left(1- \frac{C\gc^{\frac{1}{\alpha}}}{\tc}\times \frac{\sigma^{\frac{1}{\alpha}}}{k}\right)^{k}
\underset{k\to\infty}\sim e^{-D\, \sigma^{\frac{1}{\alpha}}}\\
\end{split}
\label{eq:nulambda}
\end{equation}
with $D$ as is \eqref{eq:formD}.
We now get the large $k$ equivalent
\begin{equation}
(\gc)^n \tc^{-k} [g^n]\left(g\, \Mc^2(g)\right)^k \underset{k\to\infty}\sim \frac{1}{k^\alpha} \frac{(-1)}{2\hbox{i}\pi} 
\int_{\mathcal{C}_+\cup\,\mathcal{C}_-}  d\sigma\, e^{\sigma y -D\, \sigma^{\frac{1}{\alpha}}}
=  \frac{1}{k^\alpha}  \times \frac{1}{y} \SF_{\frac{1}{\alpha}}\left(\frac{D}{y^{\frac{1}{\alpha}}}\right) 
\label{eq:P3}
\end{equation}
with $\SF_{\frac{1}{\alpha}}$ as in \eqref{eq:formS}. 
We end up wit the desired scaling distribution:
\begin{equation}
k^\alpha \Proba(Y_k=\lfloor y \, k^\alpha\rfloor )\underset{k \to \infty}{\rightarrow} \wp(y)\ , \quad \wp(y)=\frac{1}{y}\SF_{\frac{1}{\alpha}}
\left(\frac{D}{y^{\frac{1}{\alpha}}}\right)\ .
\label{eq:probscaling3}
\end{equation}
The function $\wp$ above is equivalently characterized by its Laplace transform (see \eqref{eq:invLaplace}):
\begin{equation}
\mathcal{L}\{\wp\}(\lambda):=\int_0^\infty dy\, \wp(y) e^{-\lambda\, y}=e^{-D\, \lambda^{\frac{1}{\alpha}}}.
\label{eq:Laplace}
\end{equation}
In particular, we have $\int_0^\infty dy \wp(y)=1$ so that $\wp$ is a probability density. Fig.~\ref{fig:ProbaYquad} displays the properly rescaled probability $\Proba(Y_k=n)$ and its limiting law \eqref{eq:probscaling3} in the case of block-weighted quadrangulations. 

\begin{figure}[h!]
  \centering
  \fig{1.}{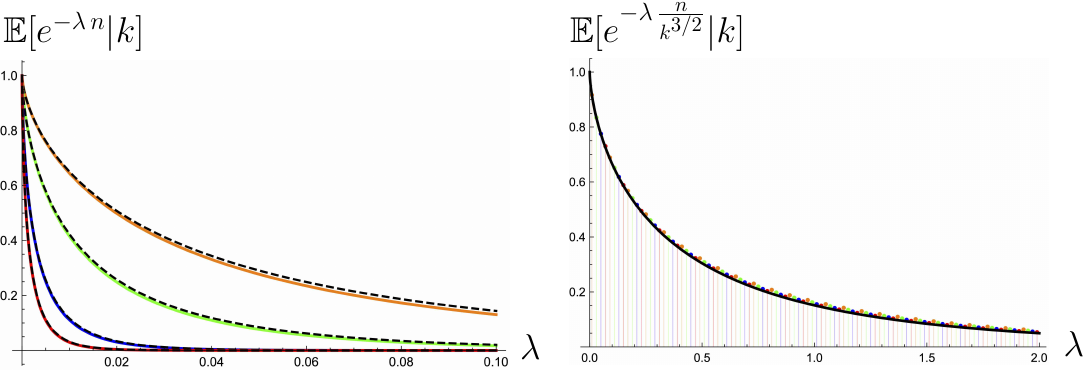}
   \caption{Left: plot of $E[e^{-\lambda\, n}|k]$ versus $\lambda$ (dashed lines) and the
   corresponding limiting estimates \eqref{eq:limitlaw} for $k=5, 10, 25, 40$ (orange, green, blue and red solid line) in the case 
   of quadrangulations ($\alpha=3/2$ and $D=3/2^{2/3}$). Right: plot of $E[e^{-\lambda\, \frac{n}{k^{3/2}}}|k]$ versus 
   $\lambda$ (dotted lines with vertical bars for a better visualization) for $k=5, 10, 25, 40$ (orange, green, blue and red)
 and the corresponding limiting law \eqref{eq:limitlawbis}.}
 \label{fig:Expoflambda}
\end{figure}

A related interesting quantity is the expectation value $\mathbb E[e^{-\lambda\, n}|k]:=\mathbb E[e^{-\lambda\, Y_k}]$. It is given by 
\begin{equation}
\mathbb E[e^{-\lambda\, n}|k] =\sum_{n\geq k}\Proba(Y_k=n) e^{-\lambda\, n}
=\tc^{-k}\left(\gc e^{-\lambda}\, \Mc^2(\gc e^{-\lambda})\right)^k\ .
\label{eq:lambdalaw}
\end{equation}
The property \eqref{eq:Laplace} allows us to estimate its large $k$ limit via
\begin{equation}
\mathbb E[e^{-\lambda\, n}|k] =\sum_{n\geq k}\Proba(Y_k=n) e^{-\lambda\, n}
\underset{k\to \infty}{\simeq} \int_0^\infty dy\, \wp(y) e^{-\lambda\, y \, k^\alpha}= e^{-D \lambda^{\frac{1}{\alpha}}k}\ .
\label{eq:limitlaw}
\end{equation}
Here we used on purpose the ill-defined symbol ``$\simeq$'' since the passage from the sum to the integral is not justified for a fixed $\lambda$. Still, as displayed in Figure~\ref{fig:Expoflambda}-left in the case of block-weighted quadrangulations, 
the right hand side gives a good estimate of the left hand side if $\lambda$ is small enough. 
A more precise statement can be made via the following, well-defined, limit:
\begin{equation}
\mathbb E[e^{-\lambda\, \frac{n}{k^\alpha}}|k] :=\mathbb E[e^{-\lambda\, \frac{Y_k}{k^\alpha}}]=\tc^{-k}\left(\gc 
e^{-\frac{\lambda}{k^\alpha}}\, \Mc^2(\gc e^{-\frac{\lambda}{k^\alpha}})\right)^k\underset{k\to \infty}{\to} 
e^{-D \lambda^{\frac{1}{\alpha}}}\ ,
\label{eq:limitlawbis}
\end{equation}
as obtained by repeating the argument that lead to \eqref{eq:nulambda} by replacing $\sigma$ by $\lambda$.

\begin{rem}
Suppose that, instead of fixing only the root block size $k$, we decide to impose a \emph{fixed root block} $\mathcal{B}$.
As before, we draw a map $\mathcal{M}$, now in the set of all rooted maps of fixed root block $\mathcal{B}$, 
with a probability proportional to
$\uc^{\mathfrak{b}(\mathcal{M})} \gc^{\mathfrak{n}(\mathcal{M})}$ where $\mathfrak{b}(\mathcal{M})$ is its number
of blocks and $\mathfrak{n}(\mathcal{M})$ its total size.  It is easily seen that
\begin{equation}
\begin{split}
& \mathbb E[e^{-\lambda\, \frac{n}{(\mathfrak{n}(\mathcal{B}))^\alpha}}|\mathcal{B}] 
=f(\mathfrak{n}(\mathcal{B}),\lambda)\\
&\hbox{with}\quad  f(k,\lambda):=\tc^{-k}\left(\gc e^{-\frac{\lambda}{k^\alpha}}\, \Mc^2(\gc e^{-\frac{\lambda}{k^\alpha}})
\right)^k\underset{k\to \infty}{\to} 
e^{-D \lambda^{\frac{1}{\alpha}}}\ ,
\end{split}
\label{eq:lawblockB}
\end{equation}
since only the size $k:=\mathfrak{n}(\mathcal{B})$ of the root block matters in the statistics for $n$. 
\end{rem}
\bigskip
\subsubsection*{Block-weighted quadrangulations}
In the case of block-weighted quadrangulations, the fonction $B(t)$ is algebraic, solution of the equation
\begin{equation}
B^3(t)-B^2(t)-18 t B(t)+27 t^2+16 t=0\ .
\end{equation}
We then deduce from \eqref{eq:mapsubst} that $\Mc(g)$ is algebraic, solution of 
\begin{equation}
  \begin{split}
27 g^2  \uc^3 \Mc^4(g)+(1-18 g \uc^2)\Mc^3(g) &+
   \left(g(18\uc^2-2\uc^3)+2 \uc-3\right)\Mc^2(g)\\
   & + (\uc-1)(\uc-3)\Mc(g)-(\uc-1)^2=0\\
   \end{split}
\end{equation}
where $\uc=9/5$, with $\Mc(g)=1+\mathcal O(g)$. This allows us to obtain a closed form for $\Mc(g)$, hence for 
$\Proba(Y_k=n)$ via \eqref{eq:probadef3} and for $ \mathbb E[e^{-\lambda\, n}|k]$ via \eqref{eq:lambdalaw}. Figure~\ref{fig:Expoflambda}-right
displays a plot of $\mathbb E[e^{-\lambda\, \frac{n}{k^\alpha}}|k]$ 
versus $\lambda$ and the corresponding limiting law \eqref{eq:limitlawbis} for values of $k$
from $k=5$ to $k=40$.

\section{Distance profile}
\label{sec:distance}
\subsection{Distance profile and duality}
\label{sec:distdual}
A quantity of interest, which characterizes the metric properties of maps,
is their asymptotic \emph{distance profile} defined as follows.
Considering doubly rooted maps as before, we may look at the \emph{graph distance} $\ell$ between
(the origin vertices of) their two marked edges. In the case of maps formed of a \emph{simple} block (as enumerated
by $C^{(2)}(t)$), it is expected that, for a large size $k$ of the map, the typical distance scales as $\ell\propto k^{1/d}$ for a Hausdorff dimension $d$ depending on the universality class at hand. 
Setting $r=\ell/k^{1/d}$, we define the distance profile as
\begin{equation}
\rho_0(r):=\lim_{k\to \infty} k^{1/d} \frac{[t^k] C^{(2)}_{\lfloor k^{1/d}\, r\rfloor}(t)}{[t^k]C^{(2)}(t)}\ ,
\label{zero51}
\end{equation}
where $C^{(2)}_\ell(t)$ enumerates those doubly rooted maps formed of a simple block for which the distance between the marked 
edges is $\ell$. Note that the quantity $C^{(2)}_\ell(t)$, as well as the Hausdorff dimension $d$, are not known
in general.

Consider now maps formed of possibly several blocks at the critical point $u=\uc$ and, more precisely, doubly rooted
such maps \emph{conditioned to have their two marked edges in the same (root) block}. We may then define
their \emph{block distance profile} as the asymptotic law for the distance between these two marked edges. For a large size $n$
of the map, the typical distance now scales as $\ell\propto n^{1/\widetilde{d}}$ for some 
new Hausdorff dimension $\widetilde{d}$, related to $d$ via \cite{DG25}
\begin{equation}
\widetilde{d}=\alpha\, d\ .
\label{ddtilde}
\end{equation}
Setting $r=\ell/n^{1/\widetilde{d}}$, the block distance profile may then be obtained via
\begin{equation}
\rhob(r)=\lim_{n\to \infty} n^{1/\widetilde{d}} \frac{[g^n] C^{(2)}_{\lfloor n^{1/\widetilde{d}}\, r\rfloor}
\left(g\, \Mc^2(g)\right)}{[g^n]C^{(2)}\left(g\, \Mc^2(g)\right)}\ .
\label{block53}
\end{equation}
Note that this expression implicitly assumes that the distance in the whole map between two vertices in the same (root) block
is the same as their distance within the root block. This simply expresses the expected fact (true for the models considered here)
that the length of a path between two vertices in the root block cannot get lowered by letting the path enter another block.
Otherwise stated, a geodesic between the vertices can be found, which stays strictly inside the root block. 
Noting that, in \eqref{block53}, we have
\begin{equation}
n^{1/\widetilde{d}} \frac{[g^n] C^{(2)}_{\lfloor n^{1/\widetilde{d}}\, r\rfloor}
\left(g\, \Mc^2(g)\right)}{[g^n]C^{(2)}\left(g\, \Mc^2(g)\right)}=\sum_{k\geq 0}
\frac{(2k-1)b_k[g^n]\left(g\, \Mc^2(g)\right)^k}{[g^n]C^{(2)}\left(g\, \Mc^2(g)\right)}\times \frac{n^{1/\widetilde{d}}}{k^{1/d}}\times 
k^{1/d}\frac{[t^k] C^{(2)}_{\lfloor n^{1/\widetilde{d}}\, r\rfloor}(t)}{[t^k]C^{(2)}(t)},
\end{equation}
where we used $[t^k]C^{(2)}(t)=(2k-1)b_k$, we recognize the building blocks defining, respectively, $\Proba(X^{(2)}_n=k)$
in \eqref{eq:probadef2} and $\rho_0(r)$ in \eqref{zero51}. Using the large $n$ scaling $k=x\, n^{1/\alpha}$, the right hand side
has a well defined scaling limit for $d$ and $\widetilde{d}$ related by \eqref{ddtilde}, and using \eqref{eq:probscaling2}, we deduce
 the identity
\begin{equation}
\rhob(r)=\int_0^\infty dx\, \sigma(x)\, \frac{1}{x^{1/d}}\rho_0\left(\frac{r}{x^{1/d}}\right)\ .
\label{eq:rho0rhob}
\end{equation}

\medskip
\begin{rem}
Using the explicit result \eqref{A7},  a direct consequence of \eqref{eq:rho0rhob} is the following identity between the moments 
$\mathbb E_{\block}[r^s]:=\int_0^\infty dr\, r^s\rhob(r)$ and 
$\mathbb E_0[r^s]:=\int_0^\infty dr\, r^s\rho_0(r)$:
\begin{equation}
\mathbb E_{\block}[r^s]=\frac{\Gamma\left((s+d-\widetilde{d}\,)/d\,\right)}
{\Gamma\left((s+d-\widetilde{d}\,)/\widetilde{d}\,\right)}\frac{\Gamma\left((d-\widetilde{d}\,)/\widetilde{d}\,\right)}
{\Gamma\left((d-\widetilde{d}\,)/d\,\right)}\frac{1}{D^{s/d}}\mathbb E_0[r^s]\ ,
\end{equation}
valid for $s>\widetilde{d}-2d$.
\end{rem}
\begin{rem}
For $r\to 0$, we expect that $\rho_0(r)\propto r^{d-1}$. We can then deduce the small $r$ behavior of $\rhob(r)$ as follows.
Changing variable from $x$ to $x\, r^d$ in \eqref{eq:rho0rhob} allows one to write
\begin{equation}
\rhob(r)=r^{d-1}\int_0^\infty dx\, \sigma(x\, r^d)\, \frac{1}{x^{1/d}}\rho_0\left(\frac{1}{x^{1/d}}\right)\ .
\label{eq:rho0rhobbis}
\end{equation}
For $r\to 0$, we may use the explicit form \eqref{eq:probscaling2} for $\sigma(x)$ and the series expansion \eqref{eq:formS}
to obtain 
\begin{equation}
\sigma(x\, r^d)\underset{r\to 0}{\sim}r^{d(1-\alpha)}\times \frac{\alpha\, \Gamma(1+1/\alpha)}{\Gamma(2-\alpha)\Gamma(1-1/\alpha)} D^{2-\alpha}
x^{1-\alpha}
\end{equation}
so that, after plugging in \eqref{eq:rho0rhobbis} and changing variable from $x$ to $y=1/x^{1/d}$, we finally get
\begin{equation}\label{eq:rhobsd}
\rhob(r)\underset{r\to 0}{\sim}r^{d(2-\alpha)-1} \frac{\alpha\, \Gamma(1+1/\alpha)}{\Gamma(2-\alpha)\Gamma(1-1/\alpha)} 
D^{2-\alpha}d\ \int_0^\infty dy\, \frac{1}{y^{d(2-\alpha)}}\, \rho_0(y)\ .
\end{equation}
Note that, since $\rho_0(y)\propto y^{d-1}$ at small $y$, the above integral converges for $\alpha>1$.

At large $r$, Fisher's law \cite{Fisher66} predicts that $\log \rho_0(r) \propto - r^\delta$
with $\delta=d/(d-1)$. Using the large $x$ behavior $\log \sigma(x) \propto - x^{\alpha/(\alpha-1)}$ (see \eqref{eq:largx}),
we deduce from a saddle point estimate of integral \eqref{eq:rho0rhob} that $\log \rhob(r) \propto - r^{\widetilde{\delta}}$
with 
\begin{equation}
\widetilde{\delta}=\frac{\frac{\alpha}{\alpha-1}\delta}{\frac{\alpha}{\alpha-1}+\frac{\delta}{d}}=\frac{\alpha d}{\alpha d-1}=\frac{\widetilde{d}}
{\widetilde{d}-1}\ ,
\end{equation}
hence again Fisher's law, as expected. 
\end{rem}

\subsection{The distance profiles of block-weighted quadrangulations}
\label{sec:distdualquad}
\begin{figure}[h]
  \centering
  \fig{.8}{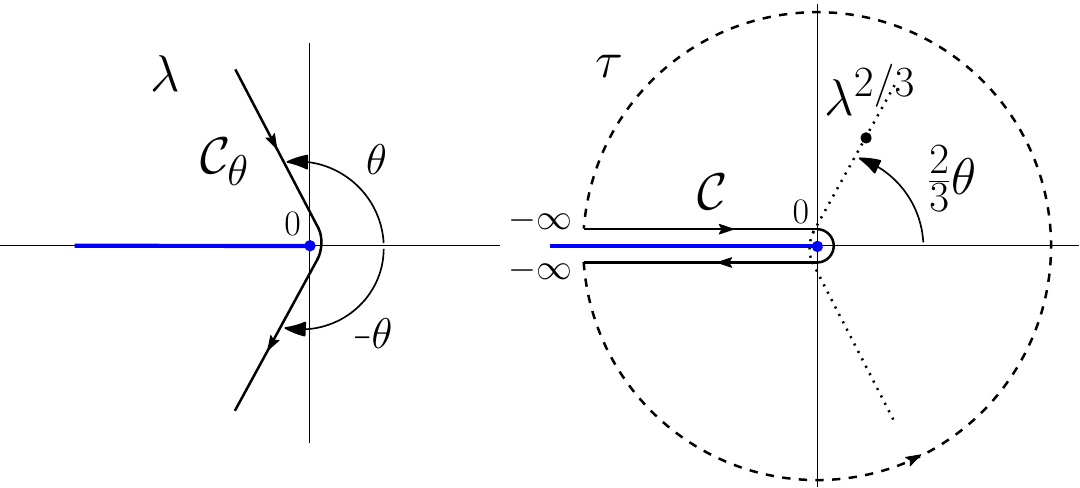}
   \caption{Left: Contour $\mathcal{C}_\theta$ for the integral over $\lambda$ in \eqref{eq:checkint} (third line).
   Right: Closure of the integral over $\tau$ in \eqref{eq:bravoure}, so as to encircle the pole at $\tau=\lambda^{2/3}$.}
  \label{fig:contourtheta}
\end{figure}

For quadrangulations, it is known that $d=4$ and $\widetilde{d}=6$ \cite{DG25}, in agreement with \eqref{ddtilde} at $\alpha=3/2$.
Moreover, the distance profile for simple planar quadrangulations was computed in \cite{Minbus}, with the following
explicit result, in condensed form:
\begin{equation}
\rho_0(r)=2 \sqrt{\frac{3}{\pi}}\int_{\mathcal{C}}d\tau\, e^\tau\, \tau^{3/4} \ \frac{1}{2 {\mathrm i}}{\mathcal F}'(\tau^{1/4}r)
\label{eq:rho0exp}
\end{equation}
where $\mathcal{C}:=\mathcal{C}_+\cup\mathcal{C}_-$ is the contour displayed in Figure~\ref{fig:contournu}-right (see also 
Figure~\ref{fig:contourtheta}-right) and where
\begin{equation}
\mathcal{F'}(r)=-4\, \omega^3 \frac{\cosh(\omega\, r)}{\sinh^3(\omega\, r)}\ , \quad \omega=\frac{3^{1/4}}{2^{1/2}}\ .
\end{equation}
Setting
$\tau=\mu^2 e^{\mathrm{i}\pi}$ along $\mathcal{C}_+$ and $\tau=\mu^2 e^{-\mathrm{i}\pi}$
along $\mathcal{C}_-$, with $\mu$ real positive, this formula can be rewritten as
\begin{equation}
\rho_0(r)=2 \sqrt{\frac{3}{\pi}}\int_0^\infty d\mu\, e^{-\mu^2}\, \mu^{5/2} \ \left\{ e^{\mathrm{i}\pi/4}{\mathcal F}'(e^{\mathrm{i}\pi/4} \mu^{1/2} r)+ e^{-\mathrm{i}\pi/4}{\mathcal F}'(e^{-\mathrm{i}\pi/4} \mu^{1/2} r) \right\}\ ,
\label{eq:rho0quad}
\end{equation}
which is precisely the expression given in \cite[(3.36)]{Minbus}.

\begin{figure}[h]
  \centering
  \fig{.45}{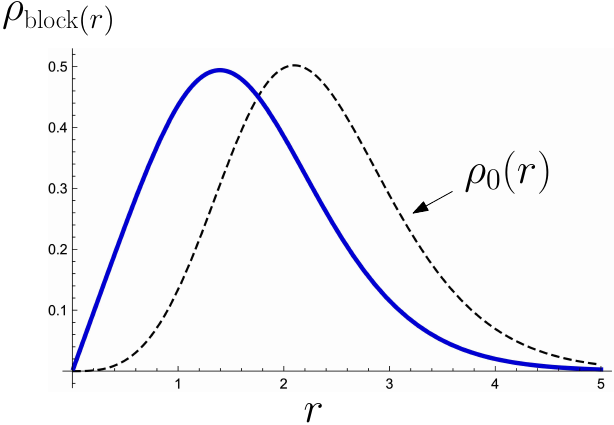}
   \caption{The block distance profile $\rhob$ (thick blue) for block-weighted quadrangulations at $u=\uc$ \eqref{eq:rhobquad}
   and its comparison with the distance profile $\rho_0$ of simple quadrangulations \eqref{eq:rho0quad}. 
   At small $r$, $\rhob(r)\propto r$, in agreement with \eqref{eq:rhobsd} at $d=4$ and $\alpha=3/2$, while $\rho_0(r)\propto r^3$.}
  \label{fig:profils}
\end{figure}

As for $\rhob(r)$, we have a very similar formula, again in condensed form \cite{DG25}:
\begin{equation}
\rhob(r)=\frac{2}{\Gamma\left(\frac{4}{3}\right)}\int_{\mathcal{C}}d\lambda\, e^\lambda\, \lambda^{1/2} \ \frac{1}{2 {\mathrm i}}D^{1/4}{\mathcal F}'(\lambda^{1/6}D^{1/4}r)
\label{eq:rhobexp}
\end{equation}
with $D=3/2^{2/3}$ as in \eqref{eq:valguDquad}.
Setting
$\lambda=\mu^3 e^{\mathrm{i}\pi}$ along $\mathcal{C}_+$ and $\lambda=\mu^3 e^{-\mathrm{i}\pi}$
along $\mathcal{C}_-$, this result can be reformulated as 
\begin{equation}
\rhob(r)=\frac{3}{\Gamma\left(\frac{4}{3}\right)}\int_0^\infty d\mu\, e^{-\mu^3}\, \mu^{7/2}
\ D^{1/4}\left\{ {\mathcal F}'(e^{\mathrm{i}\pi/6} \mu^{1/2} D^{1/4}r)+ {\mathcal F}'(e^{-\mathrm{i}\pi/6} \mu^{1/2} D^{1/4}r) \right\}\ ,
\label{eq:rhobquad}
\end{equation}
which matches exactly the expression given in \cite[Appendix B]{DG25}, after differentiating with respect to $r$.
Figure~\ref{fig:profils} displays for comparison the distance profiles $\rhob$ \eqref{eq:rhobquad} and $\rho_0$  \eqref{eq:rho0quad}.

\medskip
Let us now verify that the above explicit forms \eqref{eq:rho0exp} and \eqref{eq:rhobexp} are indeed linked by the
relation \eqref{eq:rho0rhob} with, $d=4$ and $\alpha=3/2$ here. In particular, we have (recall \eqref{eq:conti}, \eqref{eq:P1} and \eqref{eq:probscaling2})
\begin{equation}
\begin{split}
\sigma(x)&=\frac{3}{2}\frac{\Gamma\left(\frac{2}{3}\right)}{\Gamma\left(\frac{1}{2}\right)}\frac{D}{(D\, x)^{3/2}}S_{\frac{2}{3}}(Dx)\\
&=- \frac{3}{2}\frac{\Gamma\left(\frac{2}{3}\right)}{\sqrt{\pi}}\frac{D}{(D\, x)^{3/2}}
\int_\mathcal{C}\frac{d\lambda}{2\mathrm{i}\pi}e^{\lambda-\lambda^{2/3}\, D\, x}\ .
\end{split}
\label{eq:sigmaquad}
\end{equation}
The right-hand side in \eqref{eq:rho0rhob} therefore reads
\begin{equation}
\begin{split}
&-\int_0^\infty dx\, \frac{3}{2}\frac{\Gamma\left(\frac{2}{3}\right)}{\sqrt{\pi}}\frac{D}{(D\, x)^{3/2}}
\int_\mathcal{C}\frac{d\lambda}{2\mathrm{i}\pi}e^{\lambda-\lambda^{2/3}\, D\, x}\, \frac{1}{x^{1/4}}
2 \sqrt{\frac{3}{\pi}}\int_{\mathcal{C}}d\tau\, e^\tau\, \tau^{3/4} \ \frac{1}{2 {\mathrm i}}{\mathcal F}'\left(\tau^{1/4}\frac{r}{x^{1/4}}
\right)\\
&=-\int_0^\infty d x\, \frac{3}{2}\frac{\Gamma\left(\frac{2}{3}\right)}{\sqrt{\pi}}\frac{1}{x^{3/2}}
\int_\mathcal{C}\frac{d\lambda}{2\mathrm{i}\pi}e^{\lambda-\lambda^{2/3}\, x}\, \frac{D^{1/4}}{x^{1/4}}
2 \sqrt{\frac{3}{\pi}}\int_{\mathcal{C}}d\tau\, e^\tau\, \tau^{3/4} \ \frac{1}{2 {\mathrm i}}{\mathcal F}'\left(\tau^{1/4}\frac{D^{1/4}r}{x^{1/4}}
\right)\\
&=- 3\sqrt{3}\frac{\Gamma\left(\frac{2}{3}\right)}{\pi}  \int_0^\infty dx\,
\int_{\mathcal{C}_\theta}\frac{d\lambda}{2\mathrm{i}\pi}e^{\lambda-\lambda^{2/3}\, x}\,
\int_{\mathcal{C}}d\tau\, e^{\tau x}\, \tau^{3/4} \ \frac{1}{2 {\mathrm i}}D^{1/4}{\mathcal F}'\left(\tau^{1/4} D^{1/4}r\right)\\
&=-\frac{2}{\Gamma\left(\frac{4}{3}\right)}\int_{\mathcal{C}_\theta}\frac{d\lambda}{2\mathrm{i}\pi}e^{\lambda}\int_{\mathcal{C}}d\tau\tau^{3/4} \ \frac{1}{2 {\mathrm i}}D^{1/4}{\mathcal F}'\left(\tau^{1/4} D^{1/4}r\right)
 \int_0^\infty dx\, e^{(\tau-\lambda^{2/3})\, x}
\end{split}
\label{eq:checkint}
\end{equation}
after changing variables from $x$ to $D x$ (second line), then from $\tau$ to $\tau/x$ (third line). We also
deformed the contour $\mathcal{C}=\mathcal{C}_\pi$ for $\lambda$ to the contour $\mathcal{C}_\theta$ of Figure~\ref{fig:contourtheta}-left 
with $\frac{\pi}{2}<\theta<\frac{3 \pi}{4}$ so that $\Re(\lambda^{2/3})>0$ while $\Re(\lambda)<0$. Since $\Re(\tau)<0$ 
on $\mathcal{C}$, we deduce that
$\Re(\tau-\lambda^{2/3})<0$, which allows us to exchange the order of the integrals and perform the integral over $x$ first
(fourth line), with result
\begin{equation}
\begin{split}
& \frac{2}{\Gamma\left(\frac{4}{3}\right)}\int_{\mathcal{C}_\theta}d\lambda\, e^{\lambda}\int_{\mathcal{C}}
\frac{d\tau}{2\mathrm{i}\pi}
 \frac{\tau^{3/4}}{\tau-\lambda^{2/3}}\ \frac{1}{2 {\mathrm i}}D^{1/4}{\mathcal F}'\left(\tau^{1/4} D^{1/4}r\right)\\
&\qquad \qquad = \frac{2}{\Gamma\left(\frac{4}{3}\right)}\int_{\mathcal{C}_\theta}d\lambda\, e^{\lambda}\lambda^{1/2}\ 
\frac{1}{2 {\mathrm i}}D^{1/4}{\mathcal F}'\left(\lambda^{1/6} D^{1/4}r\right)\ ,\\
\end{split}
\label{eq:bravoure}
\end{equation}
where we closed the contour $\mathcal{C}$ as shown in Figure~\ref{fig:contourtheta}-right
 so as to encircle the pole at $\tau=\lambda^{2/3}$.
Note that we used the fact that $\arg(\tau^{1/4})\in [-\frac{\pi}{4},\frac{\pi}{4}]$ along the closed contour, 
hence ${\mathcal F}'\left(\tau^{1/4} D^{1/4}r\right)$ has no pole inside this contour and, moreover, 
$\vert{\mathcal F}'\left(\tau^{1/4} D^{1/4}r\right)\vert$ tends to $0$ exponentially as $e^{-2 D^{1/4}r\, \Re(\tau^{1/4})}$
for $\vert\tau\vert\to \infty$. We may finally deform $\mathcal{C}_\theta$ back to $\mathcal{C}$ to get the desired formula
\eqref{eq:rhobexp}.

\medskip
\begin{rem}
In the first line of \eqref{eq:bravoure}, we may exchange the order of the integrals and use the identity
\begin{equation}
\int_{\mathcal{C}_\theta}\frac{d\lambda}{2\mathrm{i}\pi}\, \frac{ e^{\lambda}}{\tau-\lambda^{2/3}}
= E_{\frac{2}{3},\frac{2}{3}}(\tau)
\end{equation}
where $E_{\frac{2}{3},\frac{2}{3}}(\tau)=\sum\limits_{m\geq 0} \frac{1}{\Gamma\left(\frac{2}{3}m+\frac{2}{3}\right)}\tau^m$
is the two-parameter Mittag-Leffler function. This leads to the alternative expression
\begin{equation}
\rhob(r)= \frac{2}{\Gamma\left(\frac{4}{3}\right)}\int_{\mathcal{C}} d\tau \, \tau^{3/4}\, E_{\frac{2}{3},\frac{2}{3}}(\tau) \, \frac{1}{2 {\mathrm i}}D^{1/4}{\mathcal F}'\left(\tau^{1/4} D^{1/4}r\right) 
\end{equation}
or, after deforming $\mathcal{C}$ to $\mathcal{C}_{2\pi/3}$ and changing variable from $\tau$ to $\lambda=\tau^{3/2}$,
\begin{equation}
\rhob(r)= \frac{4}{3\Gamma\left(\frac{4}{3}\right)}\int_{\mathcal{C}} d\lambda \, \lambda^{1/6}\, E_{\frac{2}{3},\frac{2}{3}}(\lambda^{2/3}) \, \frac{1}{2 {\mathrm i}}D^{1/4}{\mathcal F}'\left(\lambda^{1/6} D^{1/4}r\right)\ , 
\end{equation}
an expression reminiscent of \eqref{eq:rhobexp}.
\end{rem}

\section{Tree structures made of attached quartic maps}
\label{sec:DASetal}
\subsection{The Das-Dhar-Sengupta-Wadia model revisited}
\label{sec:DDSW} 
In 1990, Das \emph{et al.}\ introduced  in \cite{1990MPLA....5.1041D} a matrix model for the enumeration of 
tree structures made of attached quartic maps, as a particular example of a two-dimensional gravity model with a 
positive string susceptibility exponent $\gsp=1/3$ (in our notations, corresponding to $\alpha=3/2$) 
at some well-chosen value of
the attachment weight (corresponding precisely to the dual critical point, see below). 
The authors consider the matrix integral
\begin{align}
& Z(t,g,u,N)=\int d\Phi \exp(-V(\Phi))\ , \\ 
&\label{eq:Vmat} V(\Phi)=\frac{1}{2t} \Tr \Phi^2-\frac{g}{4N} \Tr \Phi^4 
-\frac{u}{4N^2} (\Tr \Phi^2)^2\ ,
\end{align}
where the integration runs over Hermitian matrices $\Phi$ of size $N$. They then discuss in detail the free energy $F_0(t,g,u)$
of the model in the so-called ``planar limit'', obtained from the large $N$ asymptotics, 
\begin{equation}
Z(t,g,u,N)\underset{N \to \infty}{\propto}e^{N^2\, F_0(t,g,u)}\ .
\end{equation} 
It is well known that, when $u=0$, $F_0(t,g,0)$ enumerates (connected) \emph{planar quartic maps}, \emph{i.e}., planar maps
whose all vertices have degree $4$, with a weight $t$ per edge and $g$ per vertex. To avoid problems of symmetry factors, 
it is customary to consider, instead of $F_0(t,g,0)$, the generating function $2t \frac{\partial}{\partial t}F_0(t,g,0)$,
which now enumerates planar quartic maps which are  \emph{rooted}, \emph{i.e.}, have a marked oriented edge. 
\begin{figure}[h]
  \centering
  \fig{.98}{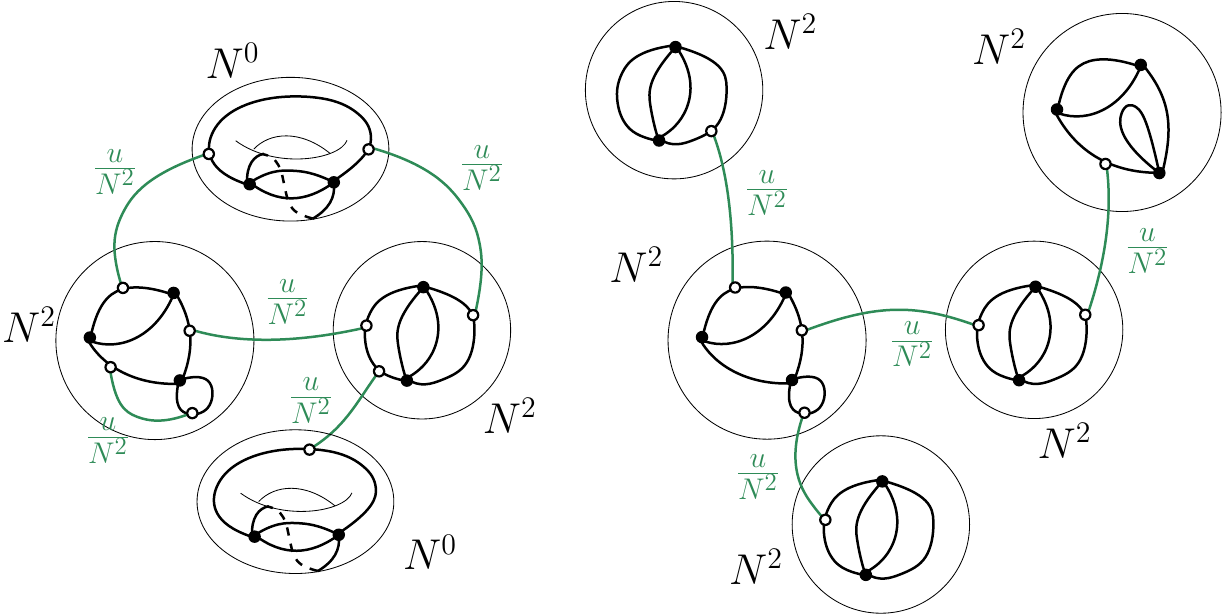}
   \caption{Objects contributing to $\log Z(t,g,u,N)$ are (non necessarily planar) quartic maps endowed with pairs of bivalent vertices (small circles) connected via extra special edges (dark green). A map with genus $h$ receives a weight 
   $N^{2-2h}$ while each special edge receives a weight $u/N^2$. The object on the left receives a global weight $u^5/N^6$
   and that on the right a global weight $u^4N^2$, hence only the second one contributes to $F_0(t,g,u)$.}
  \label{fig:objects}
\end{figure}

For $u\neq 0$, the last term of  $V(\Phi)$ in \eqref{eq:Vmat}, proportional to $ (\Tr \Phi^2)^2$, may be interpreted as describing 
a \emph{pair of (originally) bivalent vertices} further attached to each other via an extra \emph{special edge} and receiving weight $u/N^2$; see 
Figure~\ref{fig:objects}. The quantity $\log Z(t,g,u,N)$ now enumerates connected objects made of several 
(non necessarily planar) quartic maps endowed with pairs of bivalent vertices attached together via special edges.
Such an object, made out of $q\geq 1$ quartic maps of respective genera $h_1,\ldots,h_q$, attached via $p$ special edges 
(with $p\geq q-1$ so that the object is connected) receives a factor $u^p\, N^m$ with 
$m=\sum_{i=1}^q(2-2h_i)-2p$ (namely a factor $N^{2-2h_i}$ per quartic map of genus $h_i$ and a factor $N^{-2}$ per
special edge). The power of $N$ is therefore maximal and equal to $m=2$ when all the quartic maps are planar 
($h_i=0$ for all $i$) and when $p=q-1$, corresponding to quartic maps attached so as to form a \emph{tree structure}.
The quantity $F_0(t,g,u)$ therefore enumerates tree structures made of attached planar quartic maps
and the quantity $2t \frac{\partial}{\partial t}F_0(t,g,u)$ enumerates the same structures which are rooted, \emph{i.e.},
with a marked oriented regular (\emph{i.e.}, non-special) edge. These tree structures are planar graphs, but we may view them as well
as planar maps, provided we give a canonical prescription for the way we attach the pairs of vertices together. A possible 
prescription is via face colors, as we will discuss below, where we give a construction of the tree structures 
by a recursive process.  
 
As it turns out, the tree structures enumerated by $2t \frac{\partial}{\partial t}F_0(t,g,u)$ are actually block-weighted 
objects in disguise, which can be constructed in a fully combinatorial way as follows:
we start from a general rooted quartic planar map $\mathcal{Q}$. 
This map serves as the root block of the tree structure and we denote by $k$ 
its number of vertices (the map $\mathcal{Q}$ has then $2k$ edges and $k+2$ faces - we say that the tree structure
has root block size $k$). By face/vertex standard duality, the number $b_k$ of planar quartic maps is equal to the number of planar quadrangulations of size $k$, namely
\begin{equation}
b_k=2\frac{3^k}{(k+1)(k+2)}\binom{2k}{k}\ , \quad B(t):=\sum_{k\geq 0}b_k\, t^k= \frac{18t-1+(1-12 t)^{3/2}}{54 t^2}\ .
\label{eq:BDAS}
\end{equation} 
For $k=0$, we have $b_0=1$ which we interpret here as counting the unique ``rooted map with no vertex'' formed of a single oriented 
ring separating two faces on the sphere. 
The function $B(t)$ has a singular behavior of the form \eqref{eq:Bsing} with $\alpha=3/2$ and
\begin{equation}
\tc=\frac{1}{12}\ , \quad B(\tc)=\frac{4}{3} \ , \quad B'(\tc)=16 \ , \quad K_B= 2^6\sqrt{3}\ .
\label{eq:tcritDAS}
\end{equation}
\begin{figure}[h]
  \centering
  \fig{.7}{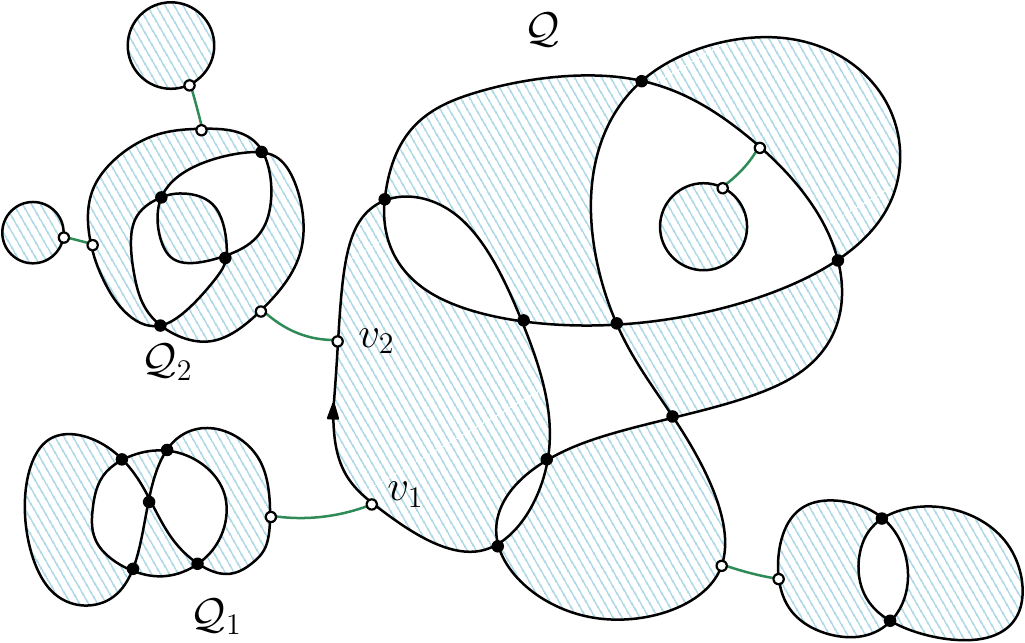}
   \caption{A rooted planar tree structure contributing $u^6 g^{19}$ to $M_u(g)$, \emph{i.e.}, with $19$ regular 
   $4-$valent vertices (here represented by filled small disks) and $6$ non-root blocks attached by $6$ special
   edges (in dark green). Black faces are here filled with blue falling lines. Here $3$ of the blocks are actually
   maps with no regular vertex (rings).}
  \label{fig:Quartic}
\end{figure}
Since $\mathcal{Q}$ is a planar map with all its vertices of even degree (here $4$), we 
may color its faces in black and white so that any two adjacent faces have different colors.
We fix the colors by demanding that the face to the left of the root edge be white. Consider now an arbitrary edge $e$ of $\mathcal{Q}$. We then attach to $e$ a (possibly empty) sequence made of an arbitrary 
number $p\geq 0$ of rooted quartic maps  $\mathcal{Q}_j, j=1,\ldots,p$ (some of them 
possibly having size $0$, \emph{i.e.}, being reduced to an oriented ring). This is done by adding $p$ bivalent vertices 
$v_1,\ldots,v_p$ along $e$ and attaching, via a special edge, the middle of the root edge of $\mathcal{Q}_j$ by its (white) 
left side to the 
$j-$th bivalent vertex $v_j$, drawing $\mathcal{Q}_j$ within the white face incident to $v_j$; see Figure~\ref{fig:Quartic}. 
We finally assign the weight $u$ to each special edge. 
The maps $\mathcal{Q}_1,\ldots,\mathcal{Q}_p$ are viewed as $p$ 
new blocks and their faces are colored in black and white. If $e$ is the root edge of $\mathcal{Q}$, we re-root the obtained map
 at one of the $p+1$ edges 
replacing $e$. We repeat this construction for each edge of $\mathcal{Q}$, leading to more special edges and more blocks. 
The operation is then repeated again recursively for all the regular (\emph{i.e.}, non-special) edges of all the new added blocks. The resulting map is a planar \emph{rooted tree structure}, canonically colored in black and white, made of quartic maps attached by special edges \emph{adjacent to white faces only}. Let us call $M_u(g)$ the partition function of such maps, 
where we assign a weight $g$ to each of the $4-$valent vertices, and of course the weight $u$ per special edge, which also
amounts to a weight $u$ per non-root block since, in the tree structure,
there are as many non-root blocks as special edges. We may moreover identify these recursively constructed 
tree structures with those enumerated by the matrix model of Das \emph{et al.}\ in \cite{1990MPLA....5.1041D} at large $N$,
so that $M_u(g)=2t \frac{\partial}{\partial t}F_0(t,g,u)\vert_{t=1}$.  A tree structure 
with a total number $n$ of $4-$valent vertices is said to have size $n$. The above recursive construction is 
encoded is the following substitution relation between $M_u(g)$ and $B(t)$: we have
\begin{equation}
M_u(g)=\frac{1}{1-u\, M_u(g)}B\left(\frac{g}{(1-u\, M_u(g))^2}\right)\ .
\label{eq:substDAS}
\end{equation}
This relation simply states that a rooted tree structure is given by the data of its root block $\mathcal{Q}$, 
a rooted quartic map
enumerated by $b_k\, g^k$ if the root block size is $k$, and by the data of $2k+1$ sequences of objects which are themselves 
rooted tree structures: each of the $2k-1$ non-root edge of $\mathcal{Q}$ indeed yields such a sequence, while the root edge
gives rise to $2$ sequences (one before and one after the root edge in the whole tree structure). Each sequence
is enumerated by $\sum_{p\geq 0} (u M_u(g))^p=1/(1-u M_u(g))$, hence the right-hand-side, equal to 
$\sum_{k\geq 0} b_k g^k /(1-u M_u(g))^{2k+1}$.  

\begin{figure}[h]
  \centering
  \fig{.55}{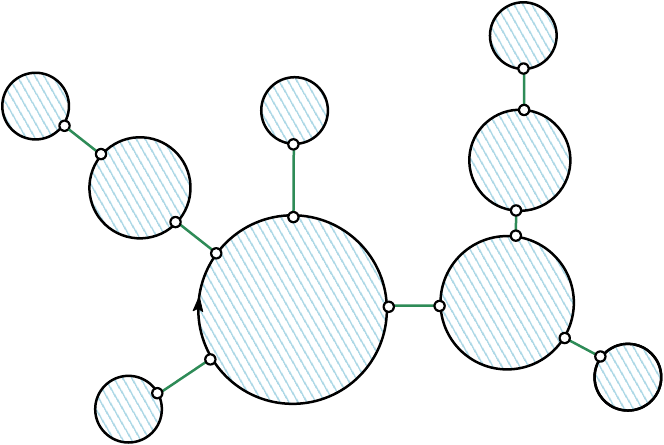}
   \caption{A rooted tree structure contributing $u^8$ to $M_u(0)$. For convenience, the size of the rings growths with
   their number of attached special edges (in dark green).}
  \label{fig:Rings}
\end{figure}

\begin{rem}
For $g=0$, Eq.~\eqref{eq:substDAS} reads $M_u(0)=1/(1-u\, M_u(0))$, hence
\begin{equation}
M_u(0)=\frac{1-\sqrt{1-4u}}{2u}=\sum_{n\geq 0} \text{Cat}(n)u^n , \quad \text{Cat}(n):=\frac{1}{n+1}\binom{2n}{n}\ ,
\end{equation}
which, as explained in \cite{1990MPLA....5.1041D}, enumerates planar (rooted) tree structures obtained by gluing rings together
by special edges receiving the weight $u$; see Figure~\ref{fig:Rings}.
\end{rem}

With the new substitution relation \eqref{eq:substDAS}, we recover the subcritical and critical behavior of 
Eqs.~\eqref{eq:subcritical} and \eqref{eq:critical}, with $\gc(u)$ now given in terms of the 
singularity $\tc$ \eqref{eq:tcritDAS} of $B$ \eqref{eq:BDAS} (instead of \eqref{eq:guB}) by
\begin{equation}
\gc(u)=\tc\, (1-u\, M_u(\gc(u))^2\ , \quad M_u(\gc(u))=\frac{1}{1-u\, M_u(\gc(u))}B(\tc)\ ,
\end{equation}
leading by elimination to
\begin{equation}
\gc(u)=\frac{\tc}{2}\left(1-2 u B(\tc)+\sqrt{1-4 u B(\tc)}\right)=\frac{1}{72} \left(3-8 u+\sqrt{9-48 u}\right)\ .
\label{eq:guDAS}
\end{equation}
As for the critical value $\uc$, it is now fixed by imposing the additional \emph{critical condition}
\begin{equation}
1=\frac{u}{(1-u\, M_u(g))^2}\left(B(\tc)+2\tc\, B'(\tc)\right)\ ,
\end{equation}
as obtained by differentiating \eqref{eq:substDAS} with respect to $M_u(g)$.
We eventually find that $\uc$ is now given (instead of \eqref{eq:equcrit}) by
\begin{equation}
\uc=\frac{B(\tc)+2\tc\, B'(\tc)}{4(B(\tc)+ \tc\, B'(\tc))^2}=\frac{9}{64}\ ,
\label{eq:equcritDAS}
\end{equation}
which is precisely the critical attachment weight\footnote{The quantity $\uc$ is called $-\alpha_0$ in 
\cite{1990MPLA....5.1041D}, while the quantity $\gc$ is called $-\beta_0$.}
obtained in \cite{1990MPLA....5.1041D}.

Focusing our study on the dual critical point upon setting $u=\uc$,
and using again the shorthand notation $\gc:=\gc(\uc)$, we obtain from
\eqref{eq:guDAS} and \eqref{eq:equcritDAS} the value
\begin{equation}
\gc=\frac{3}{64}\ ,
\end{equation}
satisfying the relation (instead of \eqref{eq:equcritbis})
\begin{equation}
\uc=\frac{2\gc-\sqrt{\gc\, \tc}}{2\tc^2\, B'(\tc)}\ .
\label{eq:equcritbisDAS}
\end{equation}
Writing 
\begin{equation}
g=\frac{t}{2}\left(1-2 \uc B(t)+\sqrt{1-4 \uc B(t)}\right)\ ,
\label{eq:gtot}
\end{equation}
as obtained by eliminating $\Mc(g)$ in the system
\begin{equation}
g=t\, (1-\uc\, \Mc(g))^2\ , \quad \Mc(g)=\frac{1}{1-\uc\, \Mc(g)}B(t)\ ,
\end{equation}
and expanding \eqref{eq:gtot} around $\tc$ via \eqref{eq:Bsing}, 
it is easily checked that, remarkably, the relation \eqref{eq:ttogcrit} remains unchanged, with in particular
the same literal expression for $C$, with the value here 
\begin{equation}
C=\left(\frac{\tc B'(\tc)}{\gc K_B}\right)^{\frac{1}{\alpha}}=\frac{2^{4/3}}{3^{5/3}}\, .
\label{eq:CDAS}
\end{equation}
The scaling estimates \eqref{eq:probscaling}, \eqref{eq:probscaling2} 
and \eqref{eq:probscaling3} remain valid, with the same literal expression $\eqref{eq:formD}$ for the scaling parameter
$D$, with the value here
\begin{equation}
D=\frac{1}{\tc}\left(\frac{\tc B'(\tc)}{K_B}\right)^{\frac{1}{\alpha}}=\frac{1}{2^{2/3}}\ ,
\label{eq:DDAS}
\end{equation}
to be compared again with the value $D=3/2^{2/3}$ in \eqref{eq:valguDquad} for block-weighted quadrangulations.

\subsection{The distance profiles of stuffed quadrangulations}
\label{sec:stuffed} 
\begin{figure}[h]
  \centering
  \fig{.7}{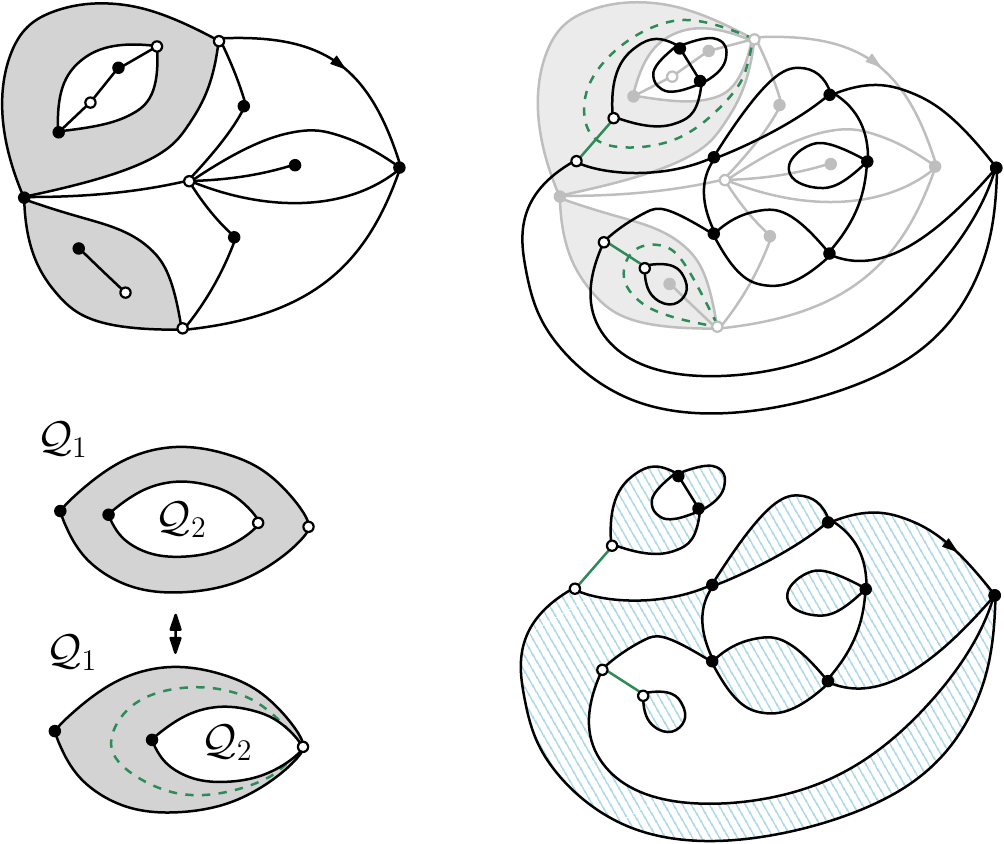}
   \caption{Upper left: an example of stuffed quadrangulation with $8$ regular quadrangles and $2$ special ones 
   (colored in gray), $20$ edges, $14$ vertices and $3$ graph connected components. 
   Lower left: a schematic picture of the deformation of a special quadrangle into two triangles by 
   merging the two incident white vertices into a single one and drawing an edge from this vertex to itself separating
   the two original $2-$gonal boundaries. Right: applying this deformation to all special quadrangles and
   the passing to the dual map (upper right) creates the desired tree structure made of attached planar quartic maps 
   (lower right).}
  \label{fig:stuffed}
\end{figure}

Stuffed maps were introduced in \cite{zbMATH06337014} as a generalization of maps with faces not only made of polygons  
(with the topology of the disk, or equivalently of the sphere with a single boundary), 
but also of more general 2-cells of type $(m,g)$ with the topology of an orientable, connected surface of arbitrary genus $g$
with $m$ boundaries of prescribed integer lengths (see  \cite{zbMATH06337014} for a precise definition). One of the simplest 
examples of stuffed maps are \emph{stuffed quadrangulations} made of two types of faces: regular quadrangles which are 
polygons with $4$ sides, and \emph{special quadrangles} which are 2-cells of type $(2,0)$, namely cylinders, with 
two $2-$gonal boundaries; see Figure~\ref{fig:stuffed} for an example.

Now we claim that rooted planar bicolored stuffed quadrangulations are in bijection with the rooted tree structures made of attached planar quartic maps described in the previous section. Here, by bicolored, we mean that the vertices of the 
stuffed map are colored in black and white so that adjacent vertices have different colors. The stuffed map is rooted
by choosing an edge and orient it away from its white incident vertex. By planar, we mean that the overall underlying surface is 
the sphere, or equivalently, that the numbers $V$, $E$, $F$, and $K$ of, 
respectively, vertices, edges, faces (regular and special),
and graph connected components of the stuffed map satisfy $V-E+F=1+K$. The bijection is illustrated in Figure~\ref{fig:stuffed}:
for each special quadrangle, we glue its two incident white vertices into a single white vertex, creating a regular quadrangle 
with a single boundary made of two cycles of length $2$. We split this quadrangle into two triangles by drawing an edge connecting the new white vertex to itself so as to separate the two cycles of length $2$. We finally take the dual of the obtained planar
map, resulting in the desired tree structures made of planar quartic maps attached by pairs of bivalent vertices
connected by special edges. The root is transferred from the original
root edge to its dual edge (canonically oriented via a counterclockwise quarter turn).

A direct consequence of this bijection is that we can interpret the quantity $M_u(g)$ in \eqref{eq:substDAS} as the
generating function of rooted planar bicolored stuffed quadrangulations, with a weight $g$ per regular quadrangle
and a weight $u$ per special quadrangle, or equivalently, per non-root block when viewing now the graph connected components of stuffed quadrangulations as blocks. 
We may also obtain straightforwardly the block distance profile $\rhobb(r)$ of stuffed quadrangulations at their critical 
point $\uc$ (given by \eqref{eq:equcritDAS}) as follows. Since stuffed maps made of a single block are nothing but regular
quadrangulations, the profile $\rhobb(r)$ is now related by Eq.~\eqref{eq:rho0rhob},
\begin{equation}
\rhobb(r)=\int_0^\infty dx\, \bar{\sigma}(x)\, \frac{1}{x^{1/4}}\bar{\rho}_0\left(\frac{r}{x^{1/4}}\right)\ ,
\label{eq:rho0rhobDAS}
\end{equation}
to the distance profile $\bar{\rho}_0(r)$ of regular quadrangulations. Here, $\bar{\sigma}(x)$ is given by 
the same formula as the quantity $\sigma(x)$ in \eqref{eq:sigmaquad} for block-weighted quadrangulations, but
with $D=3/2^{2/3}$ replaced by $\bar{D}=1/2^{2/3}$ as in \eqref{eq:DDAS}. Otherwise stated, since $D/\bar{D}=3$,
we can write
\begin{equation}
\bar{\sigma}(x)=\frac{1}{3}\sigma\left(\frac{x}{3}\right)\ .
\end{equation}
Now it is known \cite[Eq. (3.38)]{Minbus} that the distance profile $\bar{\rho}_0(r)$ of regular quadrangulations is related to 
the profile $\rho_0(r)$ \eqref{eq:rho0exp} of simple quadrangulations by 
\begin{equation}
\bar{\rho}_0(r)=3^{1/4}\rho_0\left(3^{1/4}r\right)\ ,
\end{equation}
expressing the fact that the size of the largest simple component in a quadrangulation is, on average for large maps, 
$1/3$ of its total size.
Altogether, we can write 
\begin{equation}
\begin{split}
\rhobb(r)&=\int_0^\infty dx\, \frac{1}{3}\sigma\left(\frac{x}{3}\right)\, \frac{1}{x^{1/4}}3^{1/4}\rho_0\left(3^{1/4}\frac{r}{x^{1/4}}\right)\\
&=\int_0^\infty dx\, \sigma(x)\, \frac{1}{x^{1/4}}\rho_0\left(\frac{r}{x^{1/4}}\right)\\&=
\rhob(r)\ .\\
\end{split}
\end{equation}
Remarkably, we obtain the same, unrescaled, block distance profile $\rhob(r)$ as in \eqref{eq:rhobexp} for block-weighted quadrangulations 
at their critical point.

\section{Liouville quantum duality}
\label{sec:LQD}
\subsection{Liouville quantum measure}
\label{sec:LQM}
We first recall some basic notions on the LQG measure (for a more thorough presentation, see Part I of this work \cite{DG25}). Let $h$ be a Gaussian free field on some planar domain $\mathcal D \subset \mathbb C$, with (formal) Gaussian action $\frac{1}{2\pi}\int_{\mathcal D} (\nabla h(z))^2 dz$. For simplicity, we generally consider Dirichlet boundary conditions, where $h$ is defined to be
zero in $\mathbb C \setminus \mathcal D$. The specific case of the Liouville quantum measure on the sphere will be considered in Section \ref{LQGsphere}, when dealing with LQG partition functions.

Let 
$h_\varepsilon(z)$ denote the mean value of $h$ on the circle of
radius $\varepsilon$ centered at $z$. Geometrical properties of the Dirichlet GFF \cite{2009arXiv0901.0277D,2008arXiv0808.1560D} yield    
 the explicit variance 
 $
 \hbox{Var} \, h_\varepsilon (z)=\log [C(z; \mathcal D)/\varepsilon]
$
in terms of the conformal radius, $C(z;\mathcal D) = |\phi'(z)|^{-1}$,  with $\phi$ mapping conformally $\mathcal D$ to the unit disc and $z$ to the origin.
One then has the expectation $  \mathbb E\, e^{\gamma h_\varepsilon(z)}
= \left[ {C(z;D)}/{\varepsilon} \right]^{\gamma^2/2}.$
 A \emph{regularized} Liouville quantum measure is thus defined as \cite{2009arXiv0901.0277D,2008arXiv0808.1560D},
\begin{equation}
\label{liouvillemeasure}
\mu_{\gamma,\varepsilon}(dz):=\varepsilon^{\gamma^2/2} e^{\gamma h_\varepsilon(z)}dz,
\end{equation}
 whose expectation 
is finite and independent of $\varepsilon$. (See Ref. \cite{MR0292433} 
 for the earliest H{\o}egh-Krohn model). 
 One can show that for $\gamma \in [0,2)$
it is
(almost surely) the case that as $\varepsilon \to 0$, the measure \eqref{liouvillemeasure}
weakly converges to a finite, non-vanishing measure, 
 \begin{equation}\label{mug}
 \underset{\varepsilon\to 0^+}{\lim} \mu_{\gamma,\varepsilon}(dz)=\mu_{\gamma} (dz).
 \end{equation}
 In the $\gamma=2$ case, it is known  \cite{DRSV12} that the properly regularized measure, 
\begin{equation}
\label{liouvillemeasure2}
\mu_{\gamma=2,\varepsilon}(dz):=\left(\log \left({1}/{\varepsilon}\right)\right)^{1/2}\varepsilon^{2} e^{2 h_\varepsilon(z)}dz,
\end{equation}
 weakly converges as $\varepsilon \to 0$ to an atom-free and non-vanishing mesure $\mu_{\gamma=2}$ \cite{DRSV12}.
\subsection{Quantum balls}
\label{QBS}
For the quantum area measure $\mu_\gamma$ on $\mathcal D$, let $B^\delta(z)$ be the Euclidean ball $B_\varepsilon(z)$ centered at $z$
whose radius $\varepsilon$ is chosen so that 
\begin{equation}\label{qb}
\mu_\gamma(B^\delta(z)) = \delta.  
\end{equation}
(If not  unique,  take the
radius to be $\sup \{ \varepsilon: \mu_\gamma(B_\varepsilon(z)) \leq \delta
\}$.)
We refer to $B^\delta(z)$ as the \emph{quantum ball}
of area $\delta$ centered at $z$. In particular, in the 
  case of Lebesgue measure $\mu_{\gamma=0}$ one has $\delta = \pi \varepsilon^2$.
 
When $\varepsilon$ is small,
the full quantum measure of a ball is very well
approximated \cite{2009arXiv0901.0277D,2008arXiv0808.1560D} by 
\begin{eqnarray}
\label{muodot}
\mu_{\gamma\odot}(B_{\varepsilon}(z)):=\varepsilon^{\gamma Q_\gamma} e^{\gamma h_\varepsilon(z)},\quad Q_\gamma:= 2/\gamma+\gamma/2,
\end{eqnarray}
so that $\mathbb E[\mu_{\gamma}(B_{\varepsilon}(z))|h_\varepsilon(z)]\underset{\varepsilon \to 0}{\sim} \pi \mu_{\gamma\odot}(B_{\varepsilon}(z))$. In this simplified perspective,
one views $\mu_{\gamma\odot}$ as a function on balls, rather than the fully defined measure $\mu_\gamma$ \eqref{mug}. Accordingly, one defines the  quantum ball $\tilde B^\delta(z)$
  as the (largest) Euclidean ball $B_{\varepsilon}(z)$
such that, $\mu_{\gamma\odot} (B_\varepsilon(z)) = \delta.$ 

It so happens that the function \eqref{muodot} can be continued to the dual range  $\gamma'>2$,  providing a first analytical approach to duality \cite{2008ExactMethodsBD,2009arXiv0901.0277D,zbMATH06797761}\cite[Sec. 3.1]{DG25}. One defines
 \emph{dual} quantum balls $\tilde B^{\delta}$, such that 
 \begin{equation}\label{btildeprime}
 \mu_{\gamma'\odot} (\tilde B^{\delta}(z)) = \delta.
 \end{equation} 
 This method is \emph{singular}:  when $\gamma'>2$, there is a \emph{finite} probability that a quantum area of at least $\delta$ is \emph{localized} at point $z$ for
small enough $\delta$, which then yields a \emph{vanishing} Eulclidean radius for $\tilde B^\delta(z)$. This probability goes to $1$ as $\delta\to 0$, which is quantified by the probability of the complementary event  
\cite{2008ExactMethodsBD,2009arXiv0901.0277D}\cite[Eq. (2.13)]{DG25}, 
\begin{equation}\label{deltaodotprime}
\Proba(\mu_0(\tilde B^{\delta}(z))\neq 0) = \delta^{1-4/\gamma'^2},\quad \quad \delta <1.
\end{equation} 
\subsection{Dual quantum measure}
\label{sec:DQM}
A way to define such singular quantum measures $\mu_{\gamma'>2}$, which will also reproduce the combinatorial scaling properties seen above,  has been presented in Refs. \cite{2008ExactMethodsBD,2009arXiv0901.0277D,zbMATH06797761}. In this approach,
one uses the dual measure $\mu_{\gamma<2}$ \eqref{mug} 
and the additional randomness of sets of points where  \textit{finite} amounts of  quantum area
 are \textit{localized}.  
More precisely, conditionally on the  quantum measure $\mu_{\gamma}$, with $\gamma=4/\gamma' <2$, one considers a Poisson random measure $\mathcal N_{\gamma'}(dz,d\eta)$ distributed on $\mathcal D\times (0,\infty)$,  of intensity 
$\mu_{\gamma}(dz) \times \Lambda^{\alpha'}(d\eta)$, where 
$\Lambda^{\alpha'}(d \eta) := d\eta/\eta^{1+\alpha'}, \,\,\,\alpha':=4/\gamma'^2 =\gamma^2/4 \in (0,1),$ letting each point $(z,\eta)$ represent an atom of size $\eta$ located at $z$. 
The dual measure for $\gamma'>2$ is then \emph{purely atomic},
\begin{equation}\label{dualmeasure}
\mu_{\gamma'}(dz):=\int_0^{\infty}\eta\, \mathcal N_{\gamma'} (dz,d\eta),\,\, \gamma' >2.
\end{equation}
A different construction, generally called atomic Gaussian multiplicative chaos,  has been proposed by Barral \emph{et al.} in Ref. \cite{pre06228485}. However, both perspectives yield an equivalence in law, as their respective random measures satisfy the following characteristic Laplace transform identity. 
For  any Borelian  $\mathcal A\subset \mathcal D$, we have\footnote{Recall for instance that a Poisson process $X$ of intensity $\lambda$, with discrete probability
$\Proba(X=k) = \frac{\lambda^k e^{-\lambda}}{k!}$ with $ k\in \mathbb N$,
has for exponential generating function 
$
\mathbb{E}\!\left(e^{-tX}\right) =\sum_{k \ge 0}\Proba(X=k)e^{-tk}
= e^{\lambda (e^{-t} - 1)}.
$}
\begin{eqnarray}\nonumber
\mathbb E\left[\exp(-u\,\mu_{\gamma'}(\mathcal A))\right]&=&\mathbb E\left[\exp\left(\int_{\mathcal A}\mu_{\gamma}(dz)\int_0^{\infty}\frac{d\eta}{\eta^{1+\alpha'}}\left(e^{-u\eta}-1\right)\right)\right]\\
&=&\mathbb E\exp\left[-\frac{\Gamma(1-\alpha')}{\alpha'}u^{\alpha'}\mu_{\gamma}(\mathcal A)\right], \label{Laplace}
 \end{eqnarray}
 a L\'evy-Khintchine formula valid for all $u \in \mathbb R_+$, where $-\Gamma(1-\alpha')/\alpha' =\Gamma(-\alpha')<0$ is the usual Euler $\Gamma-$function, and  
 where  use was made of the identity
 \[
x^{\beta} = \frac{\beta}{\Gamma(1 - \beta)} \int_0^{\infty} (1 - e^{-\eta x}) \, \frac{d\eta}{\eta^{1 + \beta}}
\quad \text{with} \quad 0 < \beta < 1, \; x\geq 0.
\] 
On the right-hand side of \eqref{Laplace}, expectation is taken with respect to the standard Liouville random measure $\mu_{\gamma <2}$. 
It is clear that one can instead \emph{condition} the left-hand side expectation of \eqref{Laplace} on the value of the latter measure $\mu_\gamma$, to get 
\begin{eqnarray}\nonumber
\mathbb E\left[\exp(-u\,\mu_{\gamma'}(\mathcal A))\vert{\mu_{\gamma}}\right]&=&\mathbb E\left[\exp\left(\int_{\mathcal A}\mu_{\gamma}(dz)\int_0^{\infty}\frac{d\eta}{\eta^{1+\alpha'}}\left(e^{-u\eta}-1\right)\right)\big{\vert}{\mu_{\gamma}}\right]\\
&=&\exp\left[\Gamma(-\alpha')u^{\alpha'}\mu_{\gamma}(\mathcal A)\right],\label{Laplace2}
 \end{eqnarray}
  Let us now consider  the \emph{moments}  of  the mutually-dual measures. For $0< q<\alpha'<1$, we have 
 \begin{eqnarray}\label{intrep}
 \mathbb E\left[\big(\mu_{\gamma'}(\mathcal A)\big)^q\right]&=&\frac{q}{\Gamma(1 - q)} \int_0^{\infty} (1 - \mathbb E \big[e^{-u \mu_{\gamma'}(\mathcal A)}\big]) \, \frac{du}{u^{1 + q}}\\ \nonumber
 &=& \frac{q}{\Gamma(1 - q)} \int_0^{\infty} (1 - \mathbb E \big[e^{\Gamma(-\alpha')u^{\alpha'} \mu_{\gamma}(\mathcal A)}\big]) \, \frac{du}{u^{1 + q}}.
 \end{eqnarray}
With the change of variable $x=u^{\alpha'}$, we finally get 
\begin{eqnarray}\nonumber
 \mathbb E\left[\big(\mu_{\gamma'}(\mathcal A)\big)^q\right]
 &=& \frac{q}{\alpha'\Gamma(1 - q)} \int_0^{\infty} (1 - \mathbb E \big[e^{\Gamma(-\alpha')x \mu_{\gamma}(\mathcal A)}\big]) \, \frac{dx}{x^{1 + q/\alpha'}}\\\label{momentsduaux}
 &=&\frac{\Gamma(1-q/\alpha')[-\Gamma(-\alpha')]^{q/{\alpha'}}}{\Gamma(1-q)}\mathbb E\left[\big(\mu_{\gamma}(\mathcal A)\big)^{q/{\alpha'}}\right].
 \end{eqnarray}
 For $q<0$, instead of \eqref{intrep}, we simply use
  \begin{eqnarray}\nonumber
 \mathbb E\left[\big(\mu_{\gamma'}(\mathcal A)\big)^q\right]&=&\frac{1}{\Gamma(- q)} \int_0^{\infty} \mathbb E \big[e^{-u \mu_{\gamma'}(\mathcal A)}\big] \, \frac{du}{u^{1 + q}}\\ \nonumber
 &=& \frac{-q}{\Gamma(1 - q)} \int_0^{\infty} \mathbb E \big[e^{\Gamma(-\alpha')u^{\alpha'} \mu_{\gamma}(\mathcal A)}\big] \, \frac{du}{u^{1 + q}}.
 \end{eqnarray}
 With the same change of variable, $x=u^{\alpha'}$, we finally get the same equation \eqref{momentsduaux}, this time extended to $q<0$, hence to the whole range $q<\alpha'$. This relationship will be exploited later when dealing with the multifractal analysis of the Liouville measure and its dual.

For a comparison with the results obtained from the discrete models, in particular \eqref{eq:limitlawbis}, and recalling that $\alpha\alpha'=1$, we can rewrite \eqref{Laplace2} as
\begin{eqnarray}
\mathbb E\left[\exp\left(-u\,\frac{\mu_{\gamma'}(\mathcal A)}{\mu_{\gamma}^\alpha(\mathcal A)}\right)\big{\vert}{\mu_{\gamma}}\right]
=\exp\left(\Gamma(-\alpha')u^{\alpha'}\right).\label{Laplace3}
 \end{eqnarray}
Notice that the dual measure \eqref{dualmeasure} was arbitrarily defined without an overall coefficient. If instead we define, for $D>0$,
\begin{equation}\label{dualmeasurebis}
\mu_{\gamma'}(dz):=\left(\frac{D}{\alpha\Gamma(1-1/\alpha)}\right)^{\alpha} \int_0^{\infty}\eta\, \mathcal N_{\gamma'} (dz,d\eta),\,\, \gamma' >2,
\end{equation}
then identity \eqref{Laplace2} simply becomes
\begin{eqnarray}
\mathbb E\left[\exp\left(-u\,{\mu_{\gamma'}(\mathcal A)}\right)\big{\vert}{\mu_{\gamma}}\right]=\exp\left(-D \,u^{1/\alpha}{\mu_{\gamma}(\mathcal A)}\right),\label{Laplace2bis}
  \end{eqnarray}
while identity \eqref{Laplace3}  becomes
\begin{eqnarray}
\mathbb E\left[\exp\left(-u\,\frac{\mu_{\gamma'}(\mathcal A)}{\mu_{\gamma}^\alpha(\mathcal A)}\right)\big{\vert}{\mu_{\gamma}}\right]
=\exp\left(-D \,u^{1/\alpha}\right).\label{Laplace4}
  \end{eqnarray}
Upon identification of $k$ with $\mu_\gamma$ and $n$ with $\mu_{\gamma'}$, Eq.\eqref{Laplace4} is the exact continuous analogue of Eq.\eqref{eq:limitlawbis}, obtained in the scaling limit $k\to \infty, n\to \infty$ with $n/k^\alpha$ fixed. So we posit that the general definition  \eqref{dualmeasurebis} of the dual Liouville measure allows for a \emph{universal} description of the statistics of block-weighted random planar maps in the scaling limit, where a \emph{single non-universal} parameter $D$ depends on the particular discrete model at hand.   

\subsection{Liouville partition functions}\label{LQGsphere}
\subsubsection*{LQG partition function on the sphere} 
The partition function $\Pi_{\gamma,\hat g}^{(z_i\alpha_i)_i}(\lambda)$ of the Liouville conformal field theory (LCFT) on the Riemann sphere $\mathbb S$, for a random field $X$ with parameter $\gamma<2$ and spherical metric $\hat g$, has been explicitly constructed in Ref. \cite{MR3465434}, for a so-called `cosmological constant' (or `chemical potential' in statistical mechanics)   $\lambda >0$, the conjugate parameter to the sphere random Liouville measure $\mu_\gamma(\mathbb S)$. Its finiteness  requires the insertion of vertex functions $e^{\alpha_iX(z_i)}$ at  $p$ arbitrary points $z_i,i\in\{1,\cdots,p\}$, whose parameters $\alpha_i$  must obey the so-called Seiberg bounds 
\begin{equation}
\label{seiberg}
\sum_{i=1}^{p}\alpha_i-2Q>0;\quad \alpha_i<Q=2/\gamma+\gamma/2,\quad \forall i\in\{1,\cdots,p\}.
\end{equation}
As is well-known, these bounds imply the insertion of at least three points, \emph{i.e.},  $p\geq 3$. 

Another perspective, focusing on \emph{fixed-area} Liouville quantum spheres, has been proposed in Ref. \cite{DMS14}, and its equivalence to the standard Liouville CFT of \cite{MR3465434} has been established in Ref. \cite{zbMATH06796698}.   Notice that while the definition of finite `grand-canonical' partition functions $\Pi_{\gamma,\hat g}^{(z_i\alpha_i)_i}(\lambda)$ requires a number of insertions $p\geq 3$, the definition of `canonical' fixed-area ones only requires $p\geq 2$, with slightly modified Seiberg bounds, as detailed in  \cite[Lemma 2.5]{zbMATH06796698} and \cite[Lemma 3.10]{MR3465434}.

The partition function obeys the simple but fundamental scaling law \cite[Th. 3.4]{MR3465434} (see also \cite{PhysRevLett.66.2051}),  
\begin{equation}\label{scaling}
\Pi_{\gamma,\hat g}^{(z_i\alpha_i)_i}(\lambda)= \lambda^{\frac{2Q-\sum_i \alpha_i}{\gamma}}\Pi_{\gamma,\hat g}^{(z_i\alpha_i)_i}(1),
\end{equation}
where the $\lambda=1$ partition function $\Pi_{\gamma,\hat g}^{(z_i\alpha_i)_i}(1)$ is given by a finite explicit, albeit non exactly computable, expression.  Notice that due to the leftmost Seiberg bound in Eq. \eqref{seiberg}, $\Pi_{\gamma,\hat g}^{(z_i\alpha_i)_i}(\lambda)$ always diverges as $\lambda \to 0$. 
 
This partition function is standardly related to the partition function for a \emph{fixed} Liouville measure (area) $A$, denoted by $\mathcal Z_{\gamma,\hat g}^{(z_i\alpha_i)_i}(A)$,  by a Laplace transform,
$$
\Pi_{\gamma,\hat g}^{(z_i\alpha_i)_i}(\lambda)= \mathcal L\left\{\mathcal Z_{\gamma,\hat g}^{(z_i\alpha_i)_i}(A)\right\}(\lambda)=\int_0^{\infty}dA \,\mathcal Z_{\gamma,\hat g}^{(z_i\alpha_i)_i}(A)\exp(-\lambda A), 
$$
so that the fixed-area one is obtained by \emph{inverse} Laplace transform, as 
$$
\mathcal Z_{\gamma,\hat g}^{(z_i\alpha_i)_i}(A)= \mathcal L^{-1}\left\{\Pi_{\gamma,\hat g}^{(z_i\alpha_i)_i}(\lambda)\right\}(A)
=\int_{\mathcal C}\frac{d\lambda}{2\pi \mathrm{i}}\, \Pi_{\gamma,\hat g}^{(z_i\alpha_i)_i}(\lambda)\exp(\lambda A), 
$$
where the vertical contour $\mathcal C$ is taken to the right of the singularities of integrand $\Pi$. 

Recall the classical Laplace transform,
$$
\mathcal L^{-1}\left\{\frac{1}{\lambda^q}\right\}(A)=\frac{A^{q-1}}{\Gamma(q)}\vartheta(A), \quad \lambda >0, q>0,
$$
where $\vartheta$ denotes the Heaviside step function. We thus have from \eqref{scaling}
\begin{eqnarray}\nonumber
\mathcal Z_{\gamma,\hat g}^{(z_i\alpha_i)_i}(A)&=&  \Pi_{\gamma,\hat g}^{(z_i\alpha_i)_i}(1)\, \mathcal L^{-1}\left\{\lambda^{-\frac{\sum_i \alpha_i-2Q}{\gamma}}\right\}(A)\\ \label{laplaceinvP}
&=&\frac{\Pi_{\gamma,\hat g}^{(z_i\alpha_i)_i}(1)}{\Gamma\left((\sum_i \alpha_i-2Q)/\gamma\right)}A^{\frac{\sum_i \alpha_i-2Q}{\gamma} -1}\vartheta(A). 
\end{eqnarray}

\subsubsection*{Dual LQG partition function on the sphere} 
The partition function of the \emph{dual} Liouville theory on the sphere, with dual parameter $\gamma'=4/\gamma$, and a \emph{fixed} dual Liouville measure, $\mu_{\gamma'}(\mathbb S)=A'$, can then be accordingly defined as the `conditional' integral, 
\begin{equation}\label{partdual}
\widetilde{\mathcal Z}_{\gamma',\hat g}^{(z_i\alpha_i)_i}(A')= \int_0^{\infty}dA\, \mathbb E\left[\delta(\mu_{\gamma'}(\mathbb S)-A')\vert \mu_\gamma(\mathbb S)=A\right]\mathcal Z_{\gamma,\hat g}^{(z_i\alpha_i)_i}(A).
\end{equation}
Write the Dirac distribution as $\delta(x)=\int_{\mathcal C_0} d\lambda\, e^{\lambda x}$, 
where $\mathcal C_0$ is a vertical contour located to the right of the origin in the complex $\lambda-$plane. Then the above conditional measure reads
\begin{eqnarray}\nonumber
\mathbb E\left[\delta(\mu_{\gamma'}(\mathbb S)-A')\vert \mu_\gamma(\mathbb S)=A\right]&=&\int_{\mathcal C_0} d\lambda\, e^{\lambda A'} \mathbb E\left[e^{-\lambda \mu_{\gamma'}(\mathbb S)}\vert \mu_\gamma(\mathbb S)=A\right]\\\nonumber
&=& \mathcal L^{-1}\left\{ \mathbb E\left[e^{-\lambda \mu_{\gamma'}(\mathbb S)}\vert \mu_\gamma(\mathbb S)=A\right] \right\} (A')\\ \label{LaplaceinvtoS} 
&=& \mathcal L^{-1}\left\{e^{-D \lambda^{1/\alpha} A}\right\} (A'),
\end{eqnarray}
where the last expression results from identity \eqref{Laplace2bis}. 

Recalling now that function $\wp$ in \eqref{eq:probscaling3} has for \emph{direct} Laplace transform \eqref{eq:Laplace},  
$\mathcal{L}\{\wp\}(\lambda)=e^{-D\, \lambda^{\frac{1}{\alpha}}}$, we finally get the explicit form,
\begin{equation}
\mathbb E\left[\delta(\mu_{\gamma'}(\mathbb S)-A')\vert \mu_\gamma(\mathbb S)=A\right]=\frac{\vartheta(A')}{A'}\SF_{\frac{1}{\alpha}}
\left(\frac{D A}{{A'}^{\frac{1}{\alpha}}}\right)\ .
\label{eq:probscaling4}
\end{equation}

Gathering Eqs.~\eqref{laplaceinvP}, \eqref{partdual}  and \eqref{eq:probscaling4}  yields 
\begin{eqnarray}\nonumber
\widetilde{\mathcal Z}_{\gamma',\hat g}^{(z_i\alpha_i)_i}(A')&=& \int_0^{\infty}dA\,\frac{\vartheta(A')}{A'}\SF_{\frac{1}{\alpha}}
\left(\frac{D A}{{A'}^{\frac{1}{\alpha}}}\right) \mathcal Z_{\gamma,\hat g}^{(z_i\alpha_i)_i}(A)\\ \nonumber
&=& \frac{\Pi_{\gamma,\hat g}^{(z_i\alpha_i)_i}(1)}{\Gamma\left((\sum_i \alpha_i-2Q)/\gamma\right)}\frac{\vartheta(A')}{A'}\int_0^{\infty}dA\,A^{\frac{\sum_i \alpha_i-2Q}{\gamma} -1}\SF_{\frac{1}{\alpha}}
\left(\frac{D A}{{A'}^{\frac{1}{\alpha}}}\right)\\
&=& \frac{\Pi_{\gamma,\hat g}^{(z_i\alpha_i)_i}(1)}{\Gamma\left((\sum_i \alpha_i-2Q)/\alpha \gamma\right)}\frac{\vartheta(A')}{A'}\left(\frac{{A'}^{\frac{1}{\alpha}}}{D}\right)^{\frac{\sum_i \alpha_i-2Q}{\gamma}} \label{partdualbis},
\end{eqnarray}
where use was made of  \eqref{A2} and of the $\delta-$moment integral \eqref{A7}. Notice that the existence of this moment requires  $\delta>-2$, hence only $(\sum_i \alpha_i-2Q)/\gamma >-1$, which is weaker than the Seiberg bound \eqref{seiberg}, and requires the insertion of at least $p=2$ points only. This echoes the introduction of doubly-rooted maps in Sec. \ref{sec:distkfixednscal}.

Owing to $\alpha\gamma=\gamma'$, the dual partition function \eqref{partdualbis} finally takes the simple form for $\gamma'>2$,
\begin{eqnarray}\label{partdualter}
\widetilde{\mathcal Z}_{\gamma',\hat g}^{(z_i\alpha_i)_i}(A')= \frac{\Pi_{\gamma,\hat g}^{(z_i\alpha_i)_i}(1)}{\Gamma\left((\sum_i \alpha_i-2Q)/\gamma'\right)}\frac{\vartheta(A')}{A'}\left(\frac{A'}{D^\alpha}\right)^{\frac{\sum_i \alpha_i-2Q}{\gamma'}},
\end{eqnarray}
 a formula entirely similar to that of partition function \eqref{laplaceinvP} for $\gamma<2$, up to the substitution of $\gamma'$ to $\gamma$, and the presence of an  extra factor $D^{-\frac{\sum_i \alpha_i-2Q}{\gamma}}$.  
 
The \emph{dual} `grand-canonical' partition function can in turn be obtained as the Laplace transform of \eqref{partdualter},
\begin{eqnarray*}
\widetilde{\Pi}_{\gamma,\hat g}^{(z_i\alpha_i)_i}(\lambda)&=& \mathcal L\left\{\widetilde{\mathcal Z}_{\gamma',\hat g}^{(z_i\alpha_i)_i}(A')\right\}(\lambda)=\int_0^{\infty}dA' \,\widetilde{\mathcal Z}_{\gamma',\hat g}^{(z_i\alpha_i)_i}(A')\exp(-\lambda A')\\
&=&(D^\alpha\lambda)^{\frac{2Q-\sum_i \alpha_i}{\gamma'}}\Pi_{\gamma,\hat g}^{(z_i\alpha_i)_i}(1),
\end{eqnarray*}
in full agreement with \cite[Eq.(4.8)]{MR3465434}, up to the substitution of our general normalization coefficient $D$ to the particular one $\Gamma(1-\alpha')/\alpha'$ used there. 
 \subsubsection*{String susceptibility exponents}
 The so-called string susceptibility exponent $\gs$  is usually defined by looking at $p\geq 3$ punctures of identical  parameters $\alpha_i=\gamma$, so that 
 $$
 \frac{\sum_{i=1}^p \alpha_i-2Q}{\gamma}=p-1-\frac{4}{\gamma^2}=p-1-\alpha,
 $$
 \begin{eqnarray}\label{gs}
 {\mathcal Z}_{\gamma,\hat g}^{(z_i\gamma)_i}(A) \propto 
 A^{p-2-\alpha}=A^{p+\gs-3},
 \end{eqnarray}
 with  $\gs=1-\alpha=1-4/\gamma^2$, thus  recovering the former expression \eqref{eq:valgs}. 
  
As for the dual partition function \eqref{partdualter} equipped with the same $p$ punctures of parameters $\alpha_i=\gamma$, one has the scaling  
$$
 \frac{\sum_{i=1}^p \alpha_i-2Q}{\gamma'}=\frac{\gamma}{\gamma'}\big(p-1-\frac{4}{\gamma^2}\big)=\frac{1}{\alpha}(p-1-\alpha),
 $$ 
\begin{eqnarray}\label{gsp}\widetilde {\mathcal Z}_{\gamma',\hat g}^{(z_i\gamma)_i}(A') \propto 
{(A'})^{\,\frac{1}{\alpha}(p-1-\alpha)-1}={(A')}^{p(1-\Delta'_0)+\gsp-3},
\end{eqnarray}
with $\gsp=1-1/\alpha=1-4/{\gamma'}^{\,2}$, thus recovering \eqref{eq:valgsp} and \eqref{eq:dualgs}, together with the single puncture dual quantum weight $\Delta'_0=1-1/\alpha=\gsp$ (see,  \emph{e.g.}, \cite[Sec. 3.2]{DG25}).  
\subsubsection*{Dual partition function ratios} Finally, it is possible to readily extract from the two dual partition functions \eqref{laplaceinvP} and \eqref{partdualbis} a quasi-universal ratio, \emph{i.e.},
 \begin{eqnarray}\label{ratiodual}
\frac{\widetilde{\mathcal Z}_{\gamma',\hat g}^{(z_i\alpha_i)_i}(A')}{{\mathcal Z}_{\gamma,\hat g}^{(z_i\alpha_i)_i}(A)}= \frac{\Gamma\left((\sum_i \alpha_i-2Q)/\gamma\right)}{\Gamma\left((\sum_i \alpha_i-2Q)/\alpha\gamma\right)}\frac{A}{A'}\left(\frac{{A'}^{\frac{1}{\alpha}}}{D A}\right)^{\frac{\sum_i \alpha_i-2Q}{\gamma}},
\end{eqnarray}
where, in addition to the conformally invariant areas $A$ and $A'$, only appears the parameter $D$ that characterizes the particular (dual) random planar map model at hand. Specifying to the case \eqref{gs} and \eqref{gsp} above of $p\geq 3$ punctures of identical parameters $\alpha_i=\gamma$, we obtain 
 \begin{eqnarray}\label{ratiodual2}
\frac{\widetilde{\mathcal Z}_{\gamma',\hat g}^{(z_i\gamma)_i}(A')}{{\mathcal Z}_{\gamma,\hat g}^{(z_i\gamma)_i}(A)}= \frac{\Gamma(p-1-\alpha)}{\Gamma\left((p-1-\alpha)/\alpha\right)}\frac{A}{A'}\left(\frac{{A'}^{\frac{1}{\alpha}}}{D A}\right)^{p-1-\alpha},
\end{eqnarray}
a Liouville quantum duality prediction which entirely agrees with the purely combinatorial ratio $\mathcal{R}^{(p)}(n,k)$ \eqref{eq:ratio}, upon the natural identification of $(A,A')$ with $(k,n)$.

\section{Duality and multifractality}
\label{sec:multifractality}
In this section, we focus on two different types of \emph{multifractality} \cite{1986PhRvA..33.1141H,PhysRevA.34.1601,FP} associated with the Liouville quantum measures, both in the standard $\gamma<2$ phase and dual
 $\gamma'>2$ one. The first notion concerns the moments of the LQG measures of Euclidean balls, $\mu_{\gamma<2}(B_\varepsilon(z))$ and $\mu_{\gamma'>2}(B_\varepsilon(z))$, the second those of the Lebesgue measures of quantum balls, $\mu_0(B^\delta(z))$ as defined via \eqref{qb} and $\mu_0(\tilde B^\delta(z))$ via \eqref{btildeprime}. 
 
\subsection{Multifractality of the $\gamma-$LQG measure for $\gamma<2$}\label{Sec:1stspectrum}
For $\gamma <2$, and $q\in \mathbb R$, the $q-$moment of the random measure $\mu_\gamma$ of a (closed) ball $B_\varepsilon(z)$, of radius $\varepsilon$ and centered at $z\in \mathcal D$, is finite only if $q<4/\gamma^2$ \cite{MR829798}. Its scaling behavior is then given by
\begin{eqnarray}\label{moments}
\mathbb E \left[\mu_{\gamma} (B_\varepsilon(z))^q\right]\asymp \varepsilon^{\,\xi_\gamma(q)} \quad \mathrm{as}\,\,\,\varepsilon\to 0,\quad q \in \left(-\infty,{4}/{\gamma^2}\right), 
\end{eqnarray}
where the $\asymp$ symbol stands for the asymptotic equivalence of the logarithms of respective arguments, and where 
the scaling exponent $\xi_\gamma(q)$ is given  by the well-known quadratic formula (see, \emph{e.g.}, \cite[Eq. (3.1)]{RV} and \cite[Thm. 2.4]{RV3}), 
\begin{eqnarray}\label{xi}
\xi_\gamma(q)=\left(2+\frac{\gamma^2}{2}\right)q-\frac{\gamma^2}{2}q^2, \quad q \in \left(-\infty,\frac{4}{\gamma^2}\right). 
\end{eqnarray}The (concave) non-linearity of $\xi_\gamma(q)$ is the very manifestation of a \emph{multifractal} behavior associated with the moments of measure $\mu_\gamma$ \cite{mandelbrot1974multiplications,Mandelbrot1974331,zbMATH03856061,FP,1986PhRvA..33.1141H,PhysRevA.34.1601}. 

A standard way to quantify multifractality goes via the definition of the $L^q-$\emph{spectrum} of a given measure $\mu$,
\begin{equation}\label{astau}
\tau(q)=\underset{r\to 0^+}{\lim \inf}\, \frac{\log\sup\{\sum_{i\in I} \mu(B_\varepsilon(z_i))^q\}}{\log \varepsilon},
\end{equation}
where $\left(B_\varepsilon(z_i)\right)_{i\in I}$ is a countable family of disjoint closed balls with radius $\varepsilon$ centered at $z_i\in \mathcal D$, and the supremum is taken over all such families.
In the Parisi-Frisch \cite{FP}, Halsey \emph{et al.} multifractal formalism \cite{1986PhRvA..33.1141H,PhysRevA.34.1601}, 
the $L^q-$spectrum can be related to the local singular behavior of the measure. Indeed for $\beta\geq 0$ define the fractal subset,
\begin{equation}\label{asE}
{\mathcal E}(\beta)=\{z\in \mathcal D:\underset{\varepsilon\to 0^+}{\lim}\,\frac{\log \mu(B_\varepsilon(z))}{\log \varepsilon}=\beta\},\quad \beta \geq 0.
\end{equation}
Its Hausdorff dimension, $f(\beta)=\dim_{\mathcal H}{\mathcal E}(\beta)$, provides the so-called \emph{multifractal dimension spectrum}. When the multifractal formalism applies (see, \emph{e.g.}, \cite{BARRAL202376,BARRAL2023281}), $f(\beta)$  can be obtained via the Legendre transform (for concave functions),  truncated at $0$,
\begin{equation}\label{asf0}
f(\beta)=\underset{q\in \mathbb R}{\inf}\{\beta q-\tau(q)\}\vee 0.
\end{equation}

For \emph{random} fractals, one usually distinguishes \emph{almost sure} spectra (sometimes also called \emph{quenched} in statistical mechanics), such as  \eqref{astau} and \eqref{asf0}, from \emph{expected} ones (also called \emph{annealed}). 
For a random measure such as the LQG measure $\mu_\gamma$ on the sphere $\mathbb S$, or in a domain $\mathcal D$,\footnote{In a domain $\mathcal D$ with  boundary  $\partial \mathcal D$, one uses a distance restriction, $\forall i, \mathrm{dist}(x_i, \partial \mathcal D) \geq r_0 >r \to 0^+$. Another spectrum can be associated with the Liouville boundary measure (for a GFF with Neumann boundary conditions \cite{2008arXiv0808.1560D}), by using a boundary covering whose balls are instead centered on $\partial \mathcal D$.} the \emph{expected} $L^q-$spectrum, $\bar \tau_\gamma$,  is then defined as,
\begin{equation}\label{tauq}
\overline{\tau}_\gamma(q)=\underset{\varepsilon\to 0^+}{\lim} \frac{\log\sum_{i\in I} \mathbb E [\mu_\gamma(B_\varepsilon(z_i))^q]}{\log \varepsilon},
\end{equation}
provided that $q<4/\gamma^2$. Here, all expected moments in a maximally covering ball set $I$ have same power law  \eqref{moments}, while the number of these radius-$\varepsilon$ balls  obviously scales as $\varepsilon^{-2}$. This immediately yields 
\begin{equation}\label{taubar}
\overline{\tau}_\gamma(q)=\xi_\gamma(q)-2, \quad q<4/\gamma^2.
\end{equation}
The associated \emph{expected} dimension spectrum, $\bar f_\gamma$, is then given by the Legendre transform 
\begin{equation}\label{legendre}
\bar f_\gamma(\beta)=\underset{q\in \mathbb R, \, q<4/\gamma^2}{\inf}\{\beta q-\bar \tau_\gamma(q)\}\quad \beta\geq 0.
\end{equation} Owing to the derivability of the functions, it is obtained from the symmetric Legendre equations, 
\begin{equation}\label{barf}
\begin{cases}
\bar f_\gamma(\beta)+\overline{\tau}_\gamma(q)=\beta q,&\\ 
\beta=\frac{\partial}{\partial q}\overline{\tau}_\gamma(q),\quad q=\frac{\partial}{\partial\beta}\bar f_\gamma(\beta).&
\end{cases}
\end{equation}
Eqs. \eqref{xi} and \eqref{barf}  then yield 
\begin{eqnarray}\label{fbar}
\bar f_\gamma(\beta)&=&2-\frac{1}{2\gamma^2}\left(2+\frac{\gamma^2}{2}-\beta\right)^2,\quad \beta \geq 0,\\ \label{qbeta}
\beta &=& 2+\frac{\gamma^2}{2}-\gamma^2q,\quad q<\frac{4}{\gamma^2}=\alpha.
\end{eqnarray}
 Note that the two conditions in these equations together yield here 
 \begin{equation}\label{condition}
 q\leq  \inf\left\{\frac{1}{2}+\frac{2}{\gamma^2},\frac{4}{\gamma^2}\right\} = \inf\left\{\frac{1}{2}(1+\alpha),\alpha\right\}=\frac{1}{2}(1+\alpha) \quad \mathrm{since}\quad \alpha >1.
 \end{equation}
 The \emph{expected} dimension spectrum $\bar f_\gamma$ \eqref{fbar} then develops two \emph{negative} parts, a left one for $0\leq \beta <(2-\gamma)^2/2$ corresponding to $2/\gamma<q\leq (1+\alpha)/2$, , and a right one for $ \beta>(2+\gamma)^2/2$, corresponding to $q<-2/\gamma$. 

The \emph{almost sure} (\emph{a.s.}) dimension spectrum $f_\gamma$ is then obtained by cutting out these parts, \emph {i.e.}, by the simple rule,
\begin{equation}\label{fbarf}
f_\gamma=\bar f_\gamma\vee 0.
\end{equation}
The \emph{a.s.}\ multifractal dimension spectrum of $\gamma-$LQG is thus obtained from \eqref{fbar} as
\begin{align}\label{fas}
f_\gamma(\beta)&=2-\frac{1}{2\gamma^2}\left(2+\frac{\gamma^2}{2}-\beta\right)^2\quad \mathrm{for}\quad
 \beta\in\left[\frac{1}{2}(2-\gamma)^2,\frac{1}{2}(2+\gamma)^2\right]\\ \nonumber
&=0\quad \mathrm{otherwise}.
\end{align}
The \emph{almost sure} $L^q-$spectrum is in turn obtained from the \emph{a.s.}\ dimension spectrum \eqref{fas}, by using the \emph{inverse} Legendre transform, 
\begin{equation}\label{invlegendre}
\tau_\gamma(q)=\underset{\beta \in \mathbb R^+}{\inf}\{\beta q-f_\gamma(\beta)\},
\end{equation}
 which yields
  \begin{equation} \label{tauas}
 \tau_\gamma(q)= \begin{cases}\frac{1}{2}\left(2+\gamma\right)^2 q \quad \mathrm{if} \quad q\leq -\frac{2}{\gamma},&\\
\xi_\gamma(q)-2 \quad \quad  \mathrm{if} \quad q \in[-\frac{2}{\gamma},\frac{2}{\gamma}],&\\
 \frac{1}{2}\left(2-\gamma\right)^2 q \quad \mathrm{if} \quad q\geq \frac{2}{\gamma}.&
 \end{cases}
 \end{equation}
  
\begin{figure}[h!]
\includegraphics[width=.48\linewidth]{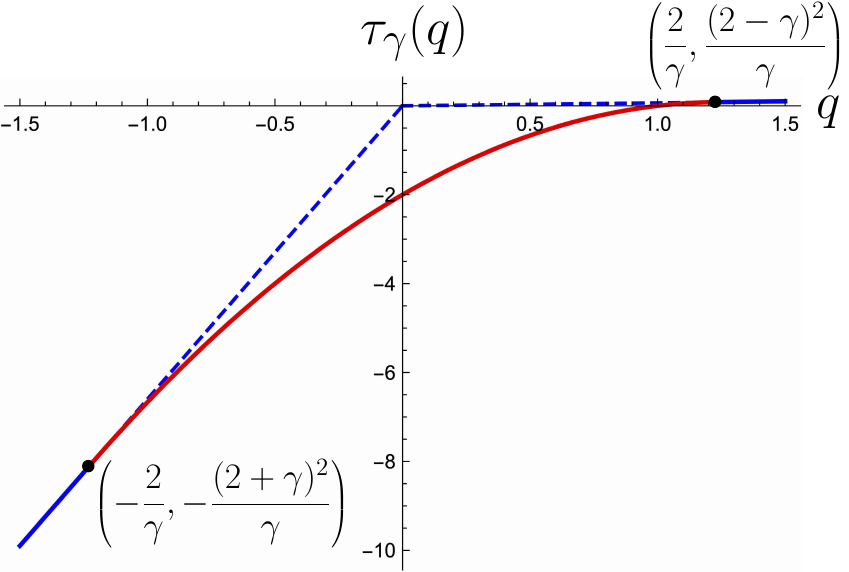}   \hfill
\includegraphics[width=.48\linewidth]{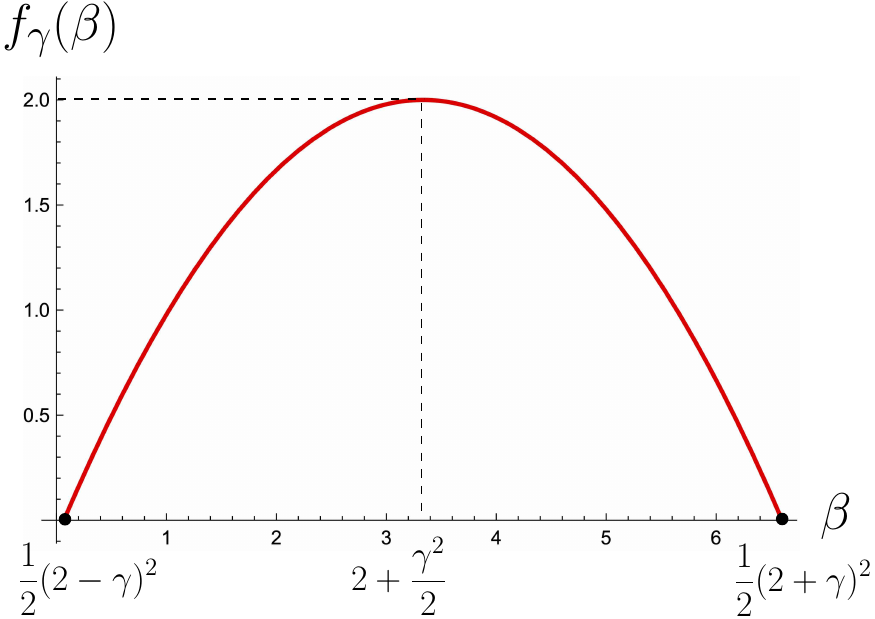}
   \caption{Left: \emph{a.s.}\ LQG $L^q-$spectrum $\tau_\gamma$. Right:  \emph{a.s.}\ LQG multifractal dimension spectrum $f_\gamma$. Both curves numerically correspond to the $\gamma=\sqrt{8/3}$ case.}
 \label{fig:tauofq}
\end{figure}

 Note that the link stated in Eq. \eqref{fbarf} between the expected and \emph{a.s.}\ dimension spectra $\bar f_\gamma$ and $f_\gamma$, although \emph{heuristic} here, is borne out by the case of the multifractal spectra of the celebrated Schramm-Loewner evolution (SLE). In the SLE bulk case, the expected multifractal spectrum, first obtained in \cite{Duplantier00} and rigorously established in \cite{BS,MR3638311}, does obey relationship \eqref{fbarf} with the \emph{a.s.}\ spectrum,  rigorously derived in \cite{gwynne2018}, with the help of Miller and Sheffield's SLE imaginary geometry.   In the case here  of the Liouville quantum measure, the  almost sure spectra \eqref{fas} and  \eqref{tauas}, which had been conjectured in \cite{RV3}, are rigorously established in the recent work by Bertacco  \cite[Theorem 3.1, Lemma 3.9]{10.1214/22-EJP893} and \cite{Bertacco_2025}. Both $\tau_\gamma$ and $f_\gamma$ \emph{a.s.}\ spectra are displayed in Fig.~\ref{fig:tauofq}. 
\subsection{Multifractality of the $\gamma'-$LQG measure for $\gamma'>2$}\label{sec:secspectrum}
Recall the relationship \eqref{momentsduaux} between moments of the LQG random measures of any Borelian subset $\mathcal A$, in the subcritical and critical dual phases,
\begin{eqnarray}\label{dualmoments} \mathbb E\left[\big(\mu_{\gamma'}(\mathcal A)\big)^q\right]
 =\frac{\Gamma(1-\alpha q)[-\Gamma(-1/\alpha)]^{\alpha q}}{\Gamma(1-q)}\mathbb E\left[\big(\mu_{\gamma}(\mathcal A)\big)^{\alpha q}\right]\quad \alpha q\in (-\infty,1),
  \end{eqnarray}
 where $\alpha=4/\gamma^2=\gamma'^2/4 >1$.
We thus find the expected moment scaling \begin{eqnarray}\nonumber 
&&\mathbb E \left[\mu_{\gamma'} (B_\varepsilon(z))^q\right]\asymp \varepsilon^{\,\xi_{\gamma'}(q)}, \quad \mathrm{as}\,\,\,\varepsilon\to 0,
\\ \label{xiprime}
&&\xi_{\gamma'}(q)=\xi_{\gamma}(\alpha q)=\left(2+\frac{\gamma'^2}{2}\right)q-\frac{\gamma'^2}{2}q^2, \quad q \in \left(-\infty,\frac{4}{\gamma'^2}\right), 
\end{eqnarray}
which has exactly the same form as $\xi_\gamma$ \eqref{xi}, up to substitution of $\gamma'$ to $\gamma$. (See also \cite[Sec. 4, Eq. (23)]{pre06228485}.) 

At the critical dual point the expected $L^q-$spectrum,  $\bar \tau_{\gamma'}=\xi_{\gamma'}-2$, defined similarly to \eqref{tauq}, obeys the duality equation
\begin{equation}\nonumber\label{tautauprime}
\bar \tau_{\gamma'}(q)=\xi_{\gamma'}(q)-2=\xi_{\gamma}(\alpha q)-2=\bar \tau_{\gamma}(\alpha q),\quad \alpha q<1,
\end{equation}
whereas the expected dimension spectrum $\bar f_{\gamma'}$, 
\begin{equation}\label{legendreprime}
\bar f_{\gamma'}(\beta)=\underset{q\in \mathbb R, \,\alpha q<1}{\inf}\{\beta q-\bar \tau_{\gamma'}(q)\}\quad \beta\geq 0,
\end{equation} 
is given by the Legendre transform \eqref{barf} as 
\begin{equation}\label{barfprime}
\begin{cases}
\bar f_{\gamma'}(\beta)=\bar f_\gamma(\beta/\alpha),\quad \beta \geq 0,&\\
\beta=\frac{\partial}{\partial q} \bar \tau_{\gamma}(\alpha q)
=2+2\alpha -4\alpha q,\quad \alpha q<1.&
\end{cases}
\end{equation}
where $\bar f_\gamma$ is given by \eqref{fbar}. In contradistinction to \eqref{condition}, we have here
 \begin{equation}\nonumber\label{conditionprime}
 q <   \inf\left\{(1+\alpha)/2\alpha,1/\alpha\right\}=1/\alpha \quad \mathrm{since}\quad \alpha >1,
 \end{equation}
 the second inequality in \eqref{barfprime} implying the first one and the range $\beta >2\alpha-2>0$.  
 At the $q^*=1/\alpha$ transition point, $\bar \tau_{\gamma'}$  reaches its  maximal value $\bar \tau_{\gamma'}(1/\alpha)=\bar \tau_\gamma (1)=0$, while its Legendre transform \eqref{legendreprime}  yields the completion of  \eqref{barfprime} by the \emph{linear part}, 
 \begin{equation}\label{linear}
 \bar f_{\gamma'}(\beta)=\beta /\alpha, \quad \mathrm{for} \quad 0\leq \beta \leq 2\alpha-2.
 \end{equation}
 As expected in the multifractal formalism, the derivative of $\bar f_{\gamma'}$ is continuous at the transition point, where the left-derivative of \eqref{linear} matches  the right-derivative of \eqref{barfprime}, since $1/\alpha=q^*=\frac{\partial \bar f_{\gamma'}}{\partial \beta}(\beta=2\alpha -2)$.
 
\begin{figure}[h!]
\includegraphics[width=.5\linewidth]{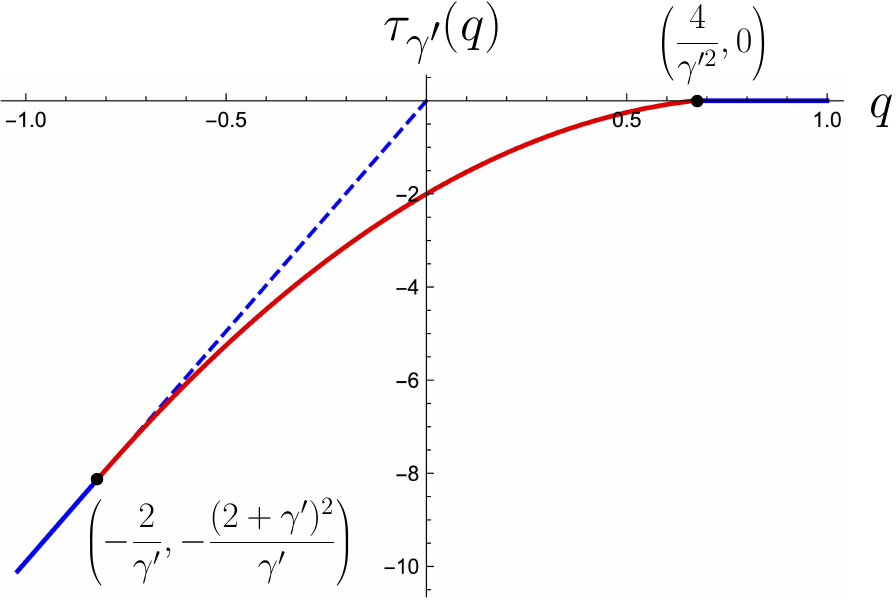}   \hfill
\includegraphics[width=.5\linewidth]{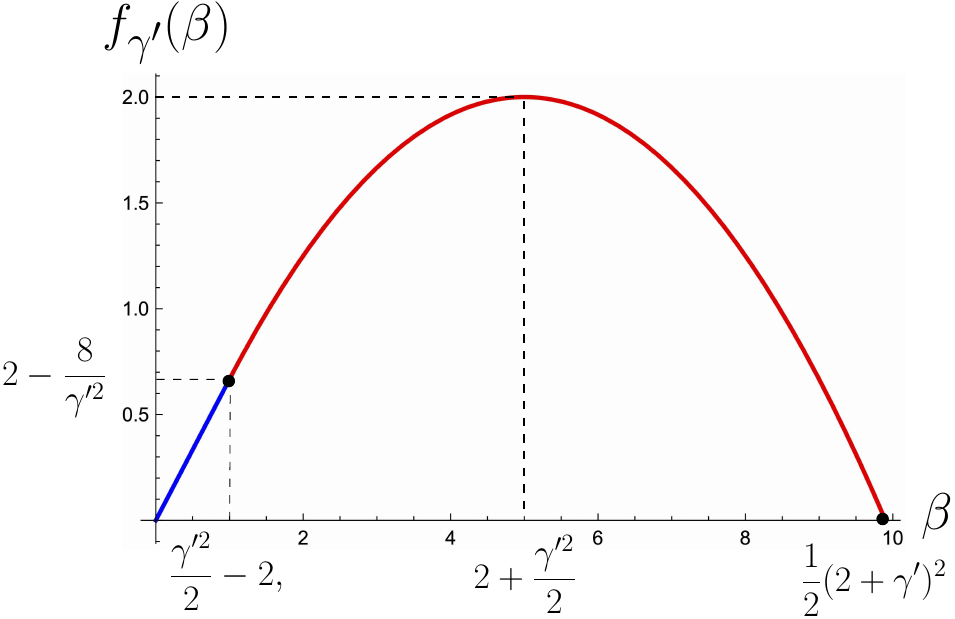}
    \caption{Left: \emph{a.s.}\ LQG $L^q-$spectrum $\tau_{\gamma'}$. Right:  \emph{a.s}. LQG multifractal dimension spectrum $f_{\gamma'}$. Both curves numerically correspond to the dual $\gamma'=\sqrt{6}$ case.}
 \label{fig:tauofqdual}
\end{figure}

 Finally, the \emph{almost sure} dimension spectrum, $f_{\gamma'}$, is predicted by the same truncation as in \eqref{fbarf}, $f_{\gamma'}(\beta)=\bar f_{\gamma'}(\beta)\vee 0$, as
 \begin{equation}  \label{fasprime}
 f_{\gamma'}(\beta)=\begin{cases}\frac{4}{\gamma'^2}\beta \quad \quad \quad \quad \quad \quad \quad \quad  \quad \quad &\mathrm{for}\quad
 \beta\in\left[0,\frac{\gamma'^2}{2}-2\right], \\ 
 2-\frac{1}{2\gamma'^2}\left(2+\frac{\gamma'^2}{2}-\beta\right)^2\quad &\mathrm{for}\quad
 \beta\in\left[\frac{\gamma'^2}{2}-2,\frac{1}{2}(2+\gamma')^2\right],\\
   0 \quad  \quad \quad \quad \quad \quad \quad \quad \quad  \quad \quad \,\,\,&\mathrm{for}\quad  \beta \in \left[\frac{1}{2}(2+\gamma')^2,+\infty\right).
  \end{cases}
\end{equation}
 In turn, the \emph{almost sure} $L^q-$spectrum, $\tau_{\gamma'}(q)$, is obtained  by an inverse Legendre transform of \eqref{fasprime} similar to  \eqref{invlegendre},  as
   \begin{equation} \label{tauasprime}
 \tau_{\gamma'}(q)= \begin{cases}
 \frac{1}{2}\left(2+\gamma'\right)^2 q \quad \quad  &\mathrm{if} \quad q\in (-\infty, -\frac{2}{\gamma'}],\\
\xi_{\gamma'}(q)-2 \quad \quad  &\mathrm{if} \quad q \in[-\frac{2}{\gamma'},\frac{4}{\gamma'^2}],\\
0 \quad  &\mathrm{if} \quad q \in[\frac{4}{\gamma'^2},+\infty).
 \end{cases}
 \end{equation}
Both  predicted \emph{a.s.}\ spectra $\tau_{\gamma'}$ and $f_{\gamma'}$  are displayed in Fig.~\ref{fig:tauofqdual}. By analogy with previous work on the multifractality of L\'evy processes \cite{BARRAL2007437}, \eqref{tauasprime} was also conjectured in \cite{pre06228485}. A rigorous derivation of these spectra in the almost sure setting, similar to that in \cite{10.1214/22-EJP893} for the multifractal spectra  of the Liouville random measure in the $\gamma<2$ case, would be highly desirable.
\subsection{Multifractality of $\gamma-$quantum balls for $\gamma<2$}\label{Sec:gammaqballs}
For fixed $z$ in $\mathcal D$, and $\delta>0$ small enough, the \emph{$\delta-$quantum ball} $B^\delta(z)$ is defined as the \emph{Euclidean} ball  centered at $z$ with  random measure $\delta$, 
$$\mu_\gamma(B^\delta(z))=\delta.$$
Its Euclidean radius yields a positive random variable, attached to point $z$, defined, up to constant factor, by
$\mu_0\,(B^\delta(z))^{1/2},$
where $\mu_{\gamma=0}$ is the (non random) Lebesgue measure. 
The moments of this random radius have been rigorously  studied in \cite{2008arXiv0808.1560D} and shown to scale as
\begin{equation}\label{epsq}
\mathbb E[\mu_0\,(B^\delta(z))^{q/2}]\asymp \delta^{\Delta_\gamma(q)} \quad \mathrm{as}\quad \delta \to 0,
\end{equation}
with the family of exponents,
\begin{align}\label{Deltagamma}
&\Delta_\gamma(q)=\frac{1}{\gamma}\left((2q+a_\gamma^2)^{1/2}-a_\gamma\right), \quad 2q \geq -a_\gamma^2,\\ \label{agamma}
&a_\gamma:=\frac{2}{\gamma}-\frac{\gamma}{2} >0\quad \mathrm{for} \quad\gamma<2.
\end{align}
This equation is the celebrated Knizhnik-Polyakov-Zamolodchikov (KPZ) equation \cite{MR947880}, where $\Delta_\gamma(q)$ \eqref{Deltagamma} is the \emph{quantum exponent} associated with an Euclidean scaling exponent $q/2$ \cite{2009arXiv0901.0277D,2008arXiv0808.1560D}, with
\begin{equation}\label{KPZ}
q=\frac{\gamma^2}{2} \Delta_\gamma^2 +\left(2-\frac{\gamma^2}{2}\right)\Delta_\gamma, \quad \Delta_\gamma=\Delta_\gamma(q).
\end{equation}
It is thus natural to define a novel $L^q-$spectrum associated with these moments, as 
\begin{equation}\label{astautilde}
\widetilde \tau_\gamma(q)=\underset{\delta\to 0^+}{\lim \inf}\, \frac{\log\sup\{\sum_{i\in I} \mu_0(B^\delta(z_i))^{q/2}\}}{\log \delta},
\end{equation}
where $\left(B^\delta(z_i)\right)_{i\in I}$ is a countable family of disjoint closed quantum balls with $\gamma-$LQG measure $\delta$ centered at $z_i\in \mathcal D$, and where the supremum is taken over all such families.

For $\beta\geq 0$, define now the fractal subset,
\begin{equation}\label{asEprime}
\widetilde{\mathcal E}(\beta)=\{z\in \mathcal D:\underset{\delta\to 0^+}{\lim}\,\frac{\log \mu_0(B^\delta(z))^{1/2}}{\log \delta}=\beta\},\quad \beta \geq 0.
\end{equation}
Its Hausdorff dimension, $\widetilde f_\gamma(\beta)=\dim_{\mathcal H}\widetilde {\mathcal E}(\beta)$, provides the \emph{multifractal dimension spectrum}. As above, the multifractal formalism posits that 
$\widetilde f_\gamma$  can be obtained via the Legendre transform of $\widetilde \tau$,  truncated at $0$,
\begin{equation}\label{asf}
\widetilde f_\gamma(\beta)=\underset{q\in \mathbb R}{\inf}\{\beta q-\widetilde\tau_\gamma(q)\}\vee 0.
\end{equation}

In complete similarity with what was done in Section \ref{Sec:1stspectrum} above, define the \emph{expected} $L^q-$spectrum  as,
\begin{equation}\label{exptildetau}
\overline{\widetilde \tau}_\gamma(q)=\underset{\delta\to 0^+}{\lim}\, \frac{\log\{\sum_{i\in I} \mathbb E\,[\mu_0(B^\delta(z_i))^{q/2}]\}}{\log \delta}.
\end{equation}
Owing to \eqref{epsq}, and since the number of $\delta-$quantum balls in a covering $I$ scales as $\delta^{-1}$, we obviously have 
 \begin{equation}\label{exptildetaug}
\overline{\widetilde \tau}_\gamma(q)=\Delta_\gamma(q)-1,\quad 2q\geq -a_\gamma^2.
\end{equation}
The associated \emph{expected} dimension spectrum, $\overline{\widetilde f_\gamma}$, is then given by the Legendre transform of \eqref{exptildetau} 
\begin{equation}\label{legendretilde}
\overline{\widetilde f_\gamma}(\beta)=\underset{q\in \mathbb R}{\inf}\{\beta q-\overline{\widetilde \tau}_\gamma(q)\},\quad 2q\geq -a_\gamma^2, \quad \beta\geq 0.
\end{equation} 
It can be computed with the help of the symmetric Legendre equations \eqref{barf}, which yield here
\begin{equation}\label{barftilde}
\overline{\widetilde f_\gamma}(\beta)=1+\frac{1}{\gamma}a_\gamma- \frac{1}{2}a_\gamma^2\beta-\frac{1}{2\gamma^2\beta},
\end{equation}
together with $\beta=\frac{\partial}{\partial q}\overline{\widetilde \tau}_\gamma(q)=\frac{1}{\gamma}(2q+a_\gamma^2)^{-1/2}\in (0,\infty)$.

\begin{figure}[h!]
\centering
\includegraphics[width=.7\linewidth]{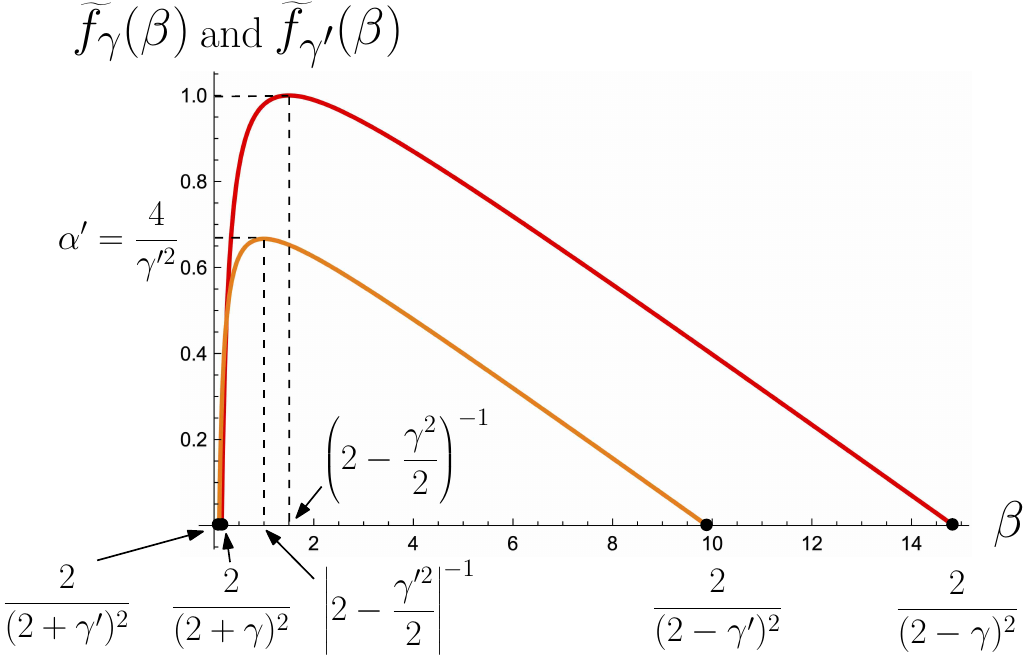}   
   \caption{Red: \emph{a.s.}\ dimension spectrum $\widetilde f_\gamma$. Its maximum value is $1$, the dimension of an LQG surface in quantum ball units. Orange: dual \emph{a.s.}\ dimension spectrum $\widetilde f_{\gamma'}$. Its maximal value, $4/\gamma'^2=\alpha'<1$, corresponds to the part of the dual LQG surface without localized area, \emph{i.e.}, to the principal bubble \cite{2009arXiv0901.0277D,2008ExactMethodsBD}, or root block \cite{DG25}. The ($\gamma=\sqrt{8/3}$, $\gamma'=\sqrt{6}$) dual cases are depicted here.}
 \label{fig:ftildeofbeta}
\end{figure}

This expected dimension spectrum possesses two zeroes, at $\beta_{\pm}=\frac{1}{\gamma}\frac{1}{Q_\gamma\pm 2}=\frac{2}{(2\pm\gamma)^2}$, corresponding respectively to exponents $q_{\pm}=2(2\pm Q_\gamma)$, where $Q_\gamma=a_\gamma+\gamma=2/\gamma+\gamma/2$. Assuming as before that the \emph{almost sure} dimension spectrum $\widetilde f_\gamma$ is obtained by the rule
\begin{equation}\label{tildefbartildef}
\widetilde f_\gamma=\overline{\widetilde f_\gamma}\vee 0,
\end{equation}
we get
\begin{align}\label{tildefas}
\widetilde f_\gamma(\beta)&=1+\frac{1}{\gamma}a_\gamma- \frac{1}{2}a_\gamma^2\beta-\frac{1}{2\gamma^2\beta}\quad \mathrm{for}\quad
 \beta\in\left[\frac{2}{(2+\gamma)^2},\frac{2}{(2-\gamma)^2}\right]\\ \nonumber
&=0\quad \mathrm{otherwise}.
\end{align}
This spectrum is illustrated in Fig.~\ref{fig:ftildeofbeta}. 

\begin{figure}[h!]
\centering
\includegraphics[width=.7\linewidth]{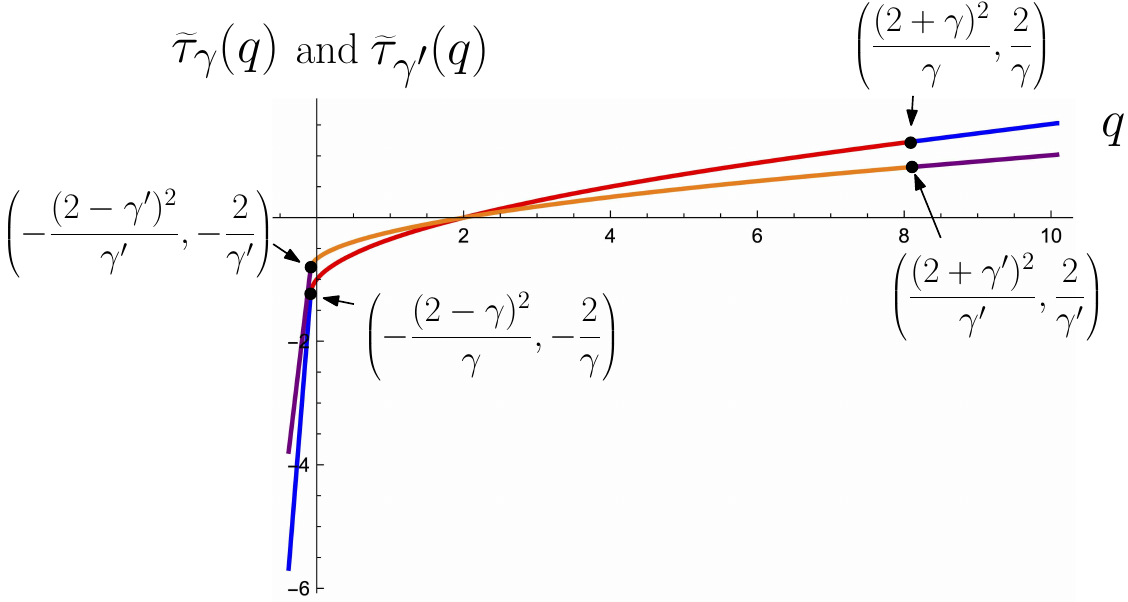}   
   \caption{Red/blue curve: \emph{a.s.}\ $L^q-$spectrum $\widetilde \tau_\gamma(q)$.  Orange/purple curve: dual \emph{a.s.}\ $L^q-$spectrum $\widetilde \tau_{\gamma'}(q)$. The blue and purple parts are linear and all their extensions pass through the origin. Because $Q_\gamma=Q_{\gamma'}$, the abscissae $q_{\pm}=2(2\pm Q_\gamma)=\pm(2\pm\gamma)^2/\gamma
   =\pm(2\pm\gamma')^2/\gamma'$ of the linear transition points are identical for  $\widetilde \tau_\gamma$ and $\widetilde \tau_{\gamma'}$. The ($\gamma=\sqrt{8/3}$, $\gamma'=\sqrt{6}$) dual cases are depicted here.}
 \label{fig:tautildeofq}
\end{figure}

The associated almost sure $L^q-$spectrum $\widetilde \tau_\gamma$ is finally obtained by taking the inverse Legendre transform \eqref{invlegendre} of $\widetilde f_\gamma$ \eqref{tildefas}. For $q\in[q_{-}=2(2-Q_\gamma), q_{+}=2(2+ Q_\gamma)]$, $\widetilde \tau(q)=\Delta_\gamma(q)-1$, whereas outside this interval, one finds two linear branches, 
 $\widetilde \tau(q)=\beta_{-}q$ for $q\leq  q_{-}$, and  $\widetilde \tau(q)=\beta_{+}q$ for $q\geq  q_{+}$. On can check that at the leftmost transition point $q_{-}<0$, the condition \eqref{Deltagamma}  $2q_{-}+a_\gamma^2\geq 0$ is still fulfilled, this self-dual quantity vanishing only at $\gamma=2$ as $(2-\gamma)^4/16$. The resulting \emph{a.s.}\ $L^q-$spectrum is predicted
as
  \begin{equation} \label{tauastilde}
\widetilde \tau_{\gamma}(q)= \begin{cases}
 \frac{2}{\left(2-\gamma\right)^2} q \quad \quad  &\mathrm{if} \quad q\in (-\infty ,-\frac{1}{\gamma}(2-\gamma)^2],\\
\Delta_{\gamma}(q)-1 \quad \quad  &\mathrm{if} \quad q \in[-\frac{1}{\gamma}(2-\gamma)^2, \frac{1}{\gamma}(2+\gamma)^2],\\
 \frac{2}{\left(2+\gamma\right)^2} q\quad  &\mathrm{if} \quad q \in[\frac{1}{\gamma}(2+\gamma)^2,+\infty),
 \end{cases}
 \end{equation} 
 and is illustrated in Fig.~\ref{fig:tautildeofq}. 
\subsection{Multifractality of $\gamma'-$quantum balls for $\gamma'>2$}\label{sec:GammaprimeQballs}
For fixed $z$ in $\mathcal D$, and $\delta>0$, recall the $\delta-$quantum ball $\tilde B^{\delta}(z)$ defined in \eqref{btildeprime} via the  function \eqref{muodot} for $\gamma'>2$, as the largest Euclidean ball  centered at $z$, such that
$\mu_{\odot \gamma'}(\tilde B^{\delta}(z))=\delta.$ 
Its Euclidean radius thus yields a random variable $\mu_0\,(\tilde B^{\delta}(z))^{1/2}$ attached to point $z$, 
where $\mu_{\gamma=0}$ is the (non random) Lebesgue measure. However, as seen in \eqref{deltaodotprime}, the probability that this radius is finite and non zero vanishes in the $\delta\to 0$ limit.  This leads one to define non-singular, conditional expectations under strict positivity of this random variable. One then finds the finite radius moments \cite{2008ExactMethodsBD,2009arXiv0901.0277D,zbMATH06797761} \cite[Sec. 3.2]{DG25}, 
\begin{equation}\label{epsqprime}
\mathbb E\left[\mu_0\,(\tilde B^{\delta}(z))^{q/2}\vert\mu_0(\tilde B^{\delta}(z))>0\right]\asymp\delta^{\Delta_{\gamma'}(q)} \quad \mathrm{as}\quad \delta \to 0,
\end{equation}
with the family of dual exponents,
\begin{align}\label{Deltagammaprime}
&\Delta_{\gamma'}(q)=\frac{1}{\gamma'}\left(2q+a_{\gamma'}^2)^{1/2}-a_{\gamma'}\right), \quad 2q \geq -a_{\gamma'}^2(=-a_{\gamma}^2),\\ \label{agammaprime}
&a_{\gamma'}=\frac{2}{\gamma'}-\frac{\gamma'}{2}=-a_\gamma <0\quad \mathrm{for} \quad\gamma'>2.
\end{align}
\begin{rem}\label{rk:d0}
Because the coefficient $a_{\gamma'}$ is negative,  $\Delta_{\gamma'>2}(0)=-2a_{\gamma'}/\gamma'=1-4/\gamma'^2$, in contradistinction to $\Delta_{\gamma<2}(0)=0$. 
\end{rem}
Note that the same relations hold for dual quantum balls $B^{\delta}$ defined in terms of the \emph{full} dual measure $\mu_{\gamma'}$ \eqref{dualmeasure} and such that $\mu_{\gamma'}(B^{\delta}(z))=\delta$, instead of balls $\tilde B^{\delta}$  \eqref{btildeprime} defined in terms of $\mu_{\gamma'\odot}$. Indeed  \eqref{epsqprime} and \eqref{Deltagammaprime} are nothing but the \emph{dual} KPZ relations \cite{2008ExactMethodsBD,2009arXiv0901.0277D}, rigorously established  in  \cite{pre06228485}. 

We thus define the expected  $L^q-$spectrum in the $\gamma'>2$ case as
\begin{align}\nonumber
\overline{\widetilde \tau}_{\gamma'}(q)&=\underset{\delta\to 0^+}{\lim}\, \frac{1}{\log \delta}\log\left\{\sum_{i\in I} \mathbb E\,\left[\mu_0(B^{\delta}(z_i))^{q/2}\vert\mu_0(B^{\delta}(z_i))>0\right]\right\}\\ \label{exptildetauprime}
&=\Delta_{\gamma'}(q)-1,\quad 2q\geq -a_{\gamma'}^2.
\end{align}
In complete parallel to the computations made in Sec. \ref{Sec:gammaqballs}, starting from the analog \eqref{exptildetaug} of \eqref{exptildetauprime}, we directly obtain the \emph{a.s.}\ dual $\gamma'-$dimension spectrum analogous to \eqref{tildefas},
\begin{align}\label{tildefasprime}
\widetilde f_{\gamma'}(\beta)&=1+\frac{1}{\gamma'}a_{\gamma'}- \frac{1}{2}a_{\gamma'}^2\beta-\frac{1}{2\gamma'^2\beta}\quad \mathrm{for}\quad
 \beta\in\left[\frac{2}{(2+\gamma')^2},\frac{2}{(2-\gamma')^2}\right]\\ \nonumber
&=0\quad \mathrm{otherwise},
\end{align}
which is illustrated in Fig.~\ref{fig:ftildeofbeta}.

Finally, the resulting \emph{a.s.}\ dual $\gamma'-$$L^q$ spectrum is predicted, in analogy to \eqref{tauastilde},  to read 
  \begin{equation} \label{tauastildeprime}
\widetilde \tau_{\gamma'}(q)= \begin{cases}
 \frac{2}{\left(2-\gamma'\right)^2} q \quad \quad  &\mathrm{if} \quad q\in (-\infty ,-\frac{1}{\gamma'}(2-\gamma')^2],\\
\Delta_{\gamma'}(q)-1 \quad \quad  &\mathrm{if} \quad q \in[-\frac{1}{\gamma'}(2-\gamma')^2, \frac{1}{\gamma'}(2+\gamma')^2],\\
 \frac{2}{\left(2+\gamma'\right)^2} q\quad  &\mathrm{if} \quad q \in[\frac{1}{\gamma'}(2+\gamma')^2,+\infty),
 \end{cases}
 \end{equation} 
 and is illustrated in Fig.~\ref{fig:tautildeofq}. 
 
According to the multifractal formalism, the maximum of the dimension spectrum corresponds to the $q=0$ moment, and given as $\underset{\beta \geq 0}{\max} \widetilde f_{\gamma'}(\beta)=-\widetilde \tau_{\gamma'}(0)=1-\Delta_{\gamma'}(0)=4/\gamma'^2<1$ (see Remark \ref{rk:d0}), in contrast to the ($\gamma<2$)-case where $\underset{\beta \geq 0}{\max} \widetilde f_{\gamma}(\beta)=-\widetilde \tau_{\gamma}(0)=1$ (see Fig.~\ref{fig:ftildeofbeta}). It precisely corresponds to the dimension of  the part of the dual LQG surface without localized area, of probability \eqref{deltaodotprime}, \emph{i.e.}, to the `principal bubble' of the quantum surface   \cite{2009arXiv0901.0277D,2008ExactMethodsBD}, or root block of the random map \cite{DG25}. 

A comparison of these LQG multifractal properties with combinatorial discrete results would require using Euclidean embeddings of random planar maps, such as a Tutte embedding \cite{10.1214/20-AOP1487}  or a Smith embedding \cite{10.1214/24-AOP1731}, a subject which goes beyond the scope of this work.

\section*{Acknowledgements} 
We thank Julien Barral, Jean-Marc Luck and Hugo Manet for 
enlightening discussions. EG is partially supported by the ANR  grant CartesEtPlus ANR-23-CE48-0018.
\appendix

\section{Properties of  function $\SF_\frac{1}{\alpha}$}
\label{app:Salpha}
The function $\SF_\frac{1}{\alpha}(x)$ defined in \eqref{eq:formS} is a particular instance of the Wright function
\begin{equation}
W_{\lambda,\mu}(z):=\sum_{m\geq 0}
\frac{z^m}{m!\, \Gamma(\lambda m+\mu)}\ , \quad \lambda >-1\ , \quad \mu\in\mathbb{C}\ .
 \end{equation}
 We have indeed 
 \begin{equation} \label{A2}
 \SF_{\frac{1}{\alpha}}(x)=F_{\frac{1}{\alpha}}(x)\ , \quad F_\nu(z):= W_{-\nu,0}(-z)\ . 
 \end{equation}
 The auxiliary function of the Wright type $F_\nu(z)$ for $0<\nu<1$ (here we have the even sharper range 
$ \frac{1}{2}< \nu=\frac{1}{\alpha}< 1$) with $z\in \mathbb{C}$ was introduced and studied in details 
by Mainardi, who gathers a number of its properties in the Appendix~F of his book \cite{Mainardi} (see also
\cite{math8060884}).
We have the series representation \cite[Eq.~(F.12)]{Mainardi}
\begin{equation}
F_{\frac{1}{\alpha}}(z)= \frac{1}{\pi}\sum_{m\geq 1}(-1)^{m-1} z^m \frac{\Gamma\left(1+\frac{m}{\alpha}\right)}{m!} \sin\left( \frac{\pi m}{\alpha} \right)\ , 
\end{equation}
and the integral representation \cite[Eq.~(F.14)]{Mainardi}
\begin{equation}
F_{\frac{1}{\alpha}}(z)=-\frac{1}{2\hbox{i}\pi} \int_{\mathcal{C}_+\cup\,\mathcal{C}_-}  d\sigma\, e^{\sigma -z\, \sigma^{\frac{1}{\alpha}}}\ ,
\end{equation}
valid for $z\in \mathbb{C}$, with $\mathcal{C}_+\cup\mathcal{C}_-$ representing the negative real contours  as in Fig.~\ref{fig:contournu}. 
The function $F_{\frac{1}{\alpha}}$ is also characterized by the inverse Laplace transform property  \cite[p.249]{Mainardi}
  \begin{equation}
 \begin{split}
\mathcal{L}^{-1}\{e^{-\lambda^{\frac{1}{\alpha}}}\}(y)
&=\frac{1}{2\hbox{i}\pi} \int_{\gamma-\hbox{i}\infty}^{\gamma+\hbox{i}\infty}
d\lambda\, e^{\lambda y -\lambda^{\frac{1}{\alpha}}}\\
&=-\frac{\vartheta(y)}{2\hbox{i}\pi\, y} \int_{\mathcal{C}_+\cup\,\mathcal{C}_-}  
d\sigma\, e^{\sigma -\, (\sigma/y)^{\frac{1}{\alpha}}}\\
&=\frac{\vartheta(y)}{y}F_{\frac{1}{\alpha}}\left(\frac{1}{y^{\frac{1}{\alpha}}}\right)\ .
\end{split}
\label{eq:invLaplace}
\end{equation}
For large real $x$, we have the equivalent \cite[Eqs.~(F.20)-(F.21) with $F_\nu(x)=\nu x\, M_\nu(x)$]{Mainardi}
\begin{equation}
F_{\frac{1}{\alpha}}(x)\underset{x\to \infty}{\sim} \sqrt{\frac{x}{2\pi(\alpha-1)}}\left(\frac{x}{\alpha}\right)^{\frac{1}{2(\alpha-1)}}e^{-(\alpha-1)\left(\frac{x}{\alpha}\right)^{\frac{\alpha}{\alpha-1}}}\ .
\label{eq:largx}
\end{equation}
Finally, we have \cite[Eq.~(F.33)]{Mainardi}
\begin{equation}\label{A7}
\int_0^\infty dx\, x^s F_{\frac{1}{\alpha}}(x)  =\frac{\Gamma(s+2)}{\alpha\, \Gamma\left(\frac{1}{\alpha}(s+1)+1\right)}=\frac{\Gamma(s+1)}{\Gamma\left(\frac{1}{\alpha}(s+1)\right)}
\ , \quad s>-2\ ,
\end{equation}
and in particular, for $s=-\alpha$
\begin{equation}
\int_0^\infty dx\, \frac{1}{x^\alpha} F_{\frac{1}{\alpha}}(x)  =\frac{\Gamma(2-\alpha)}{\alpha\, \Gamma\left(\frac{1}{\alpha}\right)}\ .
\label{eq:normal}
\end{equation}

\section{Block-weighted bicubic maps}
\label{app:bicubic}
\begin{figure}[h]
  \centering
  \fig{.65}{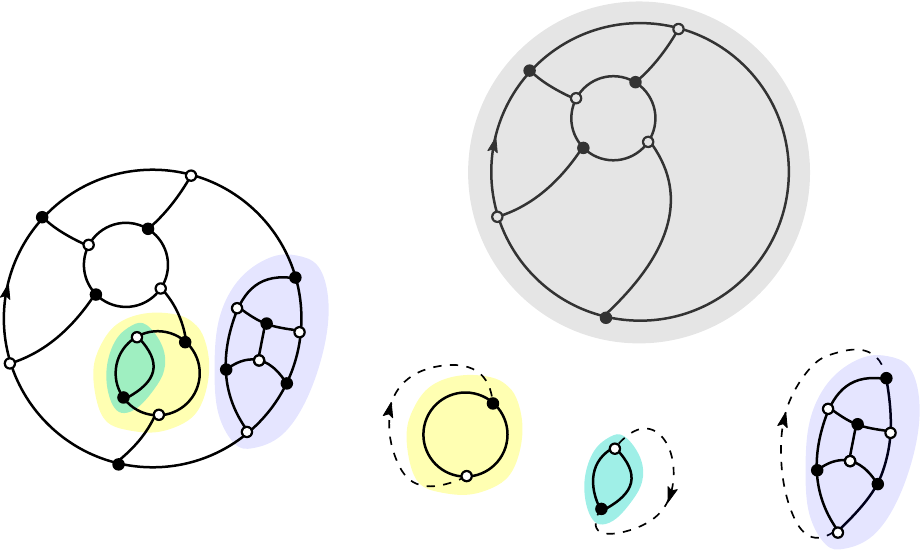}
   \caption{Decomposition of a rooted planar bicubic map (left) into $4$ blocks (right) which are rooted planar \emph{3-connected} 
   bicubic maps, after reconnecting the pending legs of inner blocks by root edges (dashed lines) oriented from white to black vertices. The original map has size $n=10$ and root block size $k=4$. }
  \label{fig:Bicubicmap}
\end{figure}

 A planar \emph{bicubic map} is a planar map whose all vertices have degree $3$ and such that 
its vertex-set can be partitioned into two disjoint subsets so that each edge has its two endpoints in different 
subsets. The enumeration of planar bicubic map is already achieved by Tutte in \cite{Tutte63}, 
which gives the number
$m_n^{(1)}$ of planar rooted bicubic maps with $2n$ vertices (we then say that the map has \emph{size} $n$)
\begin{equation}
m_n^{(1)}=\frac{3}{2}\frac{2^n}{(n+1)(n+2)}\binom{2n}{n}\ , \quad n\geq 1\ .
\end{equation}
Setting $m_0^{(1)}=1$ by convention, one obtains
\begin{equation}\label{eq:B2}
M_1(g):= \sum_{n\geq 0} m_n^{(1)}g^n=1+\frac{1}{32 g^2}\left(-1+12g-24 g^2+(1-8g)^{3/2}\right)\ ,
\end{equation}
which is singular at $g_1=1/8$. As explained in \cite{Tutte63}, planar bicubic maps may be canonically
decomposed into \emph{3-connected blocks} (see Figure~\ref{fig:Bicubicmap} for an illustration); 
  $M_1(g)$ is then related to the generating function $B(t)$ for 3-connected 
planar rooted bicubic maps,  
with a weight $\sqrt{t}$ per vertex,  by the substitution relation
\begin{equation}
M_1(g)=B\left(g M_1(g)^3\right)\ ,
\label{eq:substbicubic}
\end{equation}
expressing the fact that a rooted bicubic map is formed of a 3-connected root block (called the core in 
\cite{Tutte63}) with a possibly empty rooted bicubic map attached to each of the edges of this root block
(in number $3k$ if the root block has $2k$ vertices - we then say that the \emph{root block size} is $k$). 
The generating function $B(t)$, solution of $B^7-B^6-12 t B^4+11t B^3+16 t^2B^2-8t^2B+t^2=0$, is singular at $t_{\crit}=g_1 M_1(g_1)^3$ and has a singular
behavior of the form \eqref{eq:Bsing} with $\alpha=3/2$ and 
\begin{equation}
t_{\crit}=\frac{5^3}{2^9}\ , \quad B(t_{\crit})=\frac{5}{4} \ , \quad B'(t_{\crit})=\frac{2^9}{5^2\, 17}\ , \quad
K_B= \frac{2^{29/2}}{5^2\, 17^{5/2}}\ ,
\end{equation}
as easily seen by developing \eqref{eq:substbicubic} via \eqref{eq:B2} in powers of $g_1-g$, with $g_1=1/8$. 

\begin{figure}[h!]
  \centering
  \fig{1.}{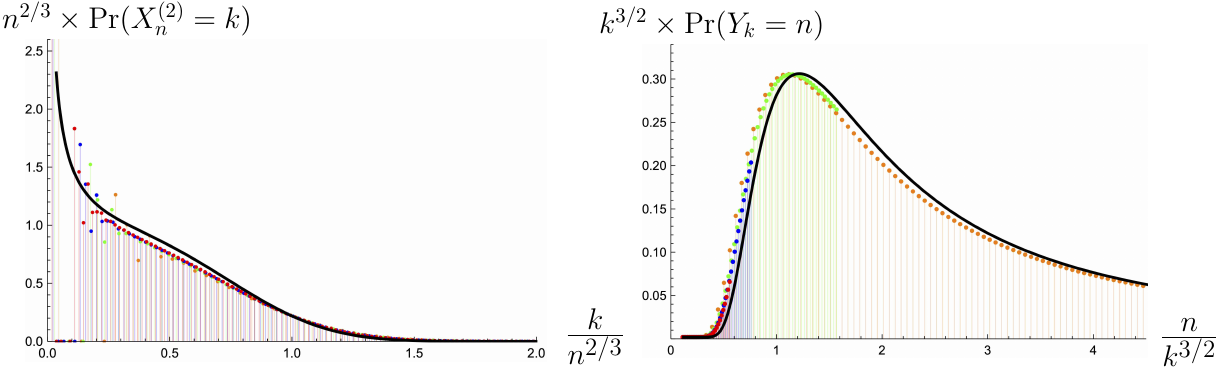}
   \caption{Left: the properly rescaled probability $\Proba(X^{(2)}_n=k)$ for $n=100, 200, 300, 400$ (orange, green, blue, 
   red) in the case of block-weighted bicubic maps and its comparaison with the probability density $
   \sigma(x)$ of \eqref{eq:probscaling2} with $\alpha=3/2$ and $D=17/(5\times 2^{2/3})$. Right: the properly rescaled probability $\Proba(Y_k=n)$ for $k=20, 40, 65, 80$ (orange, green, blue, red) and 
   $n$ limited to $400$ in the case of block-weighted bicubic maps and its comparison with the probability density $\wp(y)$ of \eqref{eq:probscaling3} for $\alpha=3/2$ and 
   $D=17/(5\times 2^{2/3})$.}
  \label{fig:Probabicubic}
\end{figure}

Denoting by $m_n^{(u)}$ 
the generating function of rooted bicubic maps with $2n$ vertices and with a weight $u$ per 3-connected block, 
the generating function $M_u(g)=\sum_{n\geq 0}m_n^{(u)} g^n$ is itself related to $B(t)$ via
\begin{equation}
M_u(g)=1+u\left[B\left(g M_u(g)^3\right)-1\right]\ .
\end{equation}
With this new substitution relation, we recover the subcritical and critical behavior of 
Eqs.~\eqref{eq:subcritical} and \eqref{eq:critical}, with $\gc(u)$ now given (instead of \eqref{eq:guB}) by
\begin{equation}
\gc(u)=\frac{\tc}{(1+u(B(\tc)-1))^3}=\frac{5^3}{8(4+u)^3}\ ,
\label{eq:gubicubic}
\end{equation}
and where the critical value $\uc$ is now given (instead of \eqref{eq:equcrit}) by
\begin{equation}
\uc=\frac{1}{1-B(\tc)+3 \tc\, B'(\tc)}=\frac{68}{43}\ .
\label{eq:equcritbicubic}
\end{equation}
Focusing our study on the dual critical point upon setting $u=\uc$,
and using again the shorthand notation $\gc:=\gc(\uc)$, we obtain from
\eqref{eq:gubicubic} and \eqref{eq:equcritbicubic} the value
\begin{equation}
\gc=\frac{43^3}{2^{15}\, 3^3}\ ,
\end{equation}
satisfying the modified relation (instead of \eqref{eq:equcritbis})
\begin{equation}
\uc=\frac{1}{3 (\tc^2 \gc)^{1/3} \, B'(\tc)}\ .
\label{eq:equcritbisbicubic}
\end{equation}
Writing $g=t/M^3_{\crit}(g)=t/(1+\uc(B(t)-1))^3$ and expanding around $\tc$ upon using \eqref{eq:Bsing}, 
it is easily checked that the relation \eqref{eq:ttogcrit} \emph{remains unchanged}, with in particular
the same literal expression for $C$, with the value here
\begin{equation}
C=\left(\frac{\tc B'(\tc)}{\gc K_B}\right)^{\frac{1}{\alpha}}=\frac{2^{1/3}\, 3^2\, 5^2\, 17}{43^2}\ .
\label{eq:Cbicubic}
\end{equation}
The scaling estimates \eqref{eq:probscaling}, \eqref{eq:probscaling2} 
and \eqref{eq:probscaling3} remain valid, with the same literal expression $\eqref{eq:formD}$ for the scaling parameter
$D$, with the value here
\begin{equation}
D=\frac{1}{\tc}\left(\frac{\tc B'(\tc)}{K_B}\right)^{\frac{1}{\alpha}}=\frac{17}{5\, 2^{2/3}}\ ,
\end{equation}
to be compared with the value $D=3/2^{2/3}$ in \eqref{eq:valguDquad} for block-weighted quadrangulations.
Figure~\ref{fig:Probabicubic} shows a numerical check of the scaling laws \eqref{eq:probscaling2} 
and \eqref{eq:probscaling3} with the new value of $D$ above for block-weighted bicubic maps.

\newpage
\bibliographystyle{plain}
\bibliography{duality-II}

\begin{thebibliography}{10}

\bibitem{AFZ24}
Marie Albenque, \'{E}ric Fusy, and Z\'{e}phyr Salvy.
\newblock {Phase Transition for Tree-Rooted Maps}.
\newblock In C\'{e}cile Mailler and Sebastian Wild, editors, {\em 35th
  International Conference on Probabilistic, Combinatorial and Asymptotic
  Methods for the Analysis of Algorithms (AofA 2024)}, volume 302 of {\em
  Leibniz International Proceedings in Informatics (LIPIcs)}, pages 6:1--6:14,
  Dagstuhl, Germany, 2024. Schloss Dagstuhl -- Leibniz-Zentrum f{\"u}r
  Informatik.

\bibitem{1993NuPhB.394..383A}
L.~{Alvarez-Gaum{\'e}}, J.~L.~F. {Barb{\'o}n}, and {\v C}.~{Crnkovi{\'c}}.
\newblock {A proposal for strings at $D > 1$}.
\newblock {\em Nucl. Phys. B}, 394:383--422, 1993.

\bibitem{1994MPLA....9.1221A}
J.~{Ambj{\o}rn}, B.~{Durhuus}, and T.~{Jonsson}.
\newblock {A Solvable 2d Gravity Model with {$\gamma > 0$}}.
\newblock {\em Mod. Phys. Lett. A}, 9:1221--1228, 1994.

\bibitem{zbMATH06796698}
Juhan Aru, Yichao Huang, and Xin Sun.
\newblock Two perspectives of the 2d unit area quantum sphere and their
  equivalence.
\newblock {\em Commun. Math. Phys.}, 356(1):261--283, 2017.

\bibitem{BFSS}
C.~Banderier, P.~Flajolet, G.~Schaeffer, and M.~Soria.
\newblock Random maps, coalescing saddles, singularity analysis, and {A}iry
  phenomena.
\newblock {\em Random Structures Algorithms}, 19(3-4):194--246, 2001.

\bibitem{1995NuPhB.440..189B}
J.~L.~F. {Barb{\'o}n}, K.~{Demeterfi}, I.~R. {Klebanov}, and C.~{Schmidhuber}.
\newblock {Correlation functions in matrix models modified by wormhole terms}.
\newblock {\em Nucl. Phys. B}, 440:189--214, 1995.

\bibitem{pre06228485}
J.~Barral, X.~Jin, R.~Rhodes, and V.~Vargas.
\newblock {Gaussian multiplicative chaos and KPZ duality.}
\newblock {\em Commun. Math. Phys.}, 323(2):451--485, 2013.

\bibitem{BARRAL2007437}
Julien Barral and Stéphane Seuret.
\newblock The singularity spectrum of {L}évy processes in multifractal time.
\newblock {\em Advances in Mathematics}, 214(1):437--468, 2007.

\bibitem{BARRAL202376}
Julien Barral and Stéphane Seuret.
\newblock The {F}risch-{P}arisi conjecture {I}: Prescribed multifractal
  behavior, and a partial solution.
\newblock {\em Journal de Mathématiques Pures et Appliquées}, 175:76--108,
  2023.

\bibitem{BARRAL2023281}
Julien Barral and Stéphane Seuret.
\newblock The {F}risch-{P}arisi conjecture {II}: Besov spaces in multifractal
  environment, and a full solution.
\newblock {\em Journal de Mathématiques Pures et Appliquées}, 175:281--329,
  2023.

\bibitem{MR3638311}
Dmitry Beliaev, Bertrand Duplantier, and Michel Zinsmeister.
\newblock Integral means spectrum of whole-plane {SLE}.
\newblock {\em Commun. Math. Phys.}, 353(1):119--133, 2017.

\bibitem{BS}
Dmitry Beliaev and Stanislas Smirnov.
\newblock Harmonic measure and {SLE}.
\newblock {\em Commun. Math. Phys.}, 290(2):577--595, 2009.

\bibitem{10.1214/22-EJP893}
Federico Bertacco.
\newblock {Multifractal analysis of Gaussian multiplicative chaos and
  applications}.
\newblock {\em Electronic Journal of Probability}, 28(none):1 -- 36, 2023.

\bibitem{Bertacco_2025}
Federico Bertacco.
\newblock On {G}aussian multiplicative chaos and random geometry - {P}h{D}
  {T}hesis, {I}mperial {C}ollege {L}ondon, July 1st, 2025.

\bibitem{10.1214/24-AOP1731}
Federico Bertacco, Ewain Gwynne, and Scott Sheffield.
\newblock {Scaling limits of planar maps under the Smith embedding}.
\newblock {\em The Annals of Probability}, 53(3):1138 -- 1196, 2025.

\bibitem{zbMATH06337014}
Ga{\"e}tan Borot.
\newblock Formal multidimensional integrals, stuffed maps, and topological
  recursion.
\newblock {\em Ann. Inst. Henri Poincar{\'e} D, Comb. Phys. Interact.},
  1(2):225--264, 2014.

\bibitem{Minbus}
J\'er\'emie Bouttier and Emmanuel Guitter.
\newblock Distance statistics in quadrangulations with no multiple edges and
  the geometry of minbus.
\newblock {\em Journal of Physics A: Mathematical and Theoretical},
  43(20):205207, apr 2010.

\bibitem{1990MPLA....5.1041D}
S.~R. {Das}, A.~{Dhar}, A.~M. {Sengupta}, and S.~R. {Wadia}.
\newblock {New critical behavior in $d = 0$ large-$N$ matrix models}.
\newblock {\em Mod. Phys. Lett. A}, 5:1041--1056, 1990.

\bibitem{MR981529}
F.~David.
\newblock Conformal field theories coupled to {$2$}-{D} gravity in the
  conformal gauge.
\newblock {\em Mod. Phys. Lett. A}, 3(17):1651--1656, 1988.

\bibitem{MR3465434}
Fran\c{c}ois David, Antti Kupiainen, R\'emi Rhodes, and Vincent Vargas.
\newblock Liouville quantum gravity on the {R}iemann sphere.
\newblock {\em Comm. Math. Phys.}, 342(3):869--907, 2016.

\bibitem{MR1320471}
P.~Di~Francesco, P.~Ginsparg, and J.~Zinn-Justin.
\newblock {$2$}{D} gravity and random matrices.
\newblock {\em Phys. Rep.}, 254:1--133, 1995.

\bibitem{MR1005268}
J.~Distler and H.~Kawai.
\newblock Conformal field theory and {$2$}{D} quantum gravity.
\newblock {\em Nucl. Phys. B}, 321:509--527, 1989.

\bibitem{2000PhRvL..84.1363D}
B.~Duplantier.
\newblock {Conformally invariant fractals and potential theory}.
\newblock {\em Phys. Rev. Lett.}, 84:1363--1367, 2000.

\bibitem{MR2112128}
B.~Duplantier.
\newblock Conformal fractal geometry \& boundary quantum gravity.
\newblock In {\em Fractal geometry and applications: a jubilee of Beno\^\i t
  Mandelbrot, Part 2}, volume~72 of {\em Proc. Sympos. Pure Math.}, pages
  365--482. Amer. Math. Soc., Providence, RI, 2004.

\bibitem{2008ExactMethodsBD}
B.~{Duplantier}.
\newblock {A rigorous perspective on Liouville quantum gravity and the KPZ
  relation}.
\newblock In S.~Ouvry, J.~Jacobsen, V.~Pasquier, D.~Serban, and L.~Cugliandolo,
  editors, {\em Exact methods in low-dimensional statistical physics and
  quantum theory (Les Houches Summer School, Session LXXXIX, 2008)}, pages
  529--561. Oxford University Press, Great Clarendon Street, 2009.

\bibitem{2009arXiv0901.0277D}
B.~{Duplantier} and S.~{Sheffield}.
\newblock {Duality and KPZ in Liouville Quantum Gravity}.
\newblock {\em Phys. Rev. Lett.}, 102:150603, 2009.

\bibitem{2008arXiv0808.1560D}
B.~{Duplantier} and S.~{Sheffield}.
\newblock {Liouville quantum gravity and KPZ}.
\newblock {\em Invent. Math.}, 185:333--393, 2011.

\bibitem{Duplantier00}
Bertrand Duplantier.
\newblock Conformally invariant fractals and potential theory.
\newblock {\em Phys. Rev. Lett.}, 84(7):1363--1367, 2000.

\bibitem{zbMATH06797761}
Bertrand Duplantier.
\newblock Liouville quantum gravity, {KPZ} and {Schramm}-{Loewner} evolution.
\newblock In {\em Proceedings of the international congress of mathematicians
  (ICM 2014), Seoul, Korea, August 13--21, 2014. Vol. III: Invited lectures},
  pages 1035--1061. Seoul: KM Kyung Moon Sa, 2014.

\bibitem{DG25}
Bertrand Duplantier and Emmanuel Guitter.
\newblock Liouville quantum duality and random planar maps.
\newblock {\em arXiv:2507.12203 [math-ph]}, 2025.

\bibitem{DMS14}
Bertrand {Duplantier}, Jason {Miller}, and Scott {Sheffield}.
\newblock {Liouville Quantum Gravity as a Mating of Trees.}
\newblock {\em {Ast\'erisque.}}, 427:1--258, 2021.

\bibitem{DRSV12}
Bertrand Duplantier, R{\'e}mi Rhodes, Scott Sheffield, and Vincent Vargas.
\newblock {Renormalization of Critical Gaussian Multiplicative Chaos and KPZ
  Relation}.
\newblock {\em Commun. Math. Phys.}, 330(1):283 -- 330, 2014.

\bibitem{1994NuPhB.426..203D}
B.~{Durhuus}.
\newblock {Multi-spin systems on a randomly triangulated surface}.
\newblock {\em Nucl. Phys. B}, 426:203--222, 1994.

\bibitem{Fisher66}
M.~E. Fisher.
\newblock Shape of a self‐avoiding walk or polymer chain.
\newblock {\em J. Chem. Phys.}, 44:616--622, 1966.

\bibitem{FSbook}
Philippe Flajolet and Robert Sedgewick.
\newblock {\em Analytic Combinatorics}.
\newblock Cambridge University Press, 2009.

\bibitem{FS24}
William Fleurat and Z{\'e}phyr Salvy.
\newblock {A phase transition in block-weighted random maps}.
\newblock {\em Electronic Journal of Probability}, 29(none):1 -- 61, 2024.

\bibitem{Ginsparg-Moore}
P.~{Ginsparg} and G.~{Moore}.
\newblock Lectures on 2{D} gravity and 2{D} string theory ({TASI} 1992).
\newblock In J.~{Harvey} and J.~{Polchinski}, editors, {\em Recent direction in
  particle theory, Proceedings of the 1992 TASI}. World Scientific, Singapore,
  1993.

\bibitem{PhysRevLett.66.2051}
M.~Goulian and M.~Li.
\newblock Correlation functions in {L}iouville theory.
\newblock {\em Phys. Rev. Lett.}, 66:2051--2055, Apr 1991.

\bibitem{10.1214/20-AOP1487}
Ewain Gwynne, Jason Miller, and Scott Sheffield.
\newblock {The Tutte embedding of the mated-CRT map converges to Liouville
  quantum gravity}.
\newblock {\em The Annals of Probability}, 49(4):1677 -- 1717, 2021.

\bibitem{gwynne2018}
Ewain Gwynne, Jason Miller, and Xin Sun.
\newblock Almost sure multifractal spectrum of {S}chramm-{L}oewner evolution.
\newblock {\em Duke Math. J.}, 167(6):1099--1237, 2018.

\bibitem{1986PhRvA..33.1141H}
T.~C. {Halsey}, M.~H. {Jensen}, L.~P. {Kadanoff}, I.~{Procaccia}, and B.~I.
  {Shraiman}.
\newblock {Fractal measures and their singularities - The characterization of
  strange sets}.
\newblock {\em Phys. Rev. A}, 33:1141--1151, 1986.

\bibitem{PhysRevA.34.1601}
T.~C. {Halsey}, M.~H. {Jensen}, L.~P. {Kadanoff}, I.~{Procaccia}, and B.~I.
  {Shraiman}.
\newblock Fractal measures and their singularities: The characterization of
  strange sets; {E}rratum: [{P}hys. {R}ev. {A} 33, 1141 (1986)].
\newblock {\em Phys. Rev. A}, 34:1601--1601, 1986.

\bibitem{zbMATH03856061}
H.~G.~E. Hentschel and Itamar Procaccia.
\newblock The infinite number of generalized dimensions of fractals and strange
  attractors.
\newblock {\em Physica D}, 8:435--444, 1983.

\bibitem{MR0292433}
R.~H{\o}egh-Krohn.
\newblock A general class of quantum fields without cut-offs in two space-time
  dimensions.
\newblock {\em Commun. Math. Phys.}, 21:244--255, 1971.

\bibitem{1992PhLB..286..239J}
S.~{Jain} and S.~D. {Mathur}.
\newblock {World-sheet geometry and baby universes in 2D quantum gravity.}
\newblock {\em Phys. Lett. B}, 286:239--246, 1992.

\bibitem{MR829798}
J.-P. Kahane.
\newblock Sur le chaos multiplicatif.
\newblock {\em Ann. Sci. Math. Qu\'ebec}, 9(2):105--150, 1985.

\bibitem{1995PhRvD..51.1836K}
I.~R. {Klebanov}.
\newblock {Touching random surfaces and Liouville gravity}.
\newblock {\em Phys. Rev. D}, 51:1836--1841, 1995.

\bibitem{1995NuPhB.434..264K}
I.~R. {Klebanov} and A.~{Hashimoto}.
\newblock {Non-perturbative solution of matrix models modified by trace-squared
  terms}.
\newblock {\em Nucl. Phys. B}, 434:264--282, 1995.

\bibitem{1996NuPhS..45..135K}
I.~R. {Klebanov} and A.~{Hashimoto}.
\newblock {Wormholes, matrix models, and Liouville gravity}.
\newblock {\em Nucl. Phys. B Proc. Suppl.}, 45:135--148, 1996.

\bibitem{MR947880}
V.~G. Knizhnik, A.~M. Polyakov, and A.~B. Zamolodchikov.
\newblock Fractal structure of {$2$}{D}-quantum gravity.
\newblock {\em Mod. Phys. Lett. A}, 3:819--826, 1988.

\bibitem{1992PhLB..296..323K}
G.~P. {Korchemsky}.
\newblock {Loops in the curvature matrix model}.
\newblock {\em Phys. Lett. B}, 296:323--334, 1992.

\bibitem{1992MPLA....7.3081K}
G.~P. {Korchemsky}.
\newblock {Matrix model perturbed by higher order curvature terms}.
\newblock {\em Mod. Phys. Lett. A}, 7:3081--3100, 1992.

\bibitem{Mainardi}
Francesco Mainardi.
\newblock {\em Fractional Calculus and Waves in Linear Viscoelasticity}.
\newblock Imperial College Press, 2010.

\bibitem{math8060884}
Francesco Mainardi and Armando Consiglio.
\newblock The {W}right functions of the second kind in mathematical physics.
\newblock {\em Mathematics}, 8(6), 2020.

\bibitem{mandelbrot1974multiplications}
B.~B. Mandelbrot.
\newblock Multiplications al{\'e}atoires it{\'e}r{\'e}es et distributions
  invariantes par moyenne pond{\'e}r{\'e}e al{\'e}atoire.
\newblock {\em {C}omptes {R}endus ({P}aris)}, 278{A}:289--292 and 355--358,
  1974.

\bibitem{Mandelbrot1974331}
Beno\^{\i}t~B. Mandelbrot.
\newblock Intermittent turbulence in self-similar cascades: Divergence of high
  moments and dimension of the carrier.
\newblock {\em Journal of Fluid Mechanics}, 62(2):331 – 358, 1974.

\bibitem{FP}
G.~{Parisi} and U.~{Frisch}.
\newblock {On the singularity structure of fully developed turbulence}.
\newblock In M.~Ghil, R.~R.~Benzi, and G.~Parisi, editors, {\em Turbulence and
  predictability in geophysical fluid dynamics and climate dynamics,
  Proceedings of the International School of Physics Enrico Fermi, course
  LXXXVIII}, pages 84--87. North Holland, New York, 1985.

\bibitem{MR623209}
A.~M. Polyakov.
\newblock Quantum geometry of bosonic strings.
\newblock {\em Phys. Lett. B}, 103:207--210, 1981.

\bibitem{PSS:8474530}
R.~{Rhodes} and V.~{Vargas}.
\newblock {KPZ} formula for log-infinitely divisible multifractal random
  measures.
\newblock {\em ESAIM: Probability and Statistics}, 15:358--371, 2011.

\bibitem{RV3}
R\'emi Rhodes and Vincent Vargas.
\newblock Gaussian multiplicative chaos and applications: a review.
\newblock {\em Probab. Surv.}, 11:315--392, 2014.

\bibitem{RV}
Raoul Robert and Vincent Vargas.
\newblock Gaussian multiplicative chaos revisited.
\newblock {\em Ann. Probab.}, 38(2):605--631, 2010.

\bibitem{ZS23}
Zéphyr Salvy.
\newblock Unified study of the phase transition for block-weighted random
  planar maps.
\newblock {\em Eurocomb’23}, page 790–798, 2023.

\bibitem{ZSPhD}
Zéphyr Salvy.
\newblock {\em Unified study of block-weighted planar maps: combinatorial and
  probabilistic properties}.
\newblock Ph{D} thesis, Universit{\'e} Gustave Eiffel, Dec 2024.
\newblock Available at \url{https://igm.univ-mlv.fr/~salvy/data/manuscrit.pdf}.

\bibitem{MR3551203}
Scott Sheffield.
\newblock Conformal weldings of random surfaces: {SLE} and the quantum gravity
  zipper.
\newblock {\em Ann. Probab.}, 44(5):3474--3545, 2016.

\bibitem{Tutte63}
W.~T. Tutte.
\newblock A census of planar maps.
\newblock {\em Canadian Journal of Mathematics}, 15:249–271, 1963.

\end{thebibliography}
\end{document}